\documentclass[english]{article}
\usepackage[T1]{fontenc}
\usepackage{color}
\usepackage{array}
\usepackage{graphicx}
\usepackage{amssymb}

\makeatletter

\providecommand{\tabularnewline}{\\}

\usepackage{a4}
\input{epsfig.sty}
\def\fnum@table{\tablename~{\bf\thetable}}
\def\fnum@figure{\figurename~{\bf\thefigure}}
\def\tablename{\footnotesize{\bf Table}}
\def\figurename{\footnotesize{\bf Figure}}

\def\be{\begin{equation}}
\def\ee{\end{equation}}

\makeatother

\usepackage{babel}

\begin{document}

\title{\textbf{Total and diffractive cross sections in enhanced Pomeron
scheme}}

\author{\textbf{S. Ostapchenko}%
\thanks{e-mail: sergey.ostapchenko@ntnu.no%
} \\
\textit{\small NTNU, Institut}{\small{} }\textit{\textcolor{black}{\small for
fysikk}}\textit{\small , 7021 Trondheim, Norway}\\
\textit{\small D.V. Skobeltsyn Institute of Nuclear Physics, Moscow
State University, 119992 Moscow, Russia}\textit{ }\\
}

\maketitle
\begin{center}
\textbf{\large Abstract}
\par\end{center}{\large \par}

For the first time, a systematic analysis of the high energy behavior
of total and diffractive proton-proton cross sections is performed
within the Reggeon Field Theory framework, based on the resummation
of all significant contributions of enhanced Pomeron diagrams to all
orders with respect to the triple-Pomeron coupling. The importance
of different classes of enhanced graphs is investigated and it is
demonstrated that absorptive corrections due to ``net''-like enhanced
diagrams and due to Pomeron ``loops'' are both significant and none
of those classes can be neglected at high energies. A comparison with
other approaches based on partial resummations of enhanced diagrams
is performed. In particular, important differences are found concerning
the predicted high energy behavior of total and single high mass diffraction
proton-proton cross sections, with our values of $\sigma_{pp}^{{\rm tot}}$
at $\sqrt{s}=14$ TeV being some $25\div40$\% higher and with the
energy rise of $\sigma_{{\rm HM}}^{{\rm SD}}$ saturating well below
the LHC energy. The main causes for those differences are analyzed
and explained.

\section{Introduction\label{intro.sec} }

Since long time theoretical investigations of high energy hadronic
interactions remain at the frontier of particle physics research.
This is especially true in the present situation, when the Large Hadron
Collider (LHC) is becoming operative and new experimental installations
are constructed for the studies of ultra-high energy cosmic rays.
The ultimate goal of the corresponding theoretical research is to
provide a fully microscopic description of the dynamics of hadronic
collisions in the framework of the perturbative quantum chromodynamics
(pQCD). However, at the present stage many important aspects of hadronic
interactions can be addressed with effective approaches only. In particular,
calculating total hadron-hadron cross section or treating diffractive
particle production, one necessarily has to deal with important contributions
from nonperturbative soft processes, which can not be described within
the pQCD framework. The traditional phenomenological approach to such
problems is provided by Gribov's Reggeon Field Theory (RFT) \cite{gri68}.
In the RFT framework, hadron-hadron scattering in the high energy
limit is dominated by multiple Pomeron exchanges, each Pomeron corresponding
to an underlying microscopic parton cascade developing between the
projectile and target hadrons. Within certain approximations, the
scheme allows one to calculate various hadronic cross sections and
to determine partial weights of hadronic final states of interest.
This provided the ground for successful theoretical models for hadron-hadron
scattering \cite{kai82} and for corresponding Monte Carlo (MC)
generators of high energy interactions \cite{wer93}.
However, proceeding to sufficiently high energies, one has to account
for nonlinear contributions to the underlying parton dynamics, described
by enhanced Pomeron diagrams \cite{kan73,car74}. The
difficulty which arises here is that enhanced graphs of more and more
complicated topologies become increasingly important with rising energy.
Thus, all-order resummation of enhanced contributions is a necessary
condition for a self-consistent model.

A procedure for an asymptotic resummation of enhanced Pomeron diagrams
has been proposed in \cite{kai86}. More recently, extensive studies
of total and diffractive proton-proton cross sections have been performed
in \cite{kmr08,kmr09}, including applications of the
formalism for calculations of rapidity gap survival (RGS) probabilities.
The corresponding treatment is based on a resummation of contributions
of ``net''-like enhanced graphs to elastic scattering amplitude, neglecting
Pomeron ``loop'' diagrams. However, the employed parametrization for
multi-Pomeron vertices implies a somewhat artificial hierarchy for
the underlying parton cascades. On the other hand, the expressions
for diffractive cross sections are derived in \cite{kmr08}
from heuristic arguments, being in an explicit conflict with the traditional
RFT treatment.

Similar studies of hadronic cross sections and of RGS probabilities
have been performed also in \cite{glm08}, taking into account
Pomeron loop contributions but neglecting net-like enhanced graphs.
Thus, the approach is internally inconsistent: while cut diagrams
of ``fan'' type (a subclass of general net-like graphs) are taken
into consideration when calculating diffractive cross sections, those
are neglected in the elastic scattering amplitude. In addition, the
authors of Ref.~\cite{glm08} restricted themselves with the
triple-Pomeron vertex only, neglecting other types of multi-Pomeron
vertices. Such an approach is known to predict total hadron-hadron
cross sections which become constant in the very high energy limit.

A general approach to the resummation of enhanced Pomeron diagrams
has been proposed in \cite{ost06,ost06a,ost08}, both for the corresponding
contributions to elastic scattering amplitude and to partial cross
sections of particular final states. In this paper, we apply the formalism
for calculations of total and diffractive proton-proton cross sections
at high energies. The principal differences of the present investigation
compared to previous ones \cite{kmr08,kmr09,glm08}
 are i) complete resummation of all important enhanced contributions
to cross sections of interest, ii) self-consistent analysis of the
structure of diffractive final states, based on the Abramovskii-Gribov-Kancheli
(AGK) \cite{agk} cutting rules. The structure of the paper is as
follows. In Section \ref{model.sec}, we describe our model. In Section
\ref{sec:High-mass-diffraction} and in the Appendix, we obtain expressions
for high mass diffraction cross sections. In Section \ref{sec:Numerical-results},
we present our results for total and diffractive proton-proton cross
sections and investigate relative importance of various classes of
enhanced graphs. In Section \ref{sec:Comparison-with-other}, the
differences with other approaches are analyzed. Finally, the conclusions
are presented in Section \ref{sec:Outlook}.

\section{The model\label{model.sec} }

\subsection{Multi-channel eikonal approach}

\begin{figure}[htb]
\begin{centering}
\includegraphics[width=7cm,height=3cm]{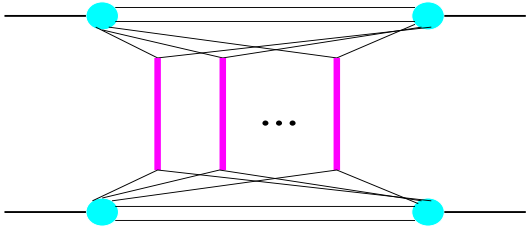}
\par\end{centering}

\caption{General multi-Pomeron contribution to hadron-hadron scattering amplitude;
elementary scattering processes (vertical thick lines) are described
as Pomeron exchanges.\label{multiple}}

\end{figure}

In the RFT framework, hadron-hadron scattering is dominated in the
high energy limit by multiple Pomeron exchanges between the projectile
and target hadrons, as depicted in Fig.~\ref{multiple}.%
 Using eikonal Pomeron-hadron vertices and employing multi-channel
scheme to account for contributions of inelastic intermediate hadronic
states of small masses%
\footnote{Here we restrict ourselves with low mass inelastic states in order
to use an energy-independent decomposition of proton wave function
in terms of the corresponding elastic scattering eigenstates.%
} between Pomeron emissions, elastic proton-proton scattering amplitude
$T(s,t)$ can be expressed as \cite{kai79,kai01}
\begin{eqnarray}
T(s,t)=2s\,\int\! d^{2}b\; e^{i\vec{q}\vec{b}}\, f(s,b)=2is\,\int\! d^{2}b\; e^{i\vec{q}\vec{b}}\,\sum_{j,k}C_{j}C_{k}\,\left[1-e^{-\chi_{jk}^{\mathbb{P}}(s,b)}\right].\label{f_ad}\end{eqnarray}
Here $s$, $t=q^{2}$, and $b$ are correspondingly c.m.~energy squared,
momentum transfer, and impact parameter for the interaction, $f(s,b)$
- elastic scattering amplitude in the impact parameter representation.
$C_{j}$ defines partial weight for proton elastic scattering eigenstate
$|j\rangle$ ($|p\rangle=\sum_{j}\sqrt{C_{j}}|j\rangle$, $\sum_{j}C_{j}=1$)
and $\chi_{jk}^{\mathbb{P}}$ is the so-called eikonal corresponding
to Pomeron exchange between the projectile and target protons, the
latter being represented by eigenstates $|j\rangle$ and $|k\rangle$.
The eikonal $\chi_{jk}^{\mathbb{P}}$ is expressed via the Pomeron
propagator $D^{\mathbb{P}}(s,t)$ and the vertex $F_{j}^{\mathbb{P}}(t)$
for Pomeron emission by eigenstate $|j\rangle$ as \begin{eqnarray}
\chi_{jk}^{\mathbb{P}}(s,b)=\frac{1}{8\pi^{2}is}\,\int\! d^{2}q\; e^{-i\vec{q}\vec{b}}\, F_{j}^{\mathbb{P}}(q^{2})\, F_{k}^{\mathbb{P}}(q^{2})\, D^{\mathbb{P}}(s,q^{2}).\label{chi-pom}\end{eqnarray}

In this work, we use 2-component model ($j,k=1,2)$ and employ the
dipole parametrization for the $t$-dependence of $F_{j}^{\mathbb{P}}$:\begin{equation}
F_{j}^{\mathbb{P}}(t)=\frac{\gamma_{j}}{(1-\Lambda_{j}t)^{2}}\,.\label{eq:F-pom}\end{equation}

Concerning $D^{\mathbb{P}}(s,t)$, we assume it to receive contributions
from two Pomeron poles: a ``soft'' Pomeron, which corresponds to the
underlying cascade of partons of small virtualities, and a ``hard''
one, which is dominated by relatively hard partons of higher $\langle|q^{2}|\rangle$.
Neglecting small real parts of Pomeron pole amplitudes and introducing
a parameter $r_{{\rm h/s}}\ll1$ for the ratio of proton couplings
to the hard and soft Pomerons, we have\begin{equation}
D^{\mathbb{P}}(s,t)=8\pi i\left[s^{\alpha_{\mathbb{P}{\rm (s)}}}\; e^{\alpha'_{\mathbb{P}{\rm (s)}}\ln\! s\; t}+r_{{\rm h/s}}\, s^{\alpha_{\mathbb{P}{\rm (h)}}}\; e^{\alpha'_{\mathbb{P}{\rm (h)}}\ln\! s\; t}\right],\label{eq:D-pom}\end{equation}
with $\alpha_{\mathbb{P}{\rm (s/h)}}$ and $\alpha_{\mathbb{P}{\rm (s/h)}}'$
being intercepts and slopes of the Pomeron Regge trajectories. In
the following we assume both Pomeron trajectories to be overcritical,
$\Delta_{{\rm s/h}}=\alpha_{\mathbb{P}{\rm (s/h)}}-1>0$, with the
hard Pomeron characterized by a steeper energy dependence, $\Delta_{h}>\Delta_{s}$,
and a smaller slope, $\alpha_{\mathbb{P}{\rm (h)}}'<\alpha_{\mathbb{P}{\rm (s)}}'$.

Knowing the elastic amplitude, one can easily calculate total and
elastic cross sections and the elastic scattering slope: \begin{eqnarray}
\sigma^{{\rm tot}}(s)=2\,{\rm Im}\int\!\! d^{2}b\, f(s,b)=2\int\! d^{2}b\;\sum_{j,k}C_{j}C_{k}\,\left(1-e^{-\chi_{jk}^{\mathbb{P}}(s,b)}\right)\label{sigma-tot}\\
\sigma^{{\rm el}}(s)=\int\!\! d^{2}b\,\left|f(s,b)\right|^{2}=\int\! d^{2}b\;\left[\sum_{j,k}C_{j}C_{k}\,\left(1-e^{-\chi_{jk}^{\mathbb{P}}(s,b)}\right)\right]^{2}\label{eq:sigma-el}\\
B^{{\rm el}}(s)=\left.\frac{d\ln\!\left(d\sigma^{{\rm el}}(s,t)/dt\right)}{dt}\right|_{t=0}=\frac{2}{\sigma^{{\rm tot}}(s)}\int\! d^{2}b\; b^{2}\,\sum_{j,k}C_{j}C_{k}\,\left(1-e^{-\chi_{jk}^{\mathbb{P}}(s,b)}\right).\label{eq:B-el}\end{eqnarray}

Moreover, making use of the optical theorem and of the AGK cutting
rules \cite{agk}, one can relate contributions of certain unitarity
cuts of elastic scattering diagrams of Fig.~\ref{multiple} to partial
cross sections of particular hadronic final states. For example, cross
sections for single and double low mass diffractive dissociation are
obtained considering cut diagrams in which the cut plane passes between
$m\geq2$ uncut Pomerons and choosing in the cut plane inelastic intermediate
states for one or for both protons correspondingly \cite{kai79}:\begin{eqnarray}
\sigma_{{\rm LM}}^{{\rm SD}}(s)=2\int\! d^{2}b\;\sum_{j,k,l,m}(C_{j}\,\delta_{jl}-C_{j}C_{l})\, C_{k}C_{m}\,\left(1-e^{-\chi_{jk}^{\mathbb{P}}(s,b)}\right)\left(1-e^{-\chi_{lm}^{\mathbb{P}}(s,b)}\right)\label{sigma-sd(lm)}\\
\sigma_{{\rm LM}}^{{\rm DD}}(s)=\int\! d^{2}b\;\sum_{j,k,l,m}(C_{j}\,\delta_{jl}-C_{j}C_{l})\,(C_{k}\,\delta_{km}-C_{k}C_{m})\,\left(1-e^{-\chi_{jk}^{\mathbb{P}}(s,b)}\right)\left(1-e^{-\chi_{lm}^{\mathbb{P}}(s,b)}\right).\label{eq:sigma-dd(lm)}\end{eqnarray}

\subsection{Enhanced diagram contributions to elastic scattering amplitude}

The above-discussed picture corresponds to multiple parton cascades
developing independently between the projectile and target hadrons.
However, proceeding to sufficiently high energies, one inevitably
faces the situation when such cascades overlap and influence each
other. Such nonlinear effects are traditionally described by enhanced
Pomeron diagrams which account for Pomeron-Pomeron interactions
 \cite{kan73,car74,kai86}.
Taking such contributions into account, hadronic cross sections are
still given by Eqs.~(\ref{sigma-tot}-\ref{eq:sigma-dd(lm)}), however,
with the simple Pomeron exchange eikonal $\chi_{jk}^{\mathbb{P}}(s,b)$
being replaced by half the total opacity\begin{equation}
\frac{1}{2}\Omega_{jk}(s,b)=\chi_{jk}^{\mathbb{P}}(s,b)+\chi_{jk}^{{\rm enh}}(s,b)\,,\label{eq:opac}\end{equation}
where the eikonal $\chi_{jk}^{{\rm enh}}(s,b)$ represents the contribution
of all irreducible enhanced Pomeron graphs exchanged between the eigenstates
$|j\rangle$ and $|k\rangle$ of the projectile and target protons. 

A general procedure for a resummation of enhanced Pomeron diagrams,
both for elastic scattering amplitude and for corresponding unitarity
cuts, has been proposed in \cite{ost06,ost08}. In this work, we consider
eikonal multi-Pomeron vertices, assuming the vertex slope to be small
and neglecting it in the following. Thus, the transition of $m$ into
$n$ Pomerons ($m+n\geq3$) is described by a vertex \cite{kai86}\begin{equation}
G^{(m,n)}=G\,\gamma_{\mathbb{P}}^{m+n},\label{eq:g_mn}\end{equation}
the constant $G$ being related to the triple-Pomeron coupling $r_{3\mathbb{P}}$
as $G=r_{3\mathbb{P}}/(4\pi\gamma_{\mathbb{P}}^{3})$.%
\footnote{In the limit $\gamma_{\mathbb{P}}\rightarrow0$, $r_{3\mathbb{P}}$
being fixed, one arrives to the scheme with the triple-Pomeron vertex
only.%
} Hence, for a Pomeron exchanged between the projectile or the target
proton (represented by diffractive eigenstate $|j\rangle$) and a
multi-Pomeron vertex separated from it by rapidity and impact parameter
distances $y$ and $b$ we obtain the eikonal\begin{eqnarray}
\chi_{j}^{\mathbb{P}}(y,b)=\frac{\gamma_{\mathbb{P}}}{8\pi^{2}ie^{y}}\,\int\! d^{2}q\; e^{-i\vec{q}\vec{b}}\, F_{j}^{\mathbb{P}}(q^{2})\, D^{\mathbb{P}}(e^{y},q^{2}),\label{chi-legpom}\end{eqnarray}
where we included the vertex factor $\gamma_{\mathbb{P}}$ in the
definition of the eikonal. Similarly, for a Pomeron exchanged between
two vertices separated by rapidity and impact parameter distances
$y$ and $b$ we have\begin{eqnarray}
\chi^{\mathbb{P}}(y,b)=\frac{\gamma_{\mathbb{P}}^{2}}{8\pi^{2}ie^{y}}\,\int\! d^{2}q\; e^{-i\vec{q}\vec{b}}\, D^{\mathbb{P}}(e^{y},q^{2})\,.\label{chi-intpom}\end{eqnarray}

As demonstrated in \cite{ost06}, a general irreducible contribution
of arbitrary net-like enhanced diagrams (i.e.~all possible enhanced
graphs with the exception of Pomeron loop contributions) can be written
down in a rather compact form, being expressed via contributions of
sub-graphs of certain structure, so-called ``net-fans''. The latter
are defined by a Schwinger-Dyson equation which resembles to some
extent the usual fan diagram equation (therefore, the name -- ``net-fans'')
and correspond to arbitrary irreducible nets of Pomerons, exchanged
between the projectile and target hadrons, with one vertex in the
net having a fixed position in the rapidity and impact parameter space.
Furthermore, it has been shown in \cite{ost08} that the scheme can
be generalized to include rather general Pomeron loop contributions
by merely replacing single Pomerons connecting neighboring net ``cells''
by arbitrary 2-point sequences of Pomerons and Pomeron loops.

\begin{figure}[t]
\begin{centering}
\includegraphics[width=6cm,height=6cm]{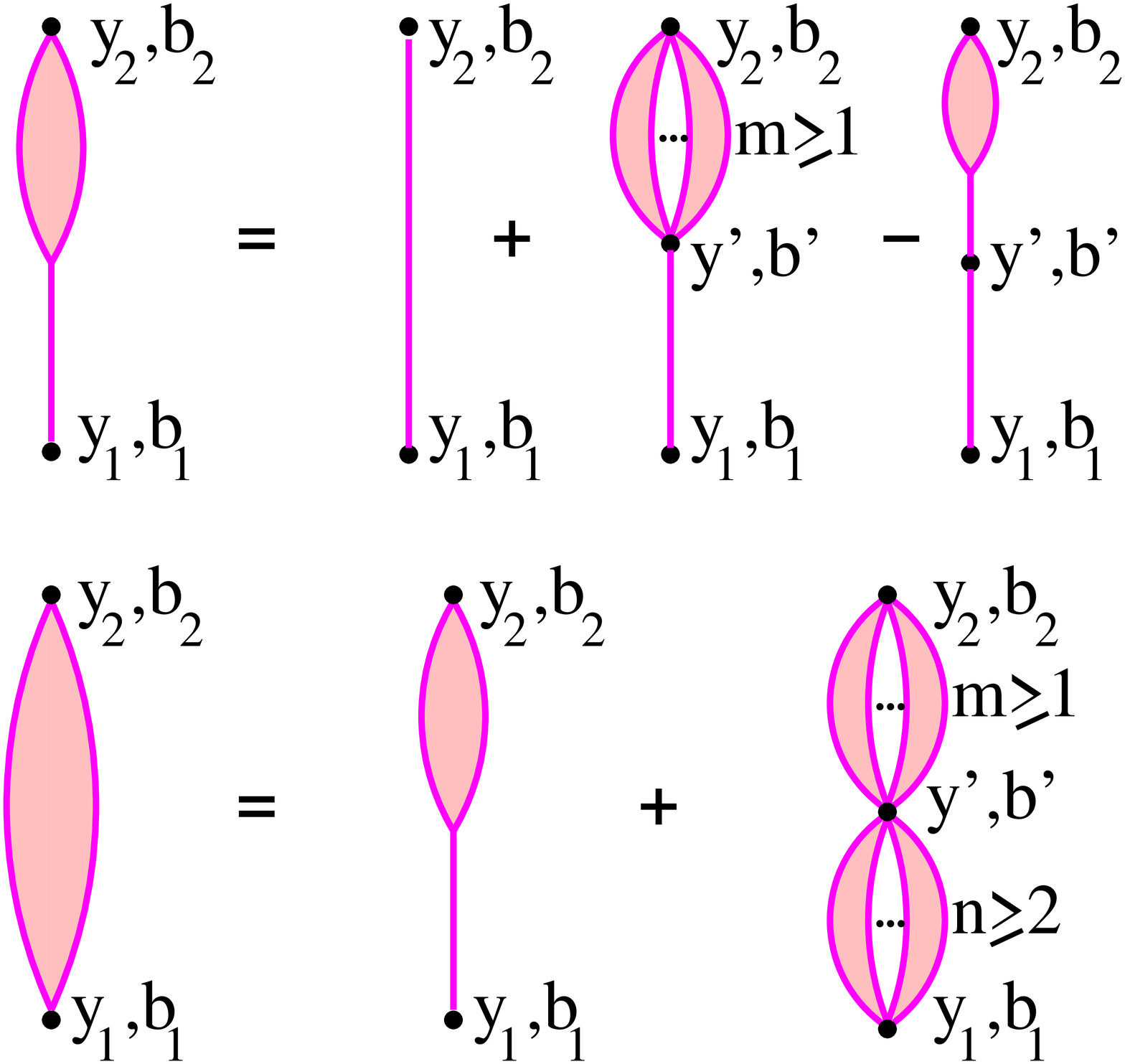}
\par\end{centering}

\caption{Recursive representations for the contributions of Pomeron loop sequences
$\chi^{{\rm loop(1)}}$ (top) and $\chi^{{\rm loop}}$ (bottom), exchanged
between the vertices $(y_{1},\vec{b}_{1})$ and $(y_{2},\vec{b}_{2})$.\label{fig: loops}}

\end{figure}

\begin{figure}[t]
\begin{centering}
\includegraphics[width=8cm,height=3cm]{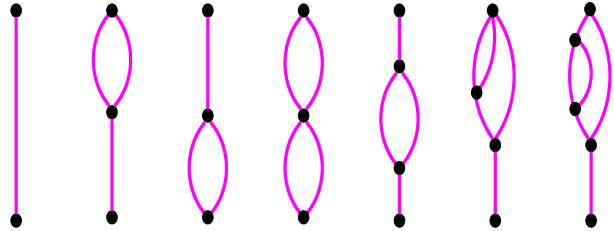}
\par\end{centering}

\caption{Examples of graphs generated by Schwinger-Dyson equations of Fig.~\ref{fig: loops}.\label{fig: loop-examples}}

\end{figure}

Let us introduce the contribution $\chi^{{\rm loop}}(y_{2}-y_{1},|\vec{b}_{2}-\vec{b}_{1}|)$
of \textit{irreducible} 2-point sequences of Pomerons and Pomeron
loops, exchanged between two vertices which are separated by rapidity
and impact parameter distances $y_{2}-y_{1}$ and $|\vec{b}_{2}-\vec{b}_{1}|$,
with the help of recursive equations of Fig.~\ref{fig: loops}%
\footnote{The present definition is somewhat more general compared to the one
in Ref.~\cite{ost08}.%
}.%
 The equation in the 1st line in Fig.~\ref{fig: loops} defines the
contribution $\chi^{{\rm loop(1)}}(y_{2}-y_{1},|\vec{b}_{2}-\vec{b}_{1}|)$
of such sub-sequences which start from a single Pomeron connected
to the vertex $(y_{1},\vec{b}_{1})$. The first term in the r.h.s.~of
the equation is just a single Pomeron exchange between the two vertices,
whereas the second consists of a single Pomeron exchanged between
the vertex $(y_{1},\vec{b}_{1})$ and an intermediate multi-Pomeron
vertex $(y',\vec{b}')$ plus $m\geq1$ irreducible loop sequences
(each one described by the eikonal $\chi^{{\rm loop}}(y_{2}-y',|\vec{b}_{2}-\vec{b}'|)$)
exchanged between the vertices $(y',\vec{b}')$ and $(y_{2},\vec{b}_{2})$.
The last term in the r.h.s.~is to subtract the Pomeron self-coupling
contribution. In turn, the equation in the 2nd line in Fig.~\ref{fig: loops}
defines the complete contribution $\chi^{{\rm loop}}(y_{2}-y_{1},|\vec{b}_{2}-\vec{b}_{1}|)$,
adding to the above-discussed terms also the 2nd graph in the r.h.s.~of
the equation, which contains $n\geq2$ irreducible loop sequences
exchanged between the vertices $(y_{1},\vec{b}_{1})$ and $(y',\vec{b}')$
plus $m\geq1$ sequences exchanged between the vertices $(y',\vec{b}')$
and $(y_{2},\vec{b}_{2})$. Examples of graphs generated by the described
equations are depicted in Fig.~\ref{fig: loop-examples}. %
Following the general rules of the Reggeon diagram technique \cite{gri68,bak76},
we obtain\begin{eqnarray}
\chi^{{\rm loop}(1)}(y_{2}-y_{1},|\vec{b}_{2}-\vec{b}_{1}|)=\chi^{\mathbb{P}}(y_{2}-y_{1},|\vec{b}_{2}-\vec{b}_{1}|)+G\int_{y_{1}+\xi}^{y_{2}-\xi}\! dy'\int\! d^{2}b'\;\chi^{\mathbb{P}}(y'-y_{1},|\vec{b}'-\vec{b}_{1}|)\nonumber \\
\times\left[1-e^{-\chi^{{\rm loop}}(y_{2}-y',|\vec{b}_{2}-\vec{b}'|)}-\chi^{{\rm loop(1)}}(y_{2}-y',|\vec{b}_{2}-\vec{b}'|)\right]\label{eq:loop(1)}\\
\chi^{{\rm loop}}(y_{2}-y_{1},|\vec{b}_{2}-\vec{b}_{1}|)=\chi^{{\rm loop}(1)}(y_{2}-y_{1},|\vec{b}_{2}-\vec{b}_{1}|)+G\int_{y_{1+}\xi}^{y_{2}-\xi}\! dy'\int\! d^{2}b'\;\nonumber \\
\times\left[1-e^{-\chi^{{\rm loop}}(y'-y_{1},|\vec{b}'-\vec{b}_{1}|)}-\chi^{{\rm loop}}(y'-y_{1},|\vec{b}'-\vec{b}_{1}|)\right]\left[1-e^{-\chi^{{\rm loop}}(y_{2}-y',|\vec{b_{2}}-\vec{b}'|)}\right].\label{eq:loop}\end{eqnarray}
It is noteworthy that the $y'$-integration in the 
r.h.s.~of (\ref{eq:loop(1)}-\ref{eq:loop}) is performed between
$y_{1}+\xi$ and $y_{2}-\xi$, with $\xi$ being the minimal rapidity
interval for the Pomeron asymptotics to be applicable.

Now, we can define the net-fan contribution via the recursive representation
of Fig.~\ref{freve} \cite{ost08}:%
\begin{figure}[t]
\begin{centering}
\includegraphics[width=7cm,height=2.5cm]{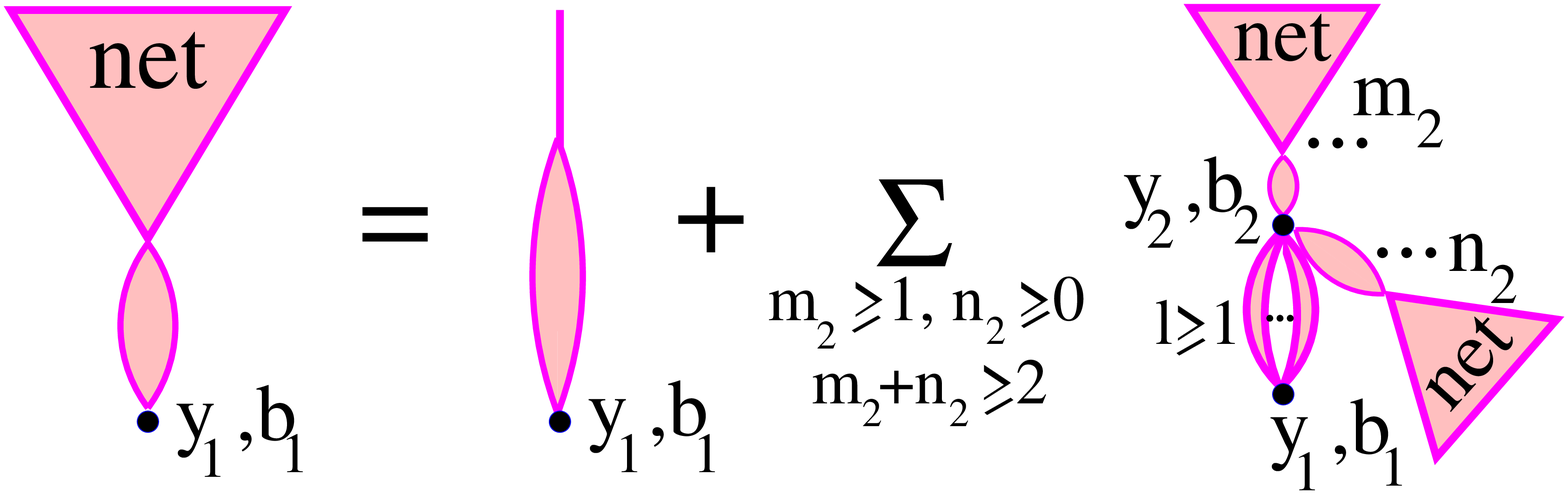}
\par\end{centering}

\caption{Recursive equation for projectile net-fan contribution $\chi_{jk}^{{\rm net}}(y_{1},\vec{b}_{1}|Y,\vec{b})$;
$y_{1}$ and $b_{1}$ are rapidity and impact parameter distances
between the projectile proton and the vertex in the ``handle'' of
the fan. The vertex $(y_{2},\vec{b}_{2})$ couples together $m_{2}$
projectile net-fans and $n_{2}$ target net-fans. In addition, there
are $l\geq1$ irreducible 2-point sequences of Pomerons and Pomeron
loops, exchanged between the vertices $(y_{1},\vec{b}_{1})$ and $(y_{2},\vec{b}_{2})$.
\label{freve}}

\end{figure}
 \begin{eqnarray}
\chi_{jk}^{{\rm net}}(y_{1},\vec{b}_{1}|Y,\vec{b})=\chi_{j}^{{\rm loop}}(y_{1},b_{1})+G\int_{\xi}^{y_{1}-\xi}\! dy_{2}\int\! d^{2}b_{2}\;\left(1-e^{-\chi^{{\rm loop}}(y_{1}-y_{2},|\vec{b}_{1}-\vec{b}_{2}|)}\right)\nonumber \\
\times\left[\left(1-e^{-\chi_{jk}^{{\rm net}}(y_{2},\vec{b}_{2}|Y,\vec{b})}\right)\; e^{-\chi_{kj}^{{\rm net}}(Y-y_{2},\vec{b}-\vec{b}_{2}|Y,\vec{b})}-\chi_{jk}^{{\rm net}}(y_{2},\vec{b}_{2}|Y,\vec{b})\right].\label{net-fan}\end{eqnarray}
By definition, $\chi_{jk}^{{\rm net}}(y_{1},\vec{b}_{1}|Y,\vec{b})$
represents the total contribution of irreducible net-like graphs exchanged
between the projectile and the target (represented by eigenstates
$|j\rangle$ and $|k\rangle$), with $Y=\ln s$ and $\vec{b}$ being
the impact parameter for the collision. The projectile (target) net-fan
starts from a given vertex and develops initially towards the projectile
(target) which is separated from the vertex by rapidity and impact
parameter distances $y_{1}$ and $b_{1}$. The 1st term in the r.h.s.~of
the equation in Fig.~\ref{freve} corresponds to a single 2-point
sequence of Pomerons and Pomeron loops exchanged between the projectile
(target) proton and the vertex $(y_{1},\vec{b}_{1})$, such that only
one Pomeron is coupled to the proton. The 2nd term describes the development
of the Pomeron net: for $m_{2}\geq2$ one has a fan-like splitting
and $m_{2}$ projectile (target) net-fans originate from the vertex
$(y_{2},\vec{b}_{2})$, developing initially in the direction of the
projectile (target); for $n_{2}\geq1$ one obtains in addition a re-scattering
on the partner proton - $n_{2}$ target (projectile) net-fans start
from the vertex and develop initially in the target (projectile) direction.
The 2-point loop sequence exchanged between the original vertex $(y_{1},\vec{b}_{1})$
and the nearest net cell (the vertex $(y_{2},\vec{b}_{2})$ in Fig.~\ref{freve})
will be referred to as the ``handle'' of the fan.

\begin{figure}[t]
\begin{centering}
\includegraphics[width=7cm,height=6cm]{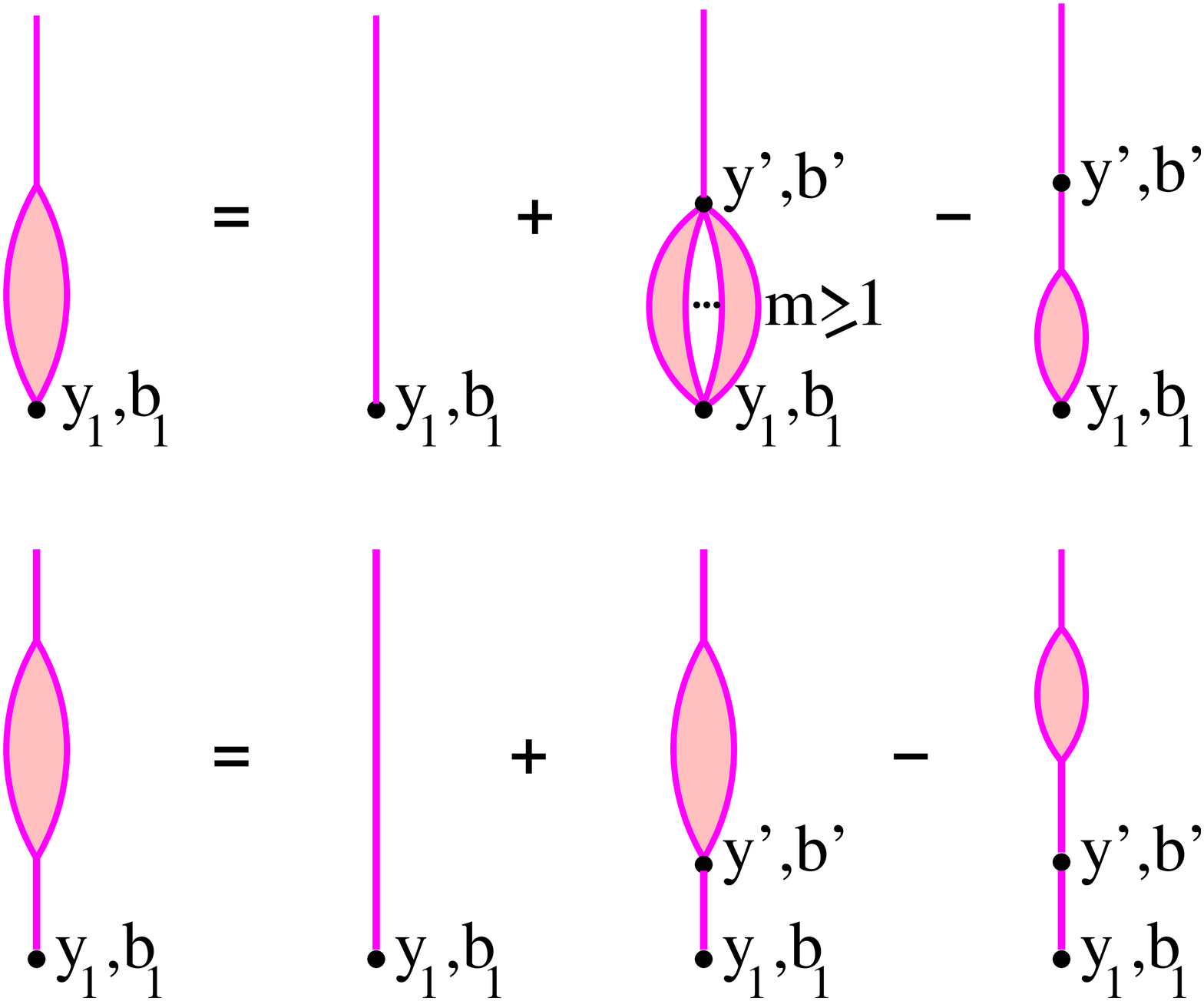}
\par\end{centering}

\caption{Contributions of 2-point sequences of Pomerons and Pomeron loops $\chi_{j}^{{\rm loop}}$
(top) and $\chi_{j}^{{\rm loop(1)}}$ (bottom), exchanged between
the projectile and the vertex $(y_{1},\vec{b}_{1})$.\label{fig: legs}}

\end{figure}

The contribution $\chi_{j}^{{\rm loop}}(y_{1},b_{1})$ of the first
graph in the r.h.s.~of the equation in Fig.~\ref{freve} and a part
of it, $\chi_{j}^{{\rm loop(1)}}(y_{1},b_{1})$, corresponding to
2-point Pomeron loop sequences which start from a single Pomeron connected
to the vertex $(y_{1},\vec{b}_{1})$, are defined in Fig.~\ref{fig: legs}
 (cf.~Fig.~\ref{fig: loops}), which gives \begin{eqnarray}
\chi_{j}^{{\rm loop}}(y_{1},b_{1})=\chi_{j}^{\mathbb{P}}(y_{1},b_{1})+G\int_{\xi}^{y_{1}-\xi}\! dy'\int\! d^{2}b'\;\chi_{j}^{\mathbb{P}}(y',b')\nonumber \\
\times\left[1-e^{-\chi^{{\rm loop}}(y_{1}-y',|\vec{b}_{1}-\vec{b}'|)}-\chi^{{\rm loop(1)}}(y_{1}-y',|\vec{b}_{1}-\vec{b}'|)\right]\label{leg}\\
\chi_{j}^{{\rm loop}(1)}(y_{1},b_{1})=\chi_{j}^{\mathbb{P}}(y_{1},b_{1})+G\int_{\xi}^{y_{1}-\xi}\! dy'\int\! d^{2}b'\;\left[\chi_{j}^{{\rm loop}}(y',b')-\chi_{j}^{{\rm loop(1)}}(y',b')\right]\nonumber \\
\times\,\chi^{\mathbb{P}}(y_{1}-y',|\vec{b}_{1}-\vec{b}'|)\,.\label{leg(1)}\end{eqnarray}

\begin{figure}[t]
\begin{centering}
\includegraphics[width=12.5cm,height=3.5cm]{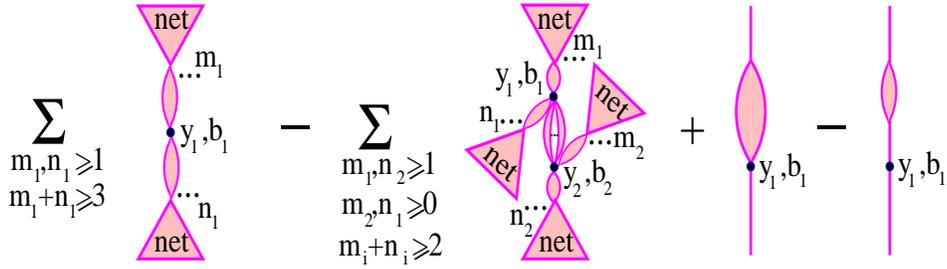}
\par\end{centering}

\caption{Total contribution of irreducible enhanced diagrams to elastic scattering
amplitude.\label{enh-full}}

\end{figure}

Finally, for the total contribution of irreducible enhanced graphs
$\chi_{jk}^{{\rm enh}}$ (exchanged between eigenstates $|j\rangle$
and $|k\rangle$ of the projectile and of the target respectively)
to elastic scattering amplitude one obtains the diagrammatic representation
of Fig.~\ref{enh-full} \cite{ost08},%
which gives \begin{eqnarray}
\chi_{jk}^{{\rm enh}}(s,b)=G\int_{\xi}^{Y-\xi}\!\! dy_{1}\!\int\!\! d^{2}b_{1}\;\left\{ \left[\left(1-e^{-\chi_{jk}^{{\rm net}}(1)}\right)\left(1-e^{-\chi_{kj}^{{\rm net}}(1)}\right)-\chi_{jk}^{{\rm net}}(1)\chi_{kj}^{{\rm net}}(1)\right]\right.\nonumber \\
-G\int_{\xi}^{y_{1}-\xi}\!\! dy_{2}\!\int\!\! d^{2}b_{2}\;\left(1-e^{-\chi^{{\rm loop}}(y_{1}-y_{2},|\vec{b}_{1}-\vec{b}_{2}|)}\right)\,\left[\left(1-e^{-\chi_{jk}^{{\rm net}}(1)}\right)e^{-\chi_{kj}^{{\rm net}}(1)}-\chi_{jk}^{{\rm net}}(1)\right]\nonumber \\
\times\left[\left(1-e^{-\chi_{kj}^{{\rm net}}(2)}\right)e^{-\chi_{jk}^{{\rm net}}(2)}-\chi_{kj}^{{\rm net}}(2)\right]\nonumber \\
\left.+\chi_{k}^{\mathbb{P}}(y_{1},b_{1})\left[\chi_{j}^{{\rm loop}}(Y-y_{1},|\vec{b}-\vec{b}_{1}|)-\chi_{j}^{{\rm loop(1)}}(Y-y_{1},|\vec{b}-\vec{b}_{1}|)\right]\right\} ,\label{chi-enh}\end{eqnarray}
where we used the abbreviations $\chi_{jk}^{{\rm net}}(i)=\chi_{jk}^{{\rm net}}(Y-y_{i},\vec{b}-\vec{b}_{i}|Y,\vec{b})$,
$\chi_{kj}^{{\rm net}}(i)=\chi_{kj}^{{\rm net}}(y_{i},\vec{b}_{i}|Y,\vec{b})$,
$i=1,2$. 

The first diagram in Fig.~\ref{enh-full} generates all possible
Pomeron nets exchanged between the projectile and target protons (with
neighboring net cells connected by 2-point sequences of Pomerons and
Pomeron loops), coupling together $m_{1}$ projectile and $n_{1}$
target net-fans ($m_{1}+n_{1}\geq3$) in the vertex $(y_{1},\vec{b}_{1})$,
whereas the second graph subtracts contributions which are generated
two or more times. The third diagram corresponds to a single 2-point
sequence of Pomerons and Pomeron loops exchanged between the projectile
and the target, each of the two being coupled to a single Pomeron
only. The last graph subtracts the Pomeron self-coupling contribution
generated by the previous diagram.

\begin{figure}[t]
\begin{centering}
\includegraphics[width=9cm,height=3cm]{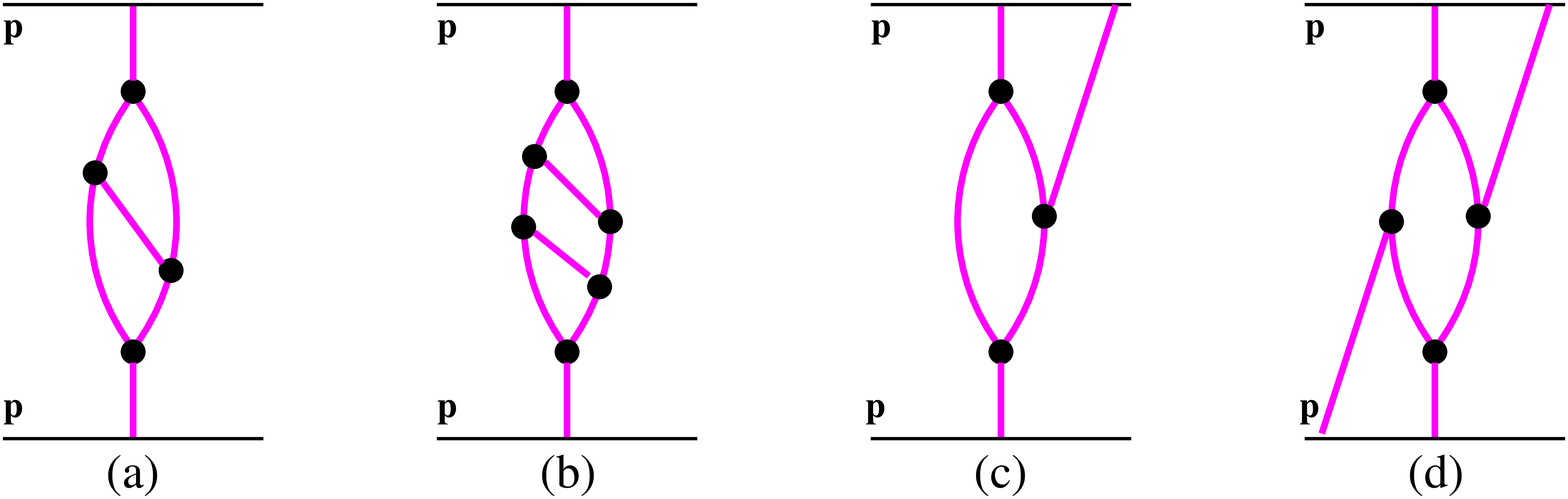}
\par\end{centering}

\caption{Examples of diagrams which are not included in the described resummation
scheme.\label{loop-ex}}

\end{figure}

The described scheme takes into consideration all possible enhanced
diagrams with the exception of two classes illustrated in Fig.~\ref{loop-ex}.%
 The graphs in Fig.~\ref{loop-ex}~(a,b) contain Pomeron loops whose
``sides'' are ``glued'' together by additional Pomeron exchanges;
such diagrams are not generated by the simple Schwinger-Dyson equations
of Fig.~\ref{fig: loops}. The graphs in Fig.~\ref{loop-ex}~(c,d)
contain loops whose ``surface'' is coupled to the projectile and/or
target by additional Pomeron exchanges. Such contributions can not
be easily resummed to all orders in the described scheme and require
a serious modification of the resummation procedure. However, both
classes of diagrams will be shown to provide negligible contributions
to elastic scattering amplitude in the described approach.

\section{High mass diffraction\label{sec:High-mass-diffraction}}

\subsection{High mass diffraction cross sections}

\begin{figure}[htb]
\begin{centering}
\includegraphics[width=12cm,height=3.5cm]{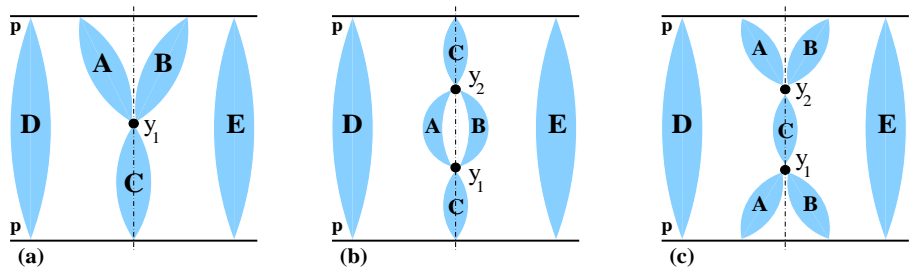}
\par\end{centering}

\caption{Classes of cut diagrams corresponding to single target diffraction
(a), double diffraction (b), and central diffraction (c) cross sections.
Thin dot-dashed lines indicate the position of the cut plane.\label{diffr-fig}}

\end{figure}

The beauty of the RFT approach is that it allows one to calculate
not only enhanced diagram contributions to elastic scattering amplitude
but also partial cross sections for various final states, particularly,
ones which contain large rapidity gaps (LRG) not covered by secondary
particle production. One derives the latter by considering unitarity
cuts of elastic scattering graphs and collecting contributions of
cuts corresponding to the desirable structure of final states. For
example, cross sections for single, double, and central diffraction
(often referred to as the double Pomeron exchange (DPE)) are defined
by cut diagrams of the kinds depicted in Fig.~\ref{diffr-fig}~(a),
(b), and (c) correspondingly.%
In Fig.~\ref{diffr-fig}~(a) particle production emerges from the
cut Pomeron block $C$, covering the rapidity range  $(0,y_{1})$.
On the other hand, in the rapidity interval $(y_{1},Y)$ no particle
is produced: the cut plane passes between the uncut Pomeron blocks
$A$ and $B$. In parallel to the cut irreducible graph $ABC$, one
may have a number of uncut irreducible graphs exchanged to the left
($D$) and to the right ($E$) of the cut plane, which represent elastic
rescattering processes. The corresponding factor may be interpreted
as the probability not to have other multiple production processes
in addition to the diffractive $ABC$ topology. Similarly, in Fig.~\ref{diffr-fig}~(b)
particles are produced in the rapidity intervals $(0,y_{1})$ and
$(y_{2},Y)$, whereas in the rapidity range $(y_{1},y_{2})$ the cut
plane passes between uncut Pomeron blocks, forming a LRG. Finally,
in Fig.~\ref{diffr-fig}~(c) particle production takes place in
the central rapidity interval $(y_{1},y_{2})$, which is separated
from the quasi-elastically scattered projectile and target protons
by large rapidity gaps $(0,y_{1})$ and $(y_{2},Y)$. 

The representation of Fig.~\ref{diffr-fig} is schematic for three
reasons. First, all the Pomeron blocks in the diffractively cut sub-graphs
may be connected both to the projectile and to the target protons
by additional \emph{uncut} Pomeron exchanges. For example, the blocks
$A$ and $B$ in Fig.~\ref{diffr-fig}~(a) may have internal multi-Pomeron
vertices, with one or a number of uncut Pomerons coupled to those
vertices and to the target proton. Secondly, contributions to diffractive
cross sections come also from multiple exchanges of diffractively
cut sub-graphs, e.g.~from multiple exchanges of the $ABC$-like Pomeron
configurations, each one with a rapidity gap of desirable size of
larger, in case of single diffraction. Third, one generally has to
choose proper intermediate states in the cut plane for the projectile
or/and target protons. For example, in Fig.~\ref{diffr-fig}~(a)
one has to choose elastic intermediate state for the projectile proton
in order to obtain single diffraction cross section. On the contrary,
choosing an inelastic intermediate state, one gets a contribution
to double diffraction cross section, with a high mass diffractive
state produced in the backward hemisphere and a small mass one in
the projectile fragmentation region.

Keeping those remarks in mind, we can express various diffractive
cross sections via contributions of irreducible cut diagrams characterized
by the desirable structure of rapidity gaps. Let us denote the contribution
of all $ABC$-like cut graphs, with a forward rapidity gap of size
$y_{{\rm gap}}$ or larger and without a LRG separating the target
proton from other particles produced, as $2\chi_{j_{1}j_{2}k}^{1-{\rm gap}}(Y,b,y_{{\rm gap}})$,
where $b$ is the impact parameter for the collision, $j_{1}$ and
$j_{2}$ are the projectile elastic scattering eigenstates to the
left and to the right of the cut plane,%
\footnote{In general, we have to consider different elastic scattering eigenstates
to the left and to the right of the cut plane, in order to project
on the elastic proton or low mass diffraction intermediate state.
Naturally, summing over all intermediate hadronic states, we obtain
a diagonal transition matrix: $\delta_{j_{1}j_{2}}=C_{j_{1}}C_{j_{2}}+(C_{j_{1}}\delta_{j_{1}j_{2}}-C_{j_{1}}C_{j_{2}})$.%
} and $k$ is the target elastic scattering eigenstate. Similarly,
we shall use the notation $2\chi_{j_{1}j_{2}k_{1}k_{2}}^{2-{\rm gap}}(Y,b,y_{{\rm gap}}^{{\rm (f)}},y_{{\rm gap}}^{{\rm (b)}})$
for the contribution of $ABCAB$-like cut graphs of Fig.~\ref{diffr-fig}~(c)
with the rapidity gaps in the forward and backward directions of the
sizes $y_{{\rm gap}}^{{\rm (f)}}$ and $y_{{\rm gap}}^{{\rm (b)}}$
or larger. Finally, the contribution of $CABC$-like cut graphs of
Fig.~\ref{diffr-fig}~(b), corresponding to high mass diffractive
states both in the forward and in the backward direction and a central
LRG of size $y_{{\rm gap}}$ or larger, is denoted as $2\chi_{jk}^{{\rm c-gap}}(Y,b,y_{{\rm gap}})$.
Then,   single high mass diffraction cross section is given as
\begin{eqnarray}
\sigma^{{\rm SD}}_{\rm HM}(s,y_{{\rm gap}})
=2\sum_{j_{1},j_{2},k}C_{j_{1}}C_{j_{2}}C_{k}\int\! d^{2}b\left[e^{2\chi_{j_{1}j_{2}k}^{1-{\rm gap}}(Y,b,y_{{\rm gap}})+2\chi_{j_{1}j_{2}kk}^{2-{\rm gap}}(Y,b,y_{{\rm gap}},\xi)}-1\right]S_{j_{1}j_{2}kk}(s,b)\nonumber \\
-\,2\sigma^{{\rm DPE}}(s,y_{{\rm gap}},\xi)\,,\label{eq:sigma-sd}
\end{eqnarray}
where for the so-called eikonal RGS factor corresponding to the contribution
of the uncut Pomeron blocks (D) and (E) in Fig.~\ref{diffr-fig}
and describing the probability not to fill the gap by secondary particles
produced in additional inelastic rescattering processes we have

\begin{equation}
S_{j_{1}j_{2}k_{1}k_{2}}(s,b)\equiv e^{-\frac{1}{2}\Omega_{j_{1}k_{1}}(s,b)-\frac{1}{2}\Omega_{j_{2}k_{2}}(s,b)},\label{eq:rgs-fac}\end{equation}
with the total opacity being defined in (\ref{eq:opac}). The factor
in the square brackets in the r.h.s.~of Eq.~(\ref{eq:sigma-sd})
comes from an exchange of an arbitrary number $(\geq 1)$ of cut graphs characterized
by a rapidity gap in the forward direction of size $y_{{\rm gap}}$
or larger. In the cut plane we choose proton (elastic) state for the
projectile (c.f.~Eq.~(\ref{sigma-sd(lm)})). Finally, from the obtained
expression we subtract the contribution of the central diffraction,
with   elastic proton states in the cut plane both for the projectile
and for the target and with rapidity gaps of sizes $y_{{\rm gap}}$
and $\xi$ correspondingly, which separate those protons from the
produced particles. The factor '2' in the two terms in the r.h.s.~accounts
for both projectile and target single diffraction contributions.

In turn, the above-discussed central diffraction cross section is\begin{eqnarray}
\sigma^{{\rm DPE}}(s,y_{{\rm gap}}^{{\rm (f)}},y_{{\rm gap}}^{{\rm (b)}})=\!\sum_{j_{1},j_{2},k_{1},k_{2}}C_{j_{1}}C_{j_{2}}C_{k_{1}}C_{k_{2}}\!\int\! d^{2}b\left[e^{2\chi_{j_{1}j_{2}k_{1}k_{2}}^{2-{\rm gap}}(Y,b,y_{{\rm gap}}^{{\rm (f)}},y_{{\rm gap}}^{{\rm (b)}})}-1\right]S_{j_{1}j_{2}k_{1}k_{2}}(s,b)\,.\label{eq:sigma-cd}\end{eqnarray}

Finally, for the double high mass diffraction cross section, with central LRG
of size $y_{{\rm gap}}$ or larger, we obtain\begin{eqnarray}
\sigma^{{\rm DD}}_{\rm HM}(s,y_{{\rm gap}})
=\sum_{j,k}C_{j}C_{k}\int\! d^{2}b\;2\chi_{jk}^{{\rm c-gap}}(Y,b,y_{{\rm gap}})\; S_{jjkk}(s,b)\nonumber \\
+\,2\sum_{j_{1},j_{2},k}(C_{j_{1}}\delta_{j_{1}j_{2}}-C_{j_{1}}C_{j_{2}})\, C_{k}\int\! d^{2}b\left[e^{2\chi_{j_{1}j_{2}k}^{1-{\rm gap}}(Y,b,y_{{\rm gap}})}-1\right]e^{2\chi_{j_{1}j_{2}kk}^{2-{\rm gap}}(Y,b,y_{{\rm gap}},\xi)}\; S_{j_{1}j_{2}kk}(s,b)\nonumber \\
+\sum_{j_{1},j_{2},k_{1},k_{2}}(C_{j_{1}}\delta_{j_{1}j_{2}}-C_{j_{1}}C_{j_{2}})\,(C_{k_{1}}\delta_{k_{1}k_{2}}-C_{k_{1}}C_{k_{2}})\nonumber \\
\times\int\! d^{2}b\left[e^{2\chi_{j_{1}j_{2}k_{1}k_{2}}^{2-{\rm gap}}(Y,b,y_{{\rm gap}},\xi)}+e^{2\chi_{j_{1}j_{2}k_{1}k_{2}}^{2-{\rm gap}}(Y,b,\xi,y_{{\rm gap}})}-e^{2\chi_{j_{1}j_{2}k_{1}k_{2}}^{2-{\rm gap}}(Y,b,y_{{\rm gap}},y_{{\rm gap}})}-1\right]\, S_{j_{1}j_{2}k_{1}k_{2}}(s,b)\nonumber \\
+\sum_{j,k}C_{j}C_{k}\int\! d^{2}b\int_{\xi}^{Y-y_{{\rm gap}}-\xi}\! dy_{1}\int_{y_{1}+y_{{\rm gap}}}^{Y-\xi}\! dy_{2}\;\frac{2d\chi_{jjk}^{1-{\rm gap}}(Y,b,Y-y_{1})}{d(Y-y_{1})}\,\frac{2d\chi_{kkj}^{1-{\rm gap}}(Y,b,y_{2})}{dy_{2}}\, S_{jjkk}(s,b)\,,\label{eq:sigma-dd}\end{eqnarray}
where in the 1st term in the r.h.s., corresponding to the production
of high mass diffractive states both in the forward and in the backward
hemisphere, we neglected multiple exchanges of cut sub-graphs with
a central rapidity gap. The 2nd term describes the production of high
mass diffractive state in the backward and a low mass one in the forward
direction (c.f.~(\ref{sigma-sd(lm)})); it is multiplied by factor
'2' to account for the opposite configuration. The 3rd term corresponds
to the low mass diffractive dissociation of both the projectile and
the target protons, which is accompanied by the production of a diffractive
bunch of secondaries in the central region. The latter is separated
from the projectile or/and target diffractive state by a LRG of size
$\geq y_{{\rm gap}}$. Finally, the last term in (\ref{eq:sigma-dd})
describes the situation when projectile and target high mass diffractive
states are produced in two different inelastic rescattering processes
(we neglect here contributions with three or more diffractively cut
sub-graphs), i.e.~to a superposition of two (projectile and target)
single diffraction processes.%
\footnote{Strictly speaking, the two contributions to double high mass diffraction
cross section, corresponding to the 1st and the last terms in the
r.h.s.~of Eq.~(\ref{eq:sigma-dd}), can not be treated separately
as they correspond to the same structure of the final state. Moreover,
as will be discussed in the following, the first contribution may
be negative in some parts of the kinematic space; it is the sum of
the two terms which provides a positively-defined result.%
} There, we consider differential contributions $2d\chi_{j_{1}j_{2}k}^{1-{\rm gap}}(Y,b,y_{{\rm gap}})/dy_{{\rm gap}}$
for a fixed size of the gap; the limits for the rapidity integrations
in the last term in (\ref{eq:sigma-dd}) are chosen such that a rapidity
gap of size $\geq y_{{\rm gap}}$ remains in the central region.

\subsection{Simplest enhanced graphs with LRG topologies\label{sub:Simplest-enhanced-graphs}}

We shall illustrate here the structure of the contributions $\chi^{{\rm 1-gap}}$,
$\chi^{{\rm 2-gap}}$, and $\chi^{{\rm c-gap}}$ by considering the
simplest enhanced graphs of the corresponding kinds, up to the second
order with respect to the triple-Pomeron coupling, while the complete
all-order treatment will be discussed in the Appendix.

\begin{figure}[t]
\begin{centering}
\includegraphics[width=15cm,height=10cm]{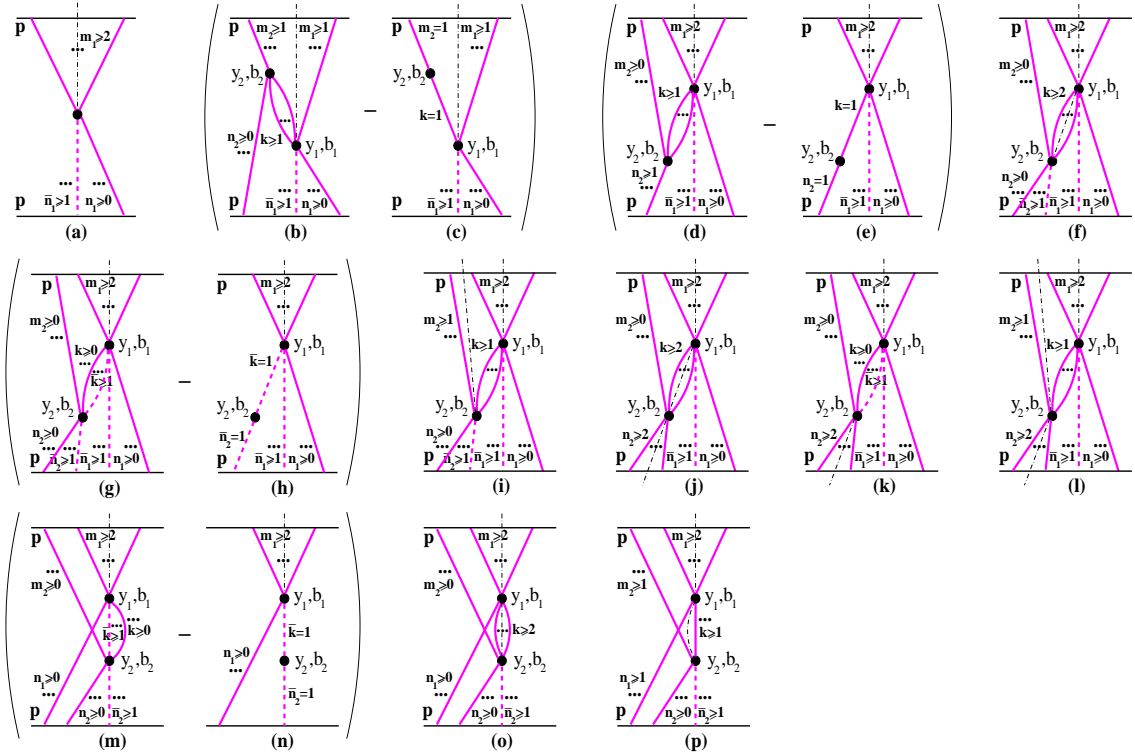}
\par\end{centering}

\caption{Simplest cut enhanced diagrams corresponding to single (target) high
mass diffraction topology. Uncut and cut Pomerons are shown respectively
by solid and dashed thick lines; thin dot-dashed lines indicate the
position of the cut plane.\label{1-gap-fig}}

\end{figure}

The simplest contribution to $2\chi_{j_{1}j_{2}k}^{1-{\rm gap}}(Y,b,y_{{\rm gap}})$,
which generates a LRG between the quasi-elastically scattered projectile
proton and a high mass diffractive state produced in the backward
direction, is depicted in Fig.~\ref{1-gap-fig}~(a),%
which corresponds to all the blocks $A$, $B$, and $C$ in Fig.~\ref{diffr-fig}~(a)
being represented by multi-Pomeron exchanges between the vertex $(y_{1},\vec{b}_{1})$
and the projectile/target protons, with at least one uncut, respectively
cut, Pomeron per block. Thus, we have $m_{1}\geq2$ uncut Pomerons
exchanged between the vertex $(y_{1},\vec{b}_{1})$ and the projectile
proton, which can be positioned both to the left and to the right
of the cut plane, with at least one Pomeron on either side of the
cut, plus $\bar{n}_{1}\geq1$ cut and $n\geq0$ uncut Pomerons exchanged
between that vertex and the target proton. Applying the Reggeon diagram
technique \cite{gri68,bak76}, we obtain the well-known result \cite{abr92}
\begin{eqnarray}
\frac{G}{2}\!\int\! d^{2}b_{1}\int_{\xi}^{\min(Y-y_{{\rm gap}},Y-\xi)}
\!\! dy_{1}\left[1-e^{-\chi_{j_{1}}^{\mathbb{P}}(Y-y_{1},|\vec{b}-\vec{b}_{1}|)}\right]\!
\left[1-e^{-\chi_{j_{2}}^{\mathbb{P}}(Y-y_{1},|\vec{b}-\vec{b}_{1}|)}\right]\!
\left[1-e^{-2\chi_{k}^{\mathbb{P}}(y_{1},b_{1})}\right]\!.\label{eq:1gap-Y-2}
\end{eqnarray}
Due to the steep fall down of the Pomeron eikonal $\chi_{j}^{\mathbb{P}}(y,b)$
for large $b$, at very large impact parameters (\ref{eq:1gap-Y-2})
reduces to the usual triple-Pomeron contribution\begin{equation}
G\int\! d^{2}b_{1}\int_{\xi}^{\min(Y-y_{{\rm gap}},Y-\xi)}\! dy_{1}\;\chi_{j_{1}}^{\mathbb{P}}(Y-y_{1},|\vec{b}-\vec{b}_{1}|)\,\chi_{j_{2}}^{\mathbb{P}}(Y-y_{1},|\vec{b}-\vec{b}_{1}|)\,\chi_{k}^{\mathbb{P}}(y_{1},b_{1})\,,\label{eq:3P}\end{equation}
which is characterized by the diffractive mass $M_{X}^{2}=e^{y_{1}}$
distribution (due to the larger slope of the soft component, it dominates
very peripheral interactions):\[
f(M_{X}^{2})\sim\left(M_{X}^{2}\right)^{-\alpha_{\mathbb{P}{\rm (s)}}}.\]
On the other hand, for central $(b\sim0)$ collisions at sufficiently
high energies a large contribution to the integral in (\ref{eq:1gap-Y-2})
comes from the kinematic region where $\chi_{j_{1}}^{\mathbb{P}}$,
$\chi_{j_{2}}^{\mathbb{P}}$, $\chi_{k}^{\mathbb{P}}$ are all large,
which leads to \[
f(M_{X}^{2})\sim1/M_{X}^{2}\,.\]

The other graphs in Fig.~\ref{1-gap-fig} describe absorptive corrections
to the simple diagram of Fig.~\ref{1-gap-fig}~(a). For example,
the diagram in Fig.~\ref{1-gap-fig}~(b) (minus the Pomeron self-coupling
contribution of Fig.~\ref{1-gap-fig}~(c)) corresponds to having
an internal multi-Pomeron vertex in the block $A$ or $B$ in Fig.~\ref{diffr-fig}~(a).
Absorptive effects arise due to additional rescatterings on the projectile
(for $m_{2}\geq2$) or/and on the target (for $n_{2}\geq1$), or/and
due to the emergence of Pomeron loops (for $k\geq2$), giving rise
to the (negative) screening contribution%
\footnote{When calculating the contributions of the graphs in Fig.~\ref{1-gap-fig}~(b,c),
we take into account that the uncut multi-Pomeron vertex $(y_{2},\vec{b}_{2})$,
along with the uncut Pomerons coupled to it, may be positioned on
either side of the cut plane, and so do $m_{1}\geq1$ uncut Pomerons
connected to the vertex $(y_{1},\vec{b}_{1})$, such that at least
one of the $m_{1}$ Pomerons is positioned on the opposite side of
the cut with respect to the vertex $(y_{2},\vec{b}_{2})$. More details
on the calculation technique can be found in the preceding publications
\cite{ost06a,ost08}.%
}\begin{eqnarray}
\frac{G^{2}}{2}\int\! d^{2}b_{1}d^{2}b_{2}\int_{\xi}^{\min(Y-y_{{\rm gap}},Y-2\xi)}\! dy_{1}\int_{y_{1}+\xi}^{Y-\xi}\! dy_{2}\left(1-e^{-2\chi_{k}^{\mathbb{P}}(y_{1},b_{1})}\right)\nonumber \\
\times\left\{ \left(1-e^{-\chi^{\mathbb{P}}(y_{2}-y_{1},|\vec{b}_{2}-\vec{b}_{1}|)}\right)e^{-\chi_{k}^{\mathbb{P}}(y_{2},b_{2})}\left[\left(1-e^{-\chi_{j_{1}}^{\mathbb{P}}(Y-y_{2},|\vec{b}-\vec{b}_{2}|)}\right)\left(1-e^{-\chi_{j_{2}}^{\mathbb{P}}(Y-y_{1},|\vec{b}-\vec{b}_{1}|)}\right)\right.\right.\nonumber \\
\times\left.e^{-\chi_{j_{1}}^{\mathbb{P}}(Y-y_{1},|\vec{b}-\vec{b}_{1}|)}+\left(1-e^{-\chi_{j_{2}}^{\mathbb{P}}(Y-y_{2},|\vec{b}-\vec{b}_{2}|)}\right)\left(1-e^{-\chi_{j_{1}}^{\mathbb{P}}(Y-y_{1},|\vec{b}-\vec{b}_{1}|)}\right)e^{-\chi_{j_{2}}^{\mathbb{P}}(Y-y_{1},|\vec{b}-\vec{b}_{1}|)}\right]\nonumber \\
-\chi^{\mathbb{P}}(y_{2}-y_{1},|\vec{b}_{2}-\vec{b}_{1}|)\left[\chi_{j_{1}}^{\mathbb{P}}(Y-y_{2},|\vec{b}-\vec{b}_{2}|)\left(1-e^{-\chi_{j_{2}}^{\mathbb{P}}(Y-y_{1},|\vec{b}-\vec{b}_{1}|)}\right)e^{-\chi_{j_{1}}^{\mathbb{P}}(Y-y_{1},|\vec{b}-\vec{b}_{1}|)}\right.\nonumber \\
+\left.\left.\chi_{j_{2}}^{\mathbb{P}}(Y-y_{2},|\vec{b}-\vec{b}_{2}|)\left(1-e^{-\chi_{j_{1}}^{\mathbb{P}}(Y-y_{1},|\vec{b}-\vec{b}_{1}|)}\right)e^{-\chi_{j_{2}}^{\mathbb{P}}(Y-y_{1},|\vec{b}-\vec{b}_{1}|)}\right]\right\} .\label{eq:1gap-A-scr}\end{eqnarray}

In turn, the diagrams in Fig.~\ref{1-gap-fig}~(d-p) correspond
to having an internal multi-Pomeron vertex in the block $C$ in the
representation of Fig.~\ref{diffr-fig}~(a). One can immediately
notice that the summary contribution of the graphs in Fig.~\ref{1-gap-fig}~(d-l)
is zero. Indeed, these diagrams differ by the structure of the sub-graph
formed by all the Pomerons coupled to the vertex $(y_{2},\vec{b}_{2})$:
while it remains uncut in Fig.~\ref{1-gap-fig}~(d,e), it is cut
in all possible ways in Fig.~\ref{1-gap-fig}~(e-l). The $s$-channel
unitarity assures a precise cancellation between these uncut and cut
contributions, which can be also verified by an explicit calculation.
Thus, non-zero corrections come only from the graphs in Fig.~\ref{1-gap-fig}~(m-p)
and read \begin{eqnarray}
\frac{G^{2}}{2}\int\! d^{2}b_{1}d^{2}b_{2}\int_{2\xi}^{\min(Y-y_{{\rm gap}},Y-\xi)}\!\! dy_{1}\int_{\xi}^{y_{1}-\xi}\!\! dy_{2}\left(1-e^{-\chi_{j_{1}}^{\mathbb{P}}(Y-y_{1},|\vec{b}-\vec{b}_{1}|)}\right)\left(1-e^{-\chi_{j_{2}}^{\mathbb{P}}(Y-y_{1},|\vec{b}-\vec{b}_{1}|)}\right)\nonumber \\
\times\left\{ \left(1-e^{-\chi^{\mathbb{P}}(y_{1}-y_{2},|\vec{b}_{1}-\vec{b}_{2}|)}\right)\left(1-e^{-2\chi_{k}^{\mathbb{P}}(y_{2},b_{2})}\right)e^{-\chi_{k}^{\mathbb{P}}(y_{1},b_{1})}\left[e^{-\chi_{j_{1}}^{\mathbb{P}}(Y-y_{2},|\vec{b}-\vec{b}_{2}|)-\chi_{j_{2}}^{\mathbb{P}}(Y-y_{2},|\vec{b}-\vec{b}_{2}|)}\right.\right.\nonumber \\
-\left.\left(e^{-\chi_{j_{1}}^{\mathbb{P}}(Y-y_{2},|\vec{b}-\vec{b}_{2}|)}+e^{-\chi_{j_{2}}^{\mathbb{P}}(Y-y_{2},|\vec{b}-\vec{b}_{2}|)}\right)\left(1-e^{-\chi_{k}^{\mathbb{P}}(y_{1},b_{1})}\right)\right]\nonumber \\
-\left.2\chi^{\mathbb{P}}(y_{1}-y_{2},|\vec{b}_{1}-\vec{b}_{2}|)\,\chi_{k}^{\mathbb{P}}(y_{2},b_{2})\, e^{-2\chi_{k}^{\mathbb{P}}(y_{1},b_{1})}\right\} .\label{eq:1gap-C-scr}\end{eqnarray}

It is noteworthy that at small impact parameters the integrands in
(\ref{eq:1gap-A-scr}), (\ref{eq:1gap-C-scr}) are damped by the exponential
factors $\exp(-\chi_{j_{1/2}}^{\mathbb{P}}(Y-y_{1},|\vec{b}-\vec{b}_{1}|))$
and $\exp(-\chi_{k}^{\mathbb{P}}(y_{1},b_{1}))$ correspondingly,
as noted in \cite{car74,kai86}, while in very peripheral interactions
these contributions are suppressed by an additional power of the triple-Pomeron
coupling, compared to the lowest order result, Eq.~(\ref{eq:1gap-Y-2}).
However, in the intermediate range they give rise to very important
absorptive effects. Let us also remark that in the graphs in Fig.~\ref{1-gap-fig}~(o,p)
there is just a narrow bunch of particles produced at $y\sim y_{1}$,
which results from the cut multi-Pomeron vertex $(y_{1},\vec{b}_{1})$.
Such low mass diffractive states produced at central rapidities are
difficult to detect experimentally. Therefore, in the next Section
we compare diffractive cross sections as calculated taking such low
mass states into account or neglecting them. In the latter case, the
limits for the $y$-integrations for these diagrams should be chosen
as $2\xi<y_{1}<Y-\xi$, $\xi<y_{2}<\min(Y-y_{{\rm gap}},y_{1}-\xi)$
(c.f.~(\ref{eq:1gap-C-scr})).

\begin{figure}[t]
\begin{centering}
\includegraphics[width=15cm,height=8cm]{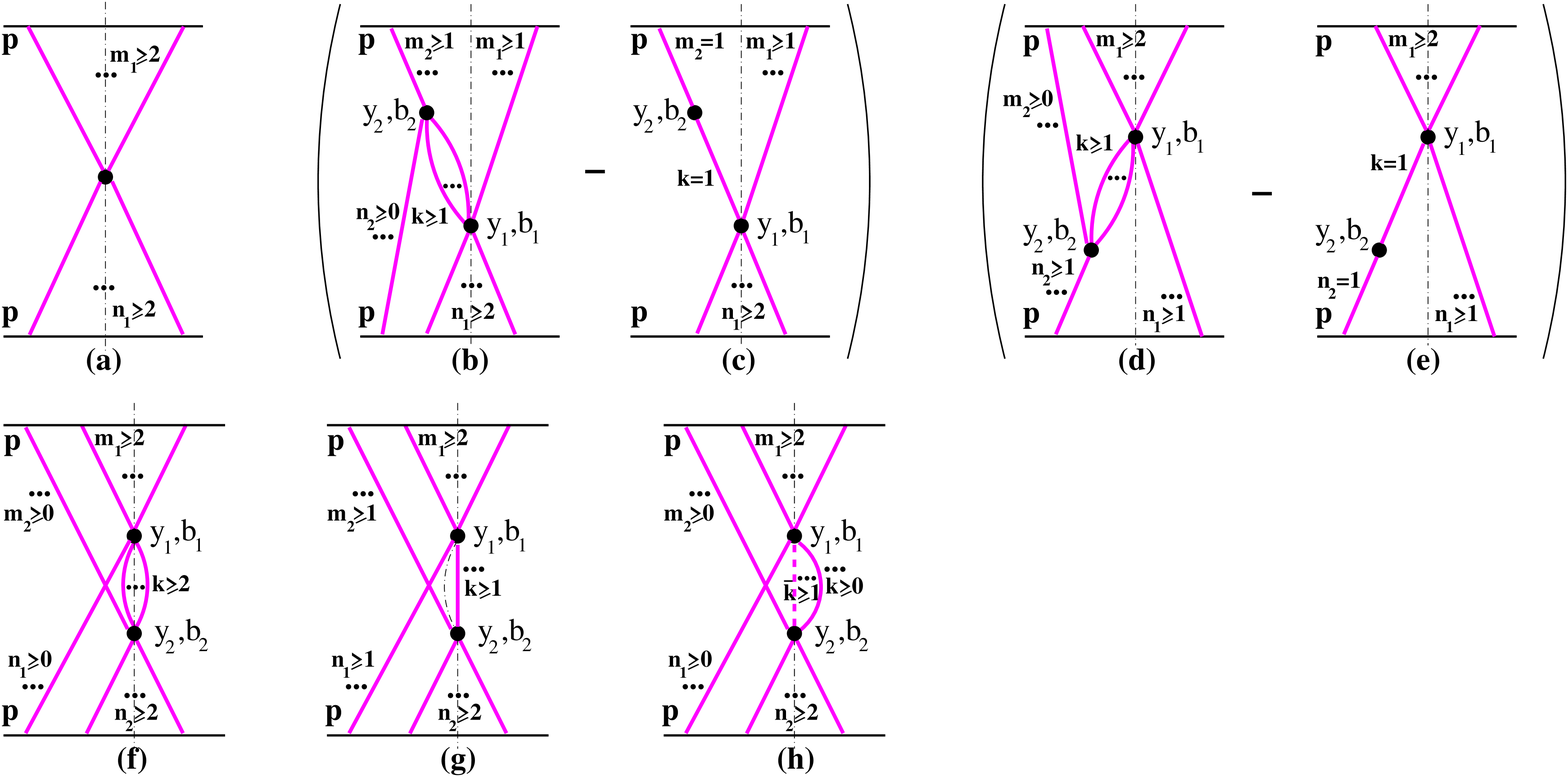}
\par\end{centering}

\caption{Simplest cut enhanced diagrams corresponding to central diffraction
(DPE) topology.\label{2-gap-fig}}

\end{figure}

Restricting oneself with enhanced diagrams of the lowest two orders,
the contribution to $2\chi_{j_{1}j_{2}k_{1}k_{2}}^{2-{\rm gap}}(Y,b,y_{{\rm gap}}^{{\rm (f)}},y_{{\rm gap}}^{{\rm (b)}})$,
corresponding to central high mass diffraction topology, comes from
the graph in Fig.~\ref{2-gap-fig}~(h) %
 only, which gives\begin{eqnarray}
\frac{G^{2}}{4}\int\! d^{2}b_{1}d^{2}b_{2}\int_{\max(2\xi,y_{{\rm gap}}^{{\rm (b)}}+\xi)}^{\min(Y-y_{{\rm gap}}^{{\rm (f)}},Y-\xi)}\!\! dy_{1}\int_{\xi}^{y_{1}-\xi}\!\! dy_{2}\left(1-e^{-\chi_{j_{1}}^{\mathbb{P}}(Y-y_{1},|\vec{b}-\vec{b}_{1}|)}\right)\nonumber \\
\times\left(1-e^{-\chi_{j_{2}}^{\mathbb{P}}(Y-y_{1},|\vec{b}-\vec{b}_{1}|)}\right)\left(1-e^{-2\chi^{\mathbb{P}}(y_{1}-y_{2},|\vec{b}_{1}-\vec{b}_{2}|)}\right)\left(1-e^{-\chi_{k_{1}}^{\mathbb{P}}(y_{2},b_{2})}\right)\left(1-e^{-\chi_{k_{2}}^{\mathbb{P}}(y_{2},b_{2})}\right)\nonumber \\
\times e^{-\chi_{j_{1}}^{\mathbb{P}}(Y-y_{2},|\vec{b}-\vec{b}_{2}|)-\chi_{j_{2}}^{\mathbb{P}}(Y-y_{2},|\vec{b}-\vec{b}_{2}|)-\chi_{k_{1}}^{\mathbb{P}}(y_{1},b_{1})-\chi_{k_{2}}^{\mathbb{P}}(y_{1},b_{1})}.\label{eq:2-gap}\end{eqnarray}
Here again we notice a damping of the contribution at small impact
parameters by the exponential factors in the 3rd line of Eq.~(\ref{eq:2-gap}).
Accounting for diffractive production of narrow bunches of secondary
particles at central rapidities, we have to add also contributions
of the graphs in Fig.~\ref{2-gap-fig}~(a-g), which we omit here
for brevity.

\begin{figure}[t]
\begin{centering}
\includegraphics[width=6cm,height=4cm]{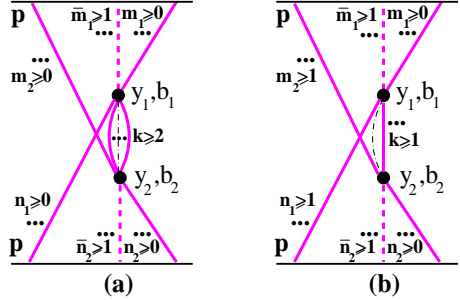}
\par\end{centering}

\caption{Simplest cut enhanced diagrams corresponding to double high mass diffraction
topology.\label{c-gap-fig}}

\end{figure}

Finally, the lowest order contribution to $2\chi_{jk}^{{\rm c-gap}}(Y,b,y_{{\rm gap}})$,
corresponding to double high mass diffraction and a central LRG, is
given by the graphs in Fig.~\ref{c-gap-fig} %
 and reads\begin{eqnarray}
\frac{G^{2}}{4}\int\! d^{2}b_{1}d^{2}b_{2}\int_{\max(\xi+y_{{\rm gap}},2\xi)}^{Y-\xi}\!\! dy_{1}\int_{\xi}^{y_{1}-\max(y_{{\rm gap}},\xi)}\!\! dy_{2}\left[\left(1-e^{-\chi^{\mathbb{P}}(y_{1}-y_{2},|\vec{b}_{1}-\vec{b}_{2}|)}\right)^{2}\right.\nonumber \\
\times\, e^{-\chi_{j}^{\mathbb{P}}(Y-y_{2},|\vec{b}-\vec{b}_{2}|)-\chi_{k}^{\mathbb{P}}(y_{1},b_{1})}-2\left(1-e^{-\chi^{\mathbb{P}}(y_{1}-y_{2},|\vec{b}_{1}-\vec{b}_{2}|)}\right)\left(1-e^{-\chi_{j}^{\mathbb{P}}(Y-y_{2},|\vec{b}-\vec{b}_{2}|)}\right)\nonumber \\
\times\left.\left(1-e^{-\chi_{k}^{\mathbb{P}}(y_{1},b_{1})}\right)\right]e^{-\chi_{j}^{\mathbb{P}}(Y-y_{2},|\vec{b}-\vec{b}_{2}|)-\chi_{k}^{\mathbb{P}}(y_{1},b_{1})}\left(1-e^{-2\chi_{j}^{\mathbb{P}}(Y-y_{1},|\vec{b}-\vec{b}_{1}|)}\right)\left(1-e^{-2\chi_{k}^{\mathbb{P}}(y_{2},b_{2})}\right).\label{eq:c-gap}\end{eqnarray}
It is easy to see that the expression in the square brackets is not
positively defined: for $b\sim b_{1}\sim b_{2}\sim0$ the first term
is strongly damped by the exponential factor $e^{-\chi_{j}^{\mathbb{P}}(Y-y_{2},|\vec{b}-\vec{b}_{2}|)-\chi_{k}^{\mathbb{P}}(y_{1},b_{1})}$.
Thus, for $b\sim0$ the whole expression (\ref{eq:c-gap}) may provide
a negative result. It is the complete contribution to the double high
mass diffraction cross section, given by the sum of the first and
the last terms in the r.h.s.~of Eq.~(\ref{eq:sigma-dd}), which
has the positive definiteness due to the dominance of the second contribution
in the small $b$ region. In fact, the diagram in Fig.~\ref{c-gap-fig}~(b)
describes the interference between the amplitudes corresponding to
a single double high mass diffraction process of Fig.~\ref{c-gap-fig}~(a)
and to a superposition of two (projectile and target) single high
mass diffraction processes of Fig.~\ref{1-gap-fig}~(a).

\section{Numerical results\label{sec:Numerical-results}}

The parameters for the above-described model approach have been fixed
to reproduce the available data on total, elastic, and diffractive
proton-proton cross sections, the elastic scattering slope, and the
differential elastic cross section. As the numerical calculations
proved to be rather time-consuming and some experimental results do
not agree well with each other, we did not perform a standard $\chi^{2}$
minimization procedure but rather tuned the parameters to obtain an
overall acceptable description of the data points.

In order to reduce the number of adjustable parameters, we used equal
weights for elastic scattering eigenstates, $C_{1}=C_{2}=1/2$. On
the other hand, due to a rather weak dependence of the results on
the $\gamma_{\mathbb{P}}$ parameter (Eq.~(\ref{eq:g_mn})), within
certain limits, we used a fixed value $\gamma_{\mathbb{P}}=0.5$ GeV$^{-1}$.
The latter point requires a special discussion. As shown in \cite{ost06},
the above-described resummation scheme approaches in the ``dense''
limit ($s\rightarrow\infty$, $b\rightarrow0$) the asymptotic result
obtained in \cite{kai86}, which corresponds to a multi-channel non-enhanced
eikonal scheme based on a ``renormalized'' Pomeron, with the intercept\begin{equation}
\alpha_{\mathbb{P}}^{{\rm ren}}=\alpha_{\mathbb{P}}-r_{3\mathbb{P}}/\gamma_{\mathbb{P}}\,.\label{eq:inter-ren}\end{equation}
Considering a single Pomeron pole contribution to $D^{\mathbb{P}}(s,t)$,
the consistency requires that the renormalized Pomeron remains an
overcritical one, i.e.~$\alpha_{\mathbb{P}}^{{\rm ren}}>1$, in order
to preserve the energy rise of total cross section. Here, using two
Pomeron poles, we investigate a more interesting option, choosing
$\gamma_{\mathbb{P}}$ such that the renormalized soft Pomeron becomes
an undercritical one, $\alpha_{\mathbb{P}(s)}^{{\rm ren}}<1$, corresponding
to the saturation of the soft physics, while the hard Pomeron remains
an overcritical one after the renormalization, $\alpha_{\mathbb{P}(h)}^{{\rm ren}}>1$.
This corresponds to the picture where central hadronic collisions
are dominated in the very high energy limit by cascades of harder
partons, while peripheral interactions remain governed by soft parton
cascades.

All other model parameters have been tuned for two choices on the
Pomeron mass cutoff, $\xi=2$ and 1.5, fitting the CDF value $\sigma^{{\rm tot}}=80.03\pm2.24$
mb \cite{abe94a} at the Tevatron, or using $\xi=2$ and fitting the
E710 value $\sigma^{{\rm tot}}=72.8\pm3.1$ mb \cite{amo92}. The
three parameter sets, referred to as (A), (B), and (C) in the following,
are listed in Table~\ref{Flo:param},%
\begin{table}[t]
\begin{centering}
{\scriptsize }\begin{tabular}{|>{\centering}p{0.37in}|>{\centering}p{0.3in}|>{\centering}p{0.3in}|>{\centering}p{0.35in}|>{\centering}p{0.35in}|>{\centering}p{0.35in}|>{\centering}p{0.15in}|>{\centering}p{0.3in}|>{\centering}p{0.35in}|>{\centering}p{0.35in}|>{\centering}p{0.35in}|>{\centering}p{0.35in}|}
\hline 
 & {\scriptsize $\begin{array}{c}
\alpha_{\mathbb{P}{\rm (s)}}\\
\\\end{array}$} & {\scriptsize $\begin{array}{c}
\alpha_{\mathbb{P}{\rm (h)}}\\
\\\end{array}$} & {\scriptsize $\begin{array}{c}
\alpha'_{\mathbb{P}{\rm (s)}},\\
{\rm GeV}^{-2}\end{array}$} & {\scriptsize $\begin{array}{c}
\alpha'_{\mathbb{P}{\rm (h)}},\\
GeV^{-2}\end{array}$} & {\scriptsize $\begin{array}{c}
\gamma,\\
{\rm GeV}^{-1}\end{array}$} & {\scriptsize $\begin{array}{c}
\eta\\
\\\end{array}$} & {\scriptsize $\begin{array}{c}
r_{{\rm h/s}}\\
\\\end{array}$} & {\scriptsize $\begin{array}{c}
\Lambda_{1},\\
{\rm GeV}^{-2}\end{array}$} & {\scriptsize $\begin{array}{c}
\Lambda_{2},\\
{\rm GeV}^{-2}\end{array}$} & {\scriptsize $\begin{array}{c}
r_{3\mathbb{P}},\\
{\rm GeV}^{-1}\end{array}$} & {\scriptsize $\begin{array}{c}
\gamma_{\mathbb{P}},\\
{\rm GeV}^{-1}\end{array}$}\tabularnewline
\hline
\hline 
{\scriptsize Set (A)} & {\scriptsize 1.145} & {\scriptsize 1.35} & {\scriptsize 0.13} & {\scriptsize 0.075} &
{\scriptsize 1.65} & {\scriptsize 0.6} & {\scriptsize 0.06} & {\scriptsize 1.06} & {\scriptsize 0.3} & {\scriptsize 0.14} & {\scriptsize 0.5}\tabularnewline
\hline 
{\scriptsize Set (B)} & {\scriptsize 1.15} & {\scriptsize 1.35} & {\scriptsize 0.165} & {\scriptsize 0.08} & {\scriptsize 1.75} & {\scriptsize 0.6} & {\scriptsize 0.065} & {\scriptsize 1.03} & {\scriptsize 0.3} & {\scriptsize 0.15} & {\scriptsize 0.5}\tabularnewline
\hline 
{\scriptsize Set (C)} & {\scriptsize 1.14} & {\scriptsize 1.31} & {\scriptsize 0.14} & {\scriptsize 0.085} &
{\scriptsize 1.6} & {\scriptsize 0.5} & {\scriptsize 0.09} & {\scriptsize 1.1} & {\scriptsize 0.4} & {\scriptsize
0.14} & {\scriptsize 0.5}\tabularnewline
\hline
\end{tabular}\caption{Model parameters.}
\label{Flo:param}
\par\end{centering}

\end{table}
 with the vertex factors $\gamma_{i}$ being expressed as $\gamma_{1/2}=\gamma(1\pm\eta)$.
Surprisingly, we got relatively large slopes both for the soft and
the hard Pomeron components, $\alpha_{\mathbb{P}(s)}'\sim0.13\div0.17$
GeV$^{-2}$ and $\alpha_{\mathbb{P}(h)}'\simeq0.08$ GeV$^{-2}$,
to be compared with $\alpha_{\mathbb{P}}'\simeq0.05$ GeV$^{-2}$
in \cite{kmr08} and $\alpha_{\mathbb{P}}'\simeq0.01$  GeV$^{-2}$ in \cite{glm08}.
On the other hand, the obtained values for the triple-Pomeron coupling
are close to the old estimates \cite{kai86,kai79}.

\begin{figure}[t]
\begin{centering}
\includegraphics[width=15cm,height=6cm]{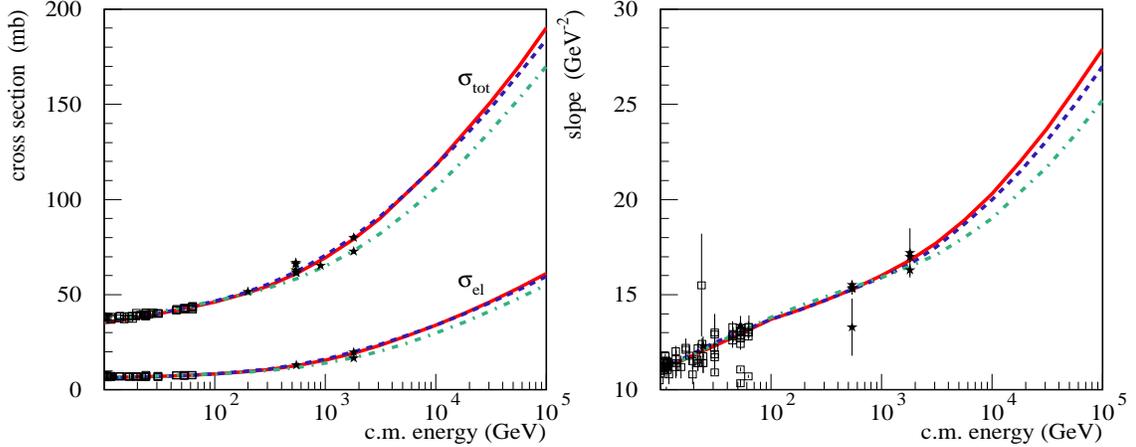}
\par\end{centering}

\caption{Total and elastic proton-proton cross sections (left) and elastic
scattering slope (right) calculated using the parameter sets (A),
(B), and (C) - solid, dashed and dot-dashed lines correspondingly.
The compilation of data is from \cite{cas98}.\label{fig:sigtot}}

\end{figure}

\begin{figure}[htb]
\begin{centering}
\includegraphics[width=8cm,height=11cm]{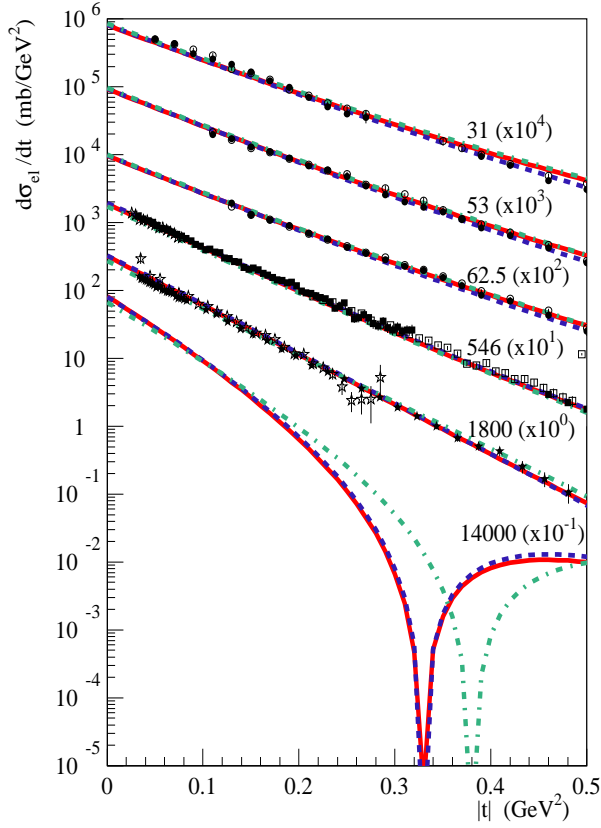}
\par\end{centering}

\caption{Calculated differential elastic proton-proton cross section for different
$\sqrt{s}$ in GeV (as indicated in the plot) compared to experimental
data \cite{bre84,bat83,boz84,amo90,abe94b}. The meaning of
the lines is the same as in Fig.~\ref{fig:sigtot}.\label{fig:dsigel}}

\end{figure}

\begin{figure}[t]
\begin{centering}
\includegraphics[width=15cm,height=6cm]{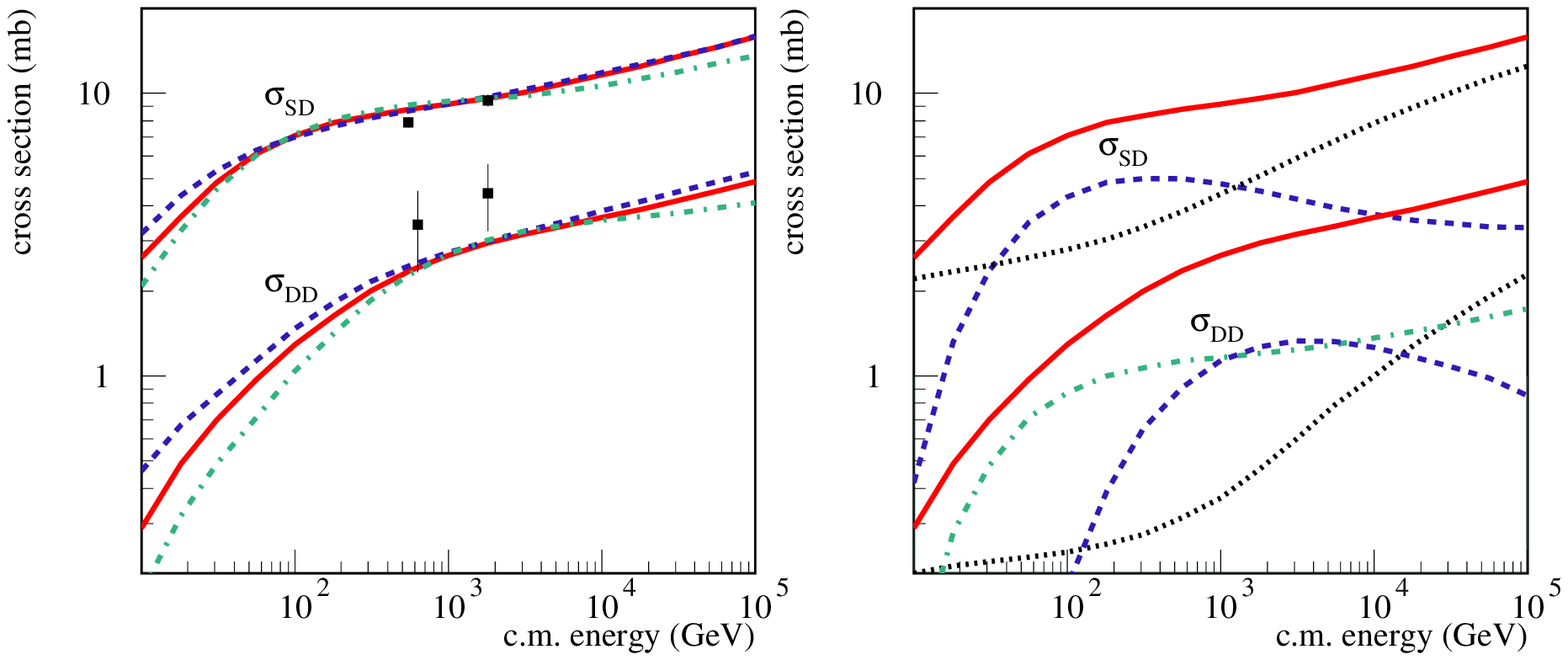}
\par\end{centering}

\caption{Left: single and double diffraction proton-proton cross sections ($\sigma^{{\rm SD}}(M_{X}^{2}/s<0.15)$,
$\sigma^{{\rm DD}}(y_{{\rm gap}}^{(0)}\geq3)$), as calculated using
the parameter sets (A), (B), and (C) - solid, dashed and dot-dashed
lines correspondingly, compared to CDF data \cite{abe94,aff01}. Right:
$\sigma^{{\rm SD}}(M_{X}^{2}/s<0.15)$ and $\sigma^{{\rm DD}}(y_{{\rm gap}}^{(0)}\geq3)$)
calculated using the parameter set (A) - solid lines, partial contributions
of high and low mass diffraction: $\sigma_{{\rm HM}}^{{\rm SD/DD}}$
and $\sigma_{{\rm LM}}^{{\rm SD/DD}}$ - dashed and dotted lines correspondingly,
$\sigma_{{\rm LHM}}^{{\rm DD}}$ - dot-dashed line.\label{fig:sigdifr}}

\end{figure}

The results of the calculations for total and elastic proton-proton
cross sections and for the forward elastic scattering slope, using
the three parameter sets obtained, are plotted in Fig.~\ref{fig:sigtot}
in comparison with experimental data. The calculated differential
elastic cross sections are presented in Fig.~\ref{fig:dsigel}. %
And in Fig.~\ref{fig:sigdifr}~(left) %
we show, in comparison with CDF measurements \cite{abe94,aff01},
single diffraction cross section $\sigma^{{\rm SD}}$ for diffractive
mass squared $M_{X}^{2}<0.15\, s$ and double diffraction cross section
$\sigma^{{\rm DD}}(y_{{\rm gap}}^{(0)}\geq3)$ corresponding to a
central rapidity gap of size $y_{{\rm gap}}\geq3$, which spans the
central rapidity $y=Y/2$ point. In addition, in Fig.~\ref{fig:sigdifr}~(right)
we plot partial contributions to $\sigma^{{\rm SD}}$ and $\sigma^{{\rm DD}}(y_{{\rm gap}}^{(0)}\geq3)$
from correspondingly single and double low ($\sigma_{{\rm LM}}^{{\rm SD}}$,
$\sigma_{{\rm LM}}^{{\rm DD}}$) and high ($\sigma_{{\rm HM}}^{{\rm SD}}$,
$\sigma_{{\rm HM}}^{{\rm DD}}$) mass diffraction, and a contribution
$\sigma_{{\rm LHM}}^{{\rm DD}}$ to $\sigma^{{\rm DD}}(y_{{\rm gap}}^{(0)}\geq3)$
from the process corresponding to a high mass diffraction of one hadron
and a low mass excitation of the other one. The numerical values for
$\sigma^{{\rm tot}}$, $\sigma^{{\rm el}}$, $\sigma^{{\rm SD}}$,
$\sigma^{{\rm DD}}$, central diffraction cross section $\sigma^{{\rm DPE}}$,
and for partial contributions to low and high mass diffraction cross
sections $\sigma_{{\rm LM}}^{{\rm SD/DD}}$, $\sigma_{{\rm HM}}^{{\rm SD/DD}}$,
$\sigma_{{\rm LHM}}^{{\rm DD}}$ for the CERN SPS, Tevatron, and LHC
energies are given in Tables~\ref{Flo:sigmas-spps}, \ref{Flo:sigmas-tevatron},
\ref{Flo:sigmas-lhc} correspondingly, with the diffraction cross
sections now being calculated for the minimal allowed rapidity gap
size $y_{{\rm gap}}=\xi$. The values in brackets for $\sigma_{{\rm HM}}^{{\rm SD}}$,
$\sigma_{{\rm LHM}}^{{\rm DD}}$ , and $\sigma^{{\rm DPE}}$ in Tables~\ref{Flo:sigmas-spps},
\ref{Flo:sigmas-tevatron}, \ref{Flo:sigmas-lhc} are obtained if
the contributions of low mass diffractive states produced at central
rapidities are neglected. %
\begin{table}[t]
\centering{}{\footnotesize }\begin{tabular}{|c|c|c|c|c|c|c|c|c|c|c|}
\hline 
 & {\footnotesize $\sigma^{{\rm tot}}$} & {\footnotesize $\sigma^{{\rm el}}$} & {\footnotesize $\sigma^{{\rm SD}}$ } & {\footnotesize $\sigma^{{\rm DD}}$ } & {\footnotesize $\sigma_{{\rm LM}}^{{\rm SD}}$ } & {\footnotesize $\sigma_{{\rm HM}}^{{\rm SD}}$ } & {\footnotesize $\sigma_{{\rm LM}}^{{\rm DD}}$ } & {\footnotesize $\sigma_{{\rm HM}}^{{\rm DD}}$ } & {\footnotesize $\sigma_{{\rm LHM}}^{{\rm DD}}$ } & {\footnotesize $\sigma^{{\rm DPE}}$}\tabularnewline
\hline
\hline 
{\footnotesize Set (A)} & {\footnotesize 61.0} & {\footnotesize 12.7} & {\footnotesize 8.79} & {\footnotesize 2.84} & {\footnotesize 3.80} & {\footnotesize 5.00 (4.82)} & {\footnotesize 0.31} & {\footnotesize 1.40 } & {\footnotesize 1.13 (1.08)} & {\footnotesize 0.24 (0.07)}\tabularnewline
\hline 
{\footnotesize Set (B)} & {\footnotesize 62.1} & {\footnotesize 13.2} & {\footnotesize 8.85} & {\footnotesize 3.32} & {\footnotesize 3.76} & {\footnotesize 5.08 (4.94)} & {\footnotesize 0.30} & {\footnotesize 1.92} & {\footnotesize 1.11 (1.07)} & {\footnotesize 0.26 (0.11)}\tabularnewline
\hline 
{\footnotesize Set (C)} & {\footnotesize 58.4} & {\footnotesize 11.7} &
{\footnotesize 9.08} & {\footnotesize
2.91} & {\footnotesize 2.60} & {\footnotesize 6.48 (6.26)} & {\footnotesize
0.14} & {\footnotesize 1.81} & {\footnotesize 0.95 (0.92)} & {\footnotesize 0.33 (0.09)}\tabularnewline
\hline
\end{tabular}\caption{Calculated total, elastic, and diffractive proton-proton cross sections
(in mb) for $\sqrt{s}=546$ GeV.}
\label{Flo:sigmas-spps}
\end{table}
\begin{table}[t]
\centering{}{\footnotesize }\begin{tabular}{|c|c|c|c|c|c|c|c|c|c|c|}
\hline 
 & {\footnotesize $\sigma^{{\rm tot}}$} & {\footnotesize $\sigma^{{\rm el}}$} & {\footnotesize $\sigma^{{\rm SD}}$ } & {\footnotesize $\sigma^{{\rm DD}}$ } & {\footnotesize $\sigma_{{\rm LM}}^{{\rm SD}}$ } & {\footnotesize $\sigma_{{\rm HM}}^{{\rm SD}}$ } & {\footnotesize $\sigma_{{\rm LM}}^{{\rm DD}}$ } & {\footnotesize $\sigma_{{\rm HM}}^{{\rm DD}}$ } & {\footnotesize $\sigma_{{\rm LHM}}^{{\rm DD}}$ } & {\footnotesize $\sigma^{{\rm DPE}}$}\tabularnewline
\hline
\hline 
{\footnotesize Set (A)} & {\footnotesize 79.3} & {\footnotesize 19.3} & {\footnotesize 9.62} & {\footnotesize 3.62} & {\footnotesize 5.10} & {\footnotesize 4.52 (4.38)} & {\footnotesize 0.48} & {\footnotesize 1.95} & {\footnotesize 1.20 (1.17)} & {\footnotesize 0.19 (0.08)}\tabularnewline
\hline 
{\footnotesize Set (B)} & {\footnotesize 80.5} & {\footnotesize 19.9} & {\footnotesize 9.84} & {\footnotesize 4.06} & {\footnotesize 5.08} & {\footnotesize 4.78 (4.66)} & {\footnotesize 0.45} & {\footnotesize 2.37} & {\footnotesize 1.24 (1.20)} & {\footnotesize 0.23 (0.11)}\tabularnewline
\hline 
{\footnotesize Set (C)} & {\footnotesize 73.0} & {\footnotesize 16.8} &
{\footnotesize 9.60} & {\footnotesize
3.93} & {\footnotesize 3.40} & {\footnotesize 6.20 (6.04)} & {\footnotesize
0.19} & {\footnotesize 2.70} & {\footnotesize 1.04 (1.01)} & {\footnotesize 0.31 (0.12)}\tabularnewline
\hline 
{\footnotesize Ref.~\cite{kmr08}} & {\footnotesize 73.7} & {\footnotesize 16.4} & {\footnotesize 13.8} &  & {\footnotesize 4.1} & {\footnotesize 9.7} &  &  &  & \tabularnewline
\hline 
{\footnotesize Ref.~\cite{glm08}} & {\footnotesize 73.3} & {\footnotesize 16.3} & {\footnotesize 9.76} & {\footnotesize 5.36} & {\footnotesize 8.56} & {\footnotesize 1.2} &  &  &  & \tabularnewline
\hline
\end{tabular}\caption{Same as in Table \ref{Flo:sigmas-spps} for $\sqrt{s}=1.8$ TeV.}
\label{Flo:sigmas-tevatron}
\end{table}
\begin{table}[t]
\centering{}{\footnotesize }\begin{tabular}{|c|c|c|c|c|c|c|c|c|c|c|}
\hline 
 & {\footnotesize $\sigma^{{\rm tot}}$} & {\footnotesize $\sigma^{{\rm el}}$} & {\footnotesize $\sigma^{{\rm SD}}$ } & {\footnotesize $\sigma^{{\rm DD}}$ } & {\footnotesize $\sigma_{{\rm LM}}^{{\rm SD}}$ } & {\footnotesize $\sigma_{{\rm HM}}^{{\rm SD}}$ } & {\footnotesize $\sigma_{{\rm LM}}^{{\rm DD}}$ } & {\footnotesize $\sigma_{{\rm HM}}^{{\rm DD}}$ } & {\footnotesize $\sigma_{{\rm LHM}}^{{\rm DD}}$ } & {\footnotesize $\sigma^{{\rm DPE}}$}\tabularnewline
\hline
\hline 
{\footnotesize Set (A)} & {\footnotesize 128} & {\footnotesize 37.5} & {\footnotesize 12.1} & {\footnotesize 4.61} & {\footnotesize 8.48} & {\footnotesize 3.62 (3.54)} & {\footnotesize 1.15} & {\footnotesize 2.06} & {\footnotesize 1.40 (1.37) } & {\footnotesize 0.10 (0.05)}\tabularnewline
\hline 
{\footnotesize Set (B)} & {\footnotesize 126} & {\footnotesize 37.3} & {\footnotesize 12.4} & {\footnotesize 5.18} & {\footnotesize 8.22} & {\footnotesize 4.24 (4.14)} & {\footnotesize 1.08} & {\footnotesize 2.50} & {\footnotesize 1.60 (1.56)} & {\footnotesize 0.14 (0.07)}\tabularnewline
\hline 
{\footnotesize Set (C)} & {\footnotesize 114} & {\footnotesize 33.0} &
{\footnotesize 11.0} & {\footnotesize 4.83}
& {\footnotesize 5.76} & {\footnotesize 5.22 (5.12)} & {\footnotesize 0.47} &
{\footnotesize 3.15} & {\footnotesize 1.22 (1.19)} & {\footnotesize 0.19 (0.09)}\tabularnewline
\hline 
{\footnotesize Ref.~\cite{kmr08}} & {\footnotesize 91.7} & {\footnotesize 21.5} & {\footnotesize 19.0} &  & {\footnotesize 4.9} & {\footnotesize 14.1} &  &  &  & \tabularnewline
\hline 
{\footnotesize Ref.~\cite{glm08}} & {\footnotesize 92.1} & {\footnotesize 20.9} & {\footnotesize 11.8} & {\footnotesize 6.08} & {\footnotesize 10.5} & {\footnotesize 1.28} &  &  &  & \tabularnewline
\hline
\end{tabular}\caption{Same as in Table \ref{Flo:sigmas-spps} for $\sqrt{s}=14$ TeV.}
\label{Flo:sigmas-lhc}
\end{table}

Clearly, both the model parameters and the cross section results depend
rather weakly on the choice of the Pomeron mass cutoff. The obtained
diffractive cross sections appeared to be rather insensitive to whether
or not low mass diffractive states produced at central rapidities
are taken into account, with the exception of $\sigma^{{\rm DPE}}$
which is dominated by such contributions, more precisely, by the one
of the graph in Fig.~\ref{fig: tree-dif2}~(c), Eq.~(\ref{eq:DPE-LM}).
The observed high energy rise of total and elastic cross sections
is qualitatively similar for all the three parameter sets, with $\sigma^{{\rm tot}}$
reaching $114\div128$ mb at the LHC. This appears to be quite different
to the results of \cite{kmr08} and \cite{glm08},
where a much flatter high energy behavior of $\sigma^{{\rm tot}}$
and $\sigma^{{\rm el}}$ has been predicted, with the total cross
section value of 92 mb only at $\sqrt{s}=14$ TeV and the elastic
one being almost a factor of two smaller compared to our results with
the parameter sets (A) and (B).%
\footnote{Possible causes for these discrepancies will be analyzed in the next
Section.%
} The calculated differential elastic cross section at $\sqrt{s}=14$
TeV manifests a diffractive peak at $-t\simeq3.3$ GeV$^{2}$ for
the parameter sets (A) and (B), which is shifted towards $-t\simeq 3.8$
GeV$^{2}$ for the set (C).

It is interesting to compare the high energy behavior of partial contributions
to $\sigma^{{\rm SD}}$ from low and high mass diffraction processes
- see Fig.~\ref{fig:sigdifr}~(right) and Tables~\ref{Flo:sigmas-spps},
\ref{Flo:sigmas-tevatron}, \ref{Flo:sigmas-lhc}. While the high
energy trend of $\sigma_{{\rm LM}}^{{\rm SD}}$ resembles, as it should,
the one of $\sigma^{{\rm el}}$ (c.f.~Eqs.~(\ref{eq:sigma-el})
and (\ref{sigma-sd(lm)})), $\sigma_{{\rm HM}}^{{\rm SD}}$ reaches
its maximal values well below the LHC and slowly dies out in the very
high energy asymptotics. Such a tendency is well expected, as the
interaction approaches the black disk limit at $s\rightarrow\infty$;
the calculated ratios $\sigma^{{\rm el}}/\sigma^{{\rm tot}}$ rise
from $0.20\div0.21$ to $0.29\div0.30$ in the c.m.~energy range
between 546 GeV and 14 TeV. The probability for a rapidity gap not
to be covered by secondary particles produced in additional inelastic
rescattering processes becomes negligible at not too large impact
parameters; the diffractive configurations of final states can thus
survive in very peripheral collisions only. This tendency is supported
by the results in Tables~\ref{Flo:sigmas-spps}, \ref{Flo:sigmas-tevatron},
\ref{Flo:sigmas-lhc}: using the parameter set (C) which corresponds
to slightly smaller $\sigma^{{\rm el}}/\sigma^{{\rm tot}}$ ratios, we obtained
bigger values for $\sigma_{{\rm HM}}^{{\rm SD}}$. However, apart
from this well-known eikonal rapidity gap suppression described by
the RGS factor $S_{j_{1}j_{2}kk}$ (Eqs.~(\ref{eq:sigma-sd}-\ref{eq:rgs-fac})),
the observed high energy trend of $\sigma_{{\rm HM}}^{{\rm SD}}$
is the consequence of the unitarization of the contribution
 $2\chi_{j_{1}j_{2}k}^{1-{\rm gap}}(Y,b,y_{{\rm gap}})$
of the irreducible diffractively cut graph itself. It is worth stressing
that the results depend crucially on whether the corresponding absorptive
corrections are properly resummed. 

\begin{figure}[htb]
\begin{centering}
\includegraphics[width=15cm,height=6cm]{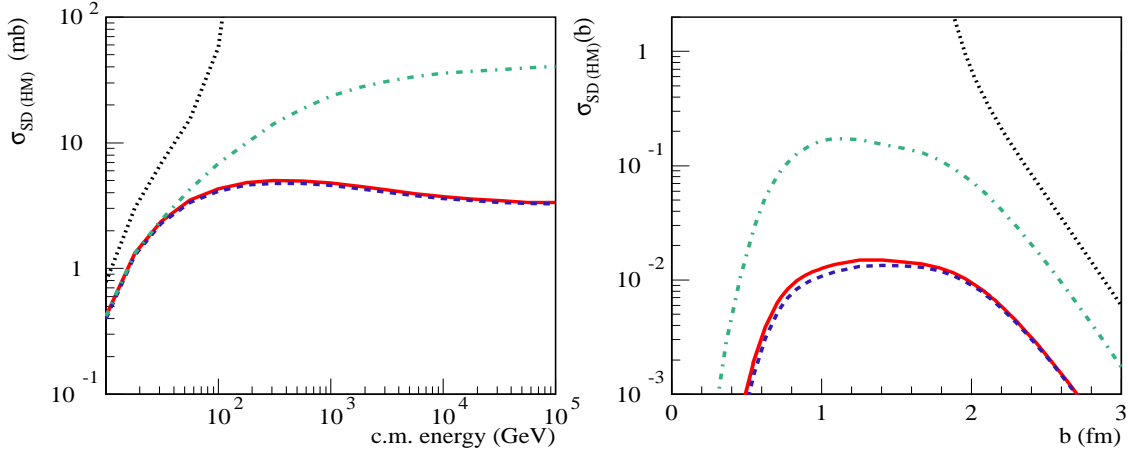}
\par\end{centering}

\caption{High energy behavior of $\sigma_{{\rm HM}}^{{\rm SD}}$ (left) and
single high mass diffraction profile at $\sqrt{s}=14$ TeV (right)
as calculated using the full resummation scheme, taking into account
the contributions of the graphs in Fig.~\ref{fig: tree-dif1}~(a,b),
using the lowest order result of Fig.~\ref{1-gap-fig}~(a), or keeping
only the triple-Pomeron contribution to $2\chi_{j_{1}j_{2}k}^{1-{\rm gap}}(Y,b,y_{{\rm gap}})$
- solid, dashed, dot-dashed, and dotted lines correspondingly (all
for the parameter set (A)).\label{fig:sigdifr-approx}}

\end{figure}

To illustrate this point, we calculated single high mass diffraction
cross sections for the parameter set (A) using Eq.~(\ref{eq:sigma-sd}),
taking into account all enhanced diagram contributions to elastic
scattering amplitude (i.e.~using $\chi_{jk}^{{\rm enh}}$ as defined
by Eq.~(\ref{chi-enh})) but employing partial resummations for the
contribution $2\chi_{j_{1}j_{2}k}^{1-{\rm gap}}(Y,b,y_{{\rm gap}})$:
i) as defined by the contributions of the graphs in Fig.~\ref{fig: tree-dif1}~(a,b),
given in Eq.~(\ref{eq:1gap-main}); ii) keeping only the lowest order
result of Fig.~\ref{1-gap-fig}~(a) - Eq.~(\ref{eq:1gap-Y-2});
iii) restricting ourselves with the triple-Pomeron contribution of
Eq.~(\ref{eq:3P}). The obtained high energy behavior of $\sigma_{{\rm HM}}^{{\rm SD}}(s)$
is plotted in Fig.~\ref{fig:sigdifr-approx}~(left)%
in comparison with the full resummation result. In addition, in Fig.~\ref{fig:sigdifr-approx}~(right)
we plot the corresponding profile functions $\sigma_{{\rm HM}}^{{\rm SD}}(s,b)$
for $\sqrt{s}=14$ TeV. While restricting oneself with the contribution
of the graphs in Fig.~\ref{fig: tree-dif1}~(a,b) to $2\chi_{j_{1}j_{2}k}^{1-{\rm gap}}(Y,b,y_{{\rm gap}})$
is a fairly good approximation, the corresponding values of $\sigma_{{\rm HM}}^{{\rm SD}}$
being less than 5\% different from the full resummation results, the
other two options considered prove to be very crude. In particular,
considering the triple-Pomeron contribution only, one violates the
$s$-channel unitarity: the high mass diffraction profile function
exceeds unity at small $b$ and $\sigma_{{\rm HM}}^{{\rm SD}}(s)$
increases more rapidly at $s\rightarrow\infty$ than the total cross
section.

\begin{figure}[htb]
\begin{centering}
\includegraphics[width=15cm,height=6cm]{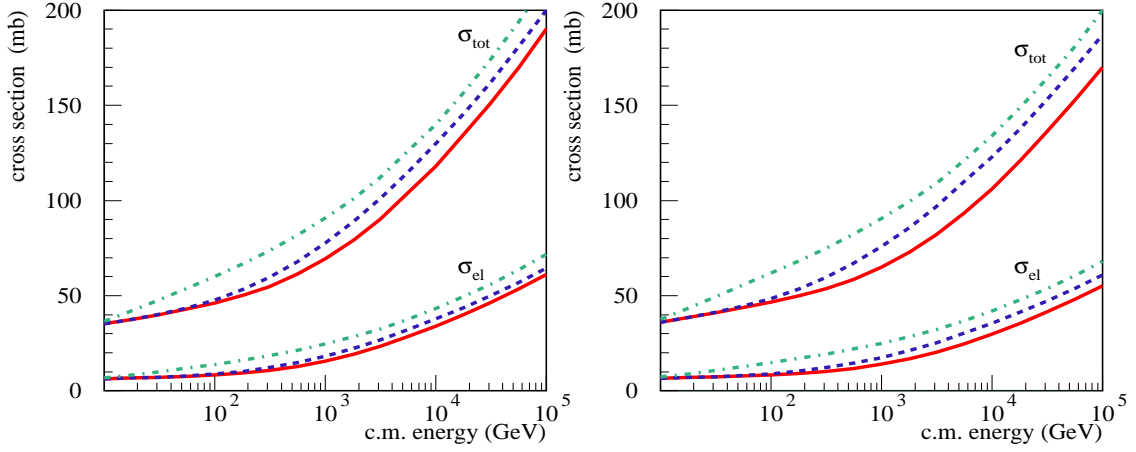}
\par\end{centering}

\caption{Total and elastic proton-proton cross sections based on the resummation
of all significant enhanced diagram contributions (solid lines), on
the resummation of net-like enhanced graphs (dashed lines), or taking
into account Pomeron loop diagrams only (dot-dashed lines), as calculated
using the  parameter set (A) -- left and (C) -- right.\label{fig:sigtcomp}}

\end{figure}

In view of partial resummations of selected classes of enhanced diagrams
in \cite{kmr08,glm08}, it is interesting
to investigate the relative importance of net-like and loop-like enhanced
graphs. Restricting ourselves with the net-like contributions and
neglecting Pomeron loops would bring us back to the approach of Ref.~\cite{ost06}.
In such a case, the contribution to elastic scattering amplitude of
all irreducible enhanced diagrams $\chi_{jk}^{{\rm enh}}$ is given
by Eq.~(\ref{chi-enh}), with the net-fan eikonal $\chi_{jk}^{{\rm net}}$
being defined in Eq.~(\ref{net-fan}), under the replacements $1-e^{-\chi^{{\rm loop}}}\rightarrow\chi^{\mathbb{P}}$,
$\chi_{j}^{{\rm loop}}\rightarrow\chi_{j}^{\mathbb{P}}$, $\chi_{j}^{{\rm loop}(1)}\rightarrow\chi_{j}^{\mathbb{P}}$
in both equations. On the other hand, taking into account Pomeron
loop diagrams only, we just have to keep the contributions of the
last two graphs in Fig.~\ref{enh-full}, i.e.~to retain just the
term in the 4th line of Eq.~(\ref{chi-enh}) in the integrand in
the r.h.s.~of the equation. The obtained energy dependencies of $\sigma^{{\rm tot}}$
and $\sigma^{{\rm el}}$ are compared in Fig.~\ref{fig:sigtcomp}
 with the full resummation results. It is easy to see that both classes
considered of enhanced diagrams provide important absorptive corrections
and none of them can be neglected in the high energy limit. Still,
 taking into consideration net-like graphs only, the calculated
cross sections come somewhat closer to the ones obtained using the
complete treatment. This is due to the fact that contributions of
Pomeron loops, when added to such a scheme, are damped at small impact
parameters by exponential factors, as noticed already in \cite{car74,kai86}
(such a suppression has been discussed in Section~\ref{sub:Simplest-enhanced-graphs}
for particular examples of diffractively cut graphs), while at large
$b$ they are suppressed by additional powers of the small triple-Pomeron
coupling.%
\footnote{In general,  relative importance of Pomeron loop corrections depends
on the parameters of the scheme, primarily, on the Pomeron slope and
on the value of the vertex parameter $\gamma_{\mathbb{P}}$ (see Eq.~(\ref{eq:g_mn})).
For smaller $\alpha'_{\mathbb{P}}$, the smallness of the triple-Pomeron
coupling $r_{3\mathbb{P}}$ is compensated at large impact parameters
by the factor $1/\alpha'_{\mathbb{P}}$. On the other hand, with $\gamma_{\mathbb{P}}\rightarrow0$
the eikonal suppression of loop contributions vanishes. However, as
discussed above, the consistency requirement severely restricts the
possible range of $\gamma_{\mathbb{P}}$ values: $\gamma_{\mathbb{P}}>r_{3\mathbb{P}}/\Delta$
(see Eq.~(\ref{eq:inter-ren})). %
} On the other hand, the resummation of just Pomeron loop graphs is
far insufficient for cross section calculations.

\begin{figure}[htb]
\begin{centering}
\includegraphics[width=12cm,height=4cm]{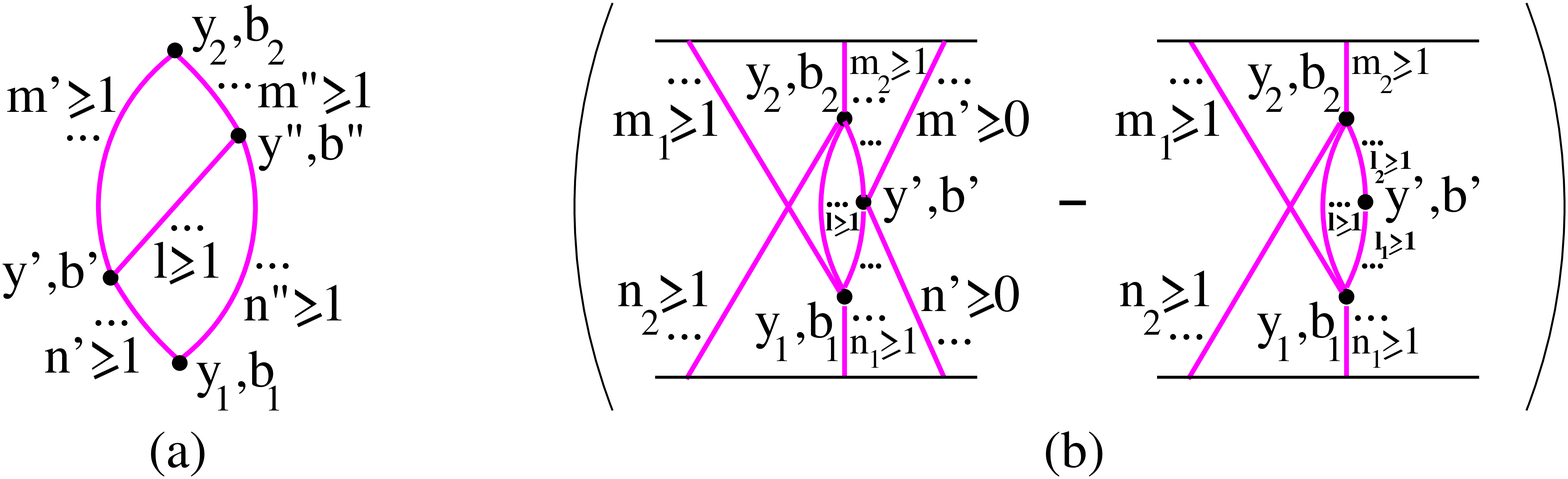}
\par\end{centering}

\caption{Considered corrections to the loop sequence contribution $\Delta\chi^{{\rm loop}}$
(a) and to the total eikonal $\Delta\chi_{jk}^{{\rm enh}}$ (b).\label{Fig: loop-corr}}

\end{figure}

From the discussion above, one may expect that corrections
due to additional loop-like graphs of the kinds depicted in Fig.~\ref{loop-ex},
which are neglected in the present treatment, are sufficiently small.
To verify that, we calculated simplest contributions of both kinds
to elastic scattering amplitude. First, we enlarged the set of graphs
corresponding to irreducible 2-point sequences of Pomerons and Pomeron
loops by adding the diagram of Fig.~\ref{Fig: loop-corr}~(a),%
 where we consider any number $(\geq 1)$ of Pomerons exchanged between
the vertices $(y_{1},\vec{b}_{1})$ and $(y',\vec{b}')$, $(y_{1},\vec{b}_{1})$
and $(y'',\vec{b}'')$, $(y',\vec{b}')$ and $(y'',\vec{b}'')$, $(y',\vec{b}')$
and $(y_{2},\vec{b}_{2})$, $(y'',\vec{b}'')$ and $(y_{2},\vec{b}_{2})$,
which resulted in a modification
\[
\chi^{{\rm loop}}(y_{1}-y_{2},|\vec{b}_{1}-\vec{b}_{2}|)\rightarrow
\chi^{{\rm loop}}(y_{1}-y_{2},|\vec{b}_{1}-\vec{b}_{2}|)+
\Delta\chi^{{\rm loop}}(y_{1}-y_{2},|\vec{b}_{1}-\vec{b}_{2}|)\,,\]
where
\begin{eqnarray*}
\Delta\chi^{{\rm loop}}(y_{1}-y_{2},|\vec{b}_{1}-\vec{b}_{2}|)
=G\int_{y_{1+}\xi}^{y_{2}-2\xi}\! dy'\int_{y'+\xi}^{y_{2}-\xi}\! dy''
\int\! d^{2}b'\, d^{2}b''\;
\left[1-e^{-\chi^{\mathbb{P}}(y''-y',|\vec{b}''-\vec{b}'|)}\right]\\
\times\left[1-e^{-\chi^{\mathbb{P}}(y'-y_{1},|\vec{b}'-\vec{b}_{1}|)}\right]
\left[1-e^{-\chi^{\mathbb{P}}(y''-y_{1},|\vec{b}''-\vec{b}_{1}|)}\right]
\left[1-e^{-\chi^{\mathbb{P}}(y_{2}-y',|\vec{b}_{2}-\vec{b}'|)}\right]
\left[1-e^{-\chi^{\mathbb{P}}(y_{2}-y'',|\vec{b}_{2}-\vec{b}''|)}\right]\!.
\end{eqnarray*}
The so-redefined loop contributions have been used in (\ref{net-fan})
and (\ref{chi-enh}) to calculate the eikonals $\chi_{jk}^{{\rm net}}$
and $\chi_{jk}^{{\rm enh}}$, the latter being then applied for cross
section calculations. Alternatively, we added the contribution of
the graphs of Fig.~\ref{Fig: loop-corr}~(b), where the surface
of the loop is coupled to the projectile or/and target protons by
additional Pomeron exchanges, directly to the eikonal $\chi_{jk}^{{\rm enh}}$,
i.e. 
\[
\chi_{jk}^{{\rm enh}}(s,b)\rightarrow
\chi_{jk}^{{\rm enh}}(s,b)+\Delta\chi_{jk}^{{\rm enh}}(s,b)\,,\]
with
\begin{eqnarray*}
\Delta\chi_{jk}^{{\rm enh}}(s,b)=G\int_{\xi}^{Y-3\xi}\! dy_{1}
\int_{y_{1}+2\xi}^{Y-\xi}\! dy_{2}\int_{y_{1+}\xi}^{y_{2}-\xi}\! dy'
\int\! d^{2}b_{1}\, d^{2}b_{2}\, d^{2}b'\;
\left[1-e^{-\chi_{j}^{\mathbb{P}}(Y-y_{2},|\vec{b}-\vec{b}_{2}|)}\right]\\
\times\left[1-e^{-\chi_{k}^{\mathbb{P}}(y_{1},b_{1})}\right]
e^{-\chi_{j}^{\mathbb{P}}(Y-y_{1},|\vec{b}-\vec{b}_{1}|)
-\chi_{k}^{\mathbb{P}}(y_{2},b_{2})}
\left[1-e^{-\chi^{\mathbb{P}}(y_{2}-y_{1},|\vec{b}_{2}-\vec{b}_{1}|)}\right]
\left[1-e^{-\chi^{\mathbb{P}}(y_{2}-y',|\vec{b}_{2}-\vec{b}'|)}\right]\\
\times\left[1-e^{-\chi^{\mathbb{P}}(y'-y_{1},|\vec{b}'-\vec{b}_{1}|)}\right]
\left[1-e^{-\chi_{j}^{\mathbb{P}}(Y-y',|\vec{b}-\vec{b}'|)
-\chi_{k}^{\mathbb{P}}(y',b')}\right]\!.
\end{eqnarray*}
In both cases, we did not observe any changes for the computed cross
sections within the calculation accuracy (few per mille).

\section{Discussion of other approaches\label{sec:Comparison-with-other}}

\begin{figure}[htb]
\begin{centering}
\includegraphics[width=5cm,height=4cm]{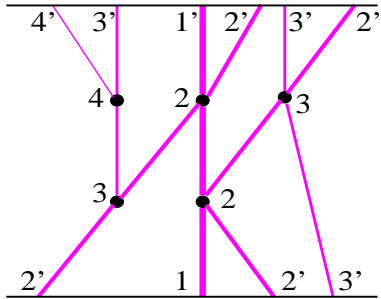}
\par\end{centering}

\caption{The structure of enhanced net-like graphs, corresponding to the parametrization
(\ref{eq:vertices-KMR}) of multi-Pomeron vertices.\label{Fig: more equal}}

\end{figure}

\begin{figure}[htb]
\begin{centering}
\includegraphics[width=10cm,height=3.5cm]{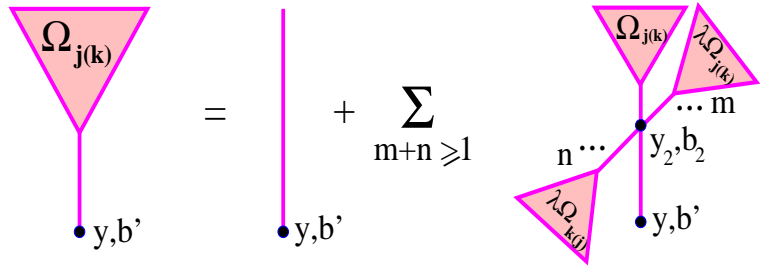}
\par\end{centering}

\caption{Recursive equation for parton opacity $\Omega_{j(k)}(y,\vec{b}',Y,\vec{b})$,
corresponding to the parametrization (\ref{eq:vertices-KMR}) of multi-Pomeron
vertices.\label{Fig: opacity}}

\end{figure}

In order to understand the differences of our treatment compared to
other approaches we are going to analyze the latter in some detail.
We shall mainly concentrate of the formalism of Ref.~\cite{kmr08},
which is to some extent similar to our treatment of Ref.~\cite{ost06}
in the sense that one takes into account multi-Pomeron vertices $G^{(m,n)}$
for arbitrary $m$, $n$ ($m+n\geq3$) and performs resummation of
net-like enhanced graphs (without Pomeron loops). The peculiarity
of the treatment of Ref.~\cite{kmr08} is that assuming the
vertices $G^{(m,n)}$ to be of the form\begin{equation}
G^{(m,n)}\sim n\, m\,\lambda^{m+n-2},\label{eq:vertices-KMR}\end{equation}
the $m$ (respectively $n$) Pomerons entering the vertex from the
projectile (target) side are no longer identical to each other, as
was the case in (\ref{eq:g_mn}). Instead, there is one ``more equal''
Pomeron above and one below the vertex, which represent the ``main
scattering'' process, while other Pomerons are related to absorptive
corrections. Thus, an arbitrary irreducible enhanced graph has a structure
which is exemplified in Fig.~\ref{Fig: more equal}.%
The main scattering process is formed by the sequence of Pomerons
shown symbolically by the thickest lines, marked (11'). Additional
rescatterings are described by Pomeron sequences drawn as less thick
lines, marked (22'). Those in turn undergo additional rescatterings
marked (33'), etc. This allows one to calculate the opacity for proton-proton
scattering $\Omega_{jk}(s,b)$ (Eq.~(\ref{eq:opac})) as a convolution
of two ``parton opacities'', which are ``glued'' together at some
rapidity $y$ \cite{kmr08}:
\begin{equation}
\Omega_{jk}(s,b)=\frac{1}{8\pi}\int\! d^{2}b'\;\Omega_{j(k)}(Y-y,|\vec{b}-\vec{b}'|,Y,\vec{b})\,\Omega_{k(j)}(y,\vec{b}',Y,\vec{b})\,.\label{eq:opacity}\end{equation}
The parton opacity $\Omega_{j(k)}$ satisfies the recursive equation
of Fig.~\ref{Fig: opacity}, %
 which is a particular case of the net-fan equation (c.f.~Fig.~\ref{freve}),
corresponding to (\ref{eq:vertices-KMR}). Fixing the normalization
in (\ref{eq:vertices-KMR}) by means of the triple-Pomeron coupling,
$G^{(m,n)}=r_{3\mathbb{P}}/(8\pi\lambda)\, n\, m\,\lambda^{m+n-2}$,
we get\begin{eqnarray}
\Omega_{j(k)}(y,\vec{b}',Y,\vec{b})=2\tilde{\chi}_{j}^{\mathbb{P}}(y,b')+\frac{r_{3\mathbb{P}}}{\lambda}\int_{\xi}^{y-\xi}\! dy_{2}\int\! d^{2}b_{2}\;\tilde{\chi}^{\mathbb{P}}(y-y_{2},|\vec{b}'-\vec{b}_{2}|)\;\Omega_{j(k)}(y_{2},\vec{b}_{2},Y,\vec{b})\nonumber \\
\times\left\{ e^{-\frac{\lambda}{2}\,\Omega_{j(k)}(y_{2},\vec{b}_{2},Y,\vec{b})-\frac{\lambda}{2}\,\Omega_{k(j)}(Y-y_{2},\vec{b}-\vec{b}_{2},Y,\vec{b})}-1\right\} ,\label{eq: p-opacity}\end{eqnarray}
where to come to conventions of Ref.~\cite{kmr08} we introduced
the eikonals $\tilde{\chi}_{j}^{\mathbb{P}}$, $\tilde{\chi}^{\mathbb{P}}$
related to $\chi_{j}^{\mathbb{P}}$, $\chi^{\mathbb{P}}$ (see Eqs.~(\ref{chi-legpom}-\ref{chi-intpom}))
as\begin{equation}
\tilde{\chi}_{j}^{\mathbb{P}}(y,b)=\frac{\chi_{j}^{\mathbb{P}}(y,b)}{\gamma_{\mathbb{P}}},\;\;\;\tilde{\chi}^{\mathbb{P}}(y,b)=\frac{\chi^{\mathbb{P}}(y,b)}{4\pi\gamma_{\mathbb{P}}^{2}}\,.\label{eq:conven}\end{equation}

Assuming now that the Pomeron propagator $D^{\mathbb{P}}(s,t)$ contains
a single Pomeron pole contribution and using \cite{ama62}\begin{equation}
\left(\frac{d}{dy}-\alpha'_{\mathbb{P}}(0)\,\Delta_{b}^{(2)}\right)\tilde{\chi}_{j}^{\mathbb{P}}(y,b)=\Delta\,\tilde{\chi}_{j}^{\mathbb{P}}(y,b)\label{eq:afs}\end{equation}
(similarly for $\tilde{\chi}^{\mathbb{P}}(y,b)$), with $\Delta=\alpha_{\mathbb{P}}-1$
and $\Delta_{b}^{(2)}$ being the 2-dimensional Laplacian, we obtain\begin{eqnarray}
\left(\frac{d}{dy}-\alpha'_{\mathbb{P}}(0)\,\Delta_{b'}^{(2)}\right)\Omega_{j(k)}(y,\vec{b}',Y,\vec{b})=\Delta\,\Omega_{j(k)}(y,\vec{b}',Y,\vec{b})+\frac{r_{3\mathbb{P}}}{\lambda}\int\! d^{2}b_{2}\nonumber \\
\times\tilde{\chi}^{\mathbb{P}}(\xi,|\vec{b}'-\vec{b}_{2}|)\;\Omega_{j(k)}(y-\xi,\vec{b}_{2},Y,\vec{b})\left\{ e^{-\frac{\lambda}{2}\,\Omega_{j(k)}(y-\xi,\vec{b}_{2},Y,\vec{b})-\frac{\lambda}{2}\,\Omega_{k(j)}(Y-y+\xi,\vec{b}-\vec{b}_{2},Y,\vec{b})}-1\right\} .\label{eq:net-deriv}\end{eqnarray}
Omitting the Pomeron mass cutoff, i.e.~considering the limit $\xi\rightarrow0$,
and using a particular relation between $r_{3\mathbb{P}}$, $\lambda$,
and $\Delta$\begin{equation}
r_{3\mathbb{P}}=\Delta\,\lambda\,,\label{eq:g3p-delta}\end{equation}
we obtain indeed the evolution equations for parton opacities as defined
in \cite{kmr08}:
\begin{eqnarray}
\left(\frac{d}{dy}-\alpha'_{\mathbb{P}}(0)\,\Delta_{b'}^{(2)}\right)\Omega_{j(k)}(y,\vec{b}',Y,\vec{b})=\Delta\,\Omega_{j(k)}(y,\vec{b}',Y,\vec{b})\; e^{-\frac{\lambda}{2}\,\Omega_{j(k)}(y,\vec{b}',Y,\vec{b})-\frac{\lambda}{2}\,\Omega_{k(j)}(Y-y,\vec{b}-\vec{b}',Y,\vec{b})}.\label{eq:evol-omega-1}\end{eqnarray}

It is worth stressing that it is the particular parametrization (\ref{eq:vertices-KMR})
for multi-Pomeron vertices which allowed one to obtain simple
relations (\ref{eq:opacity}-\ref{eq: p-opacity}) between the proton-proton
opacity and the net-fan contributions (parton opacities). The latter
are defined with respect to some point $(y,\vec{b}')$ which can be
chosen arbitrarily on the Pomeron sequence of the ``main scattering''
process (marked as (11') in Fig.~\ref{Fig: more equal}); no double
counting emerges in that case. On the contrary, using the usual ``symmetric''
multi-Pomeron vertices (\ref{eq:g_mn}), we had to arrange net-fans
with respect to some ``central'' vertex $(y_{1},\vec{b}_{1})$ in
Fig.~\ref{enh-full}. For any enhanced graph with $n$ internal vertices,
there are $n$ choices for the ``central'' vertex. Consequently, any
enhanced diagram with $n$ vertices is generated $n$ times by the
1st graph in Fig.~\ref{enh-full} and then subtracted $(n-1)$ times
by the 2nd graph of the Figure.

It is noteworthy, however, that the above-discussed simplification
has a   price to be paid, which is the artificial hierarchy of the
underlying parton cascades, depicted symbolically in Fig.~\ref{Fig: more equal}:
absorptive corrections to the main scattering process are described
by parton cascades which are distinguishable from the ``main stream'',
with the same arrangement continued in the sub-cascades. This may
happen, for example, if the main cascade is formed by harder partons
(of higher virtualities), whereas additional rescatterings are dominated
by softer partons. Keeping in mind that we discuss here contributions
to the elastic scattering amplitude, without any ``built-in trigger'',
rather than to particular final states, such an hierarchy does not
appear to be very natural.

The integral equation (\ref{eq: p-opacity}) for the opacity, when
compared to the net-fan equation (\ref{net-fan}), helps us to understand
the slower rise of $\sigma_{pp}^{{\rm tot}}$ in \cite{kmr08}
compared to our treatment. Indeed, making replacements $\chi_{j}^{{\rm loop}}(y_{1},b_{1})\rightarrow\chi_{j}^{\mathbb{P}}(y_{1},b_{1})$,
$1-e^{-\chi^{{\rm loop}}(y_{1}-y_{2},|\vec{b}_{1}-\vec{b}_{2}|)}\rightarrow\chi^{\mathbb{P}}(y_{1}-y_{2},|\vec{b}_{1}-\vec{b}_{2}|)$
in (\ref{net-fan}) in order to suppress Pomeron loop contributions
and bearing in mind the correspondence between the net-fan eikonals
and the parton opacities \[
\chi_{jk}^{{\rm net}}(y,\vec{b}'|Y,\vec{b})\rightarrow\frac{\lambda}{2}\,\Omega_{j(k)}(y,\vec{b}',Y,\vec{b})\]
as well as the changes in the normalizations (Eq.~\ref{eq:conven}),
it is easy to see that for the same choice of the Pomeron propagator
$D^{\mathbb{P}}(s,t)$, proton form factor $F_{j}^{\mathbb{P}}(t)$,
triple-Pomeron coupling $r_{3\mathbb{P}}$, and for $\gamma_{\mathbb{P}}=\lambda$,
Eq.~(\ref{eq: p-opacity}) implies stronger screening effects compared
to (\ref{net-fan}), which follows from \[
1-e^{-\frac{\lambda}{2}\,\Omega_{j(k)}(y_{2},\vec{b}_{2},Y,\vec{b})}>\frac{\lambda}{2}\,\Omega_{j(k)}(y_{2},\vec{b}_{2},Y,\vec{b})\; e^{-\frac{\lambda}{2}\,\Omega_{j(k)}(y_{2},\vec{b}_{2},Y,\vec{b})}\,.\]
This is, however, not the main reason for the above-mentioned difference.
To understand the main cause, let us notice that in the dense limit
($s\rightarrow\infty$, $b\rightarrow0$) the parton opacity $\Omega_{j(k)}(y,\vec{b}',Y,\vec{b})$,
as defined in Eq.~(\ref{eq: p-opacity}) for $\xi=0$, approaches
the asymptotic limit obtained in \cite{kai86}, which corresponds
to an exchange of a single ``renormalized'' Pomeron between the projectile
proton and the vertex $(y,\vec{b}')$:\begin{equation}
\Omega_{j(k)}(y,\vec{b}',Y,\vec{b})\rightarrow\tilde{\chi}_{j}^{\mathbb{P}_{{\rm ren}}}(y,b')\,,\label{eq:KMR-dense}\end{equation}
where $\tilde{\chi}_{j}^{\mathbb{P}_{{\rm ren}}}$ is defined by Eq.~(\ref{chi-legpom})
without the $\gamma_{\mathbb{P}}$ factor and with the renormalized
Pomeron intercept\begin{equation}
\alpha_{\mathbb{P}}^{{\rm ren}}=\alpha_{\mathbb{P}}-r_{3\mathbb{P}}/\lambda\,.\label{eq:KMR-pom-ren}\end{equation}
To see that, it is sufficient to notice that in the discussed limit
either $\Omega_{j(k)}(y_{2},\vec{b}_{2},Y,\vec{b})$ or/and $\Omega_{k(j)}(Y-y_{2},\vec{b}-\vec{b}_{2},Y,\vec{b})$
is large in the integrand in the r.h.s.~of (\ref{chi-legpom}). Hence,
the 1st term in the curly brackets vanishes and the equation has the
solution (\ref{eq:KMR-dense}). Using (\ref{eq:g3p-delta}) and taking
into account the relation (\ref{eq:opacity}) between parton opacites
and the total one, we immediately observe that the approach of 
Ref.~\cite{kmr08}
becomes equivalent in the dense limit to a non-enhanced multi-channel
eikonal scheme based on a critical Pomeron:\[
\Delta^{{\rm ren}}\equiv\alpha_{\mathbb{P}}^{{\rm ren}}-1=0\,.\]
Thus, it is the particular choice (\ref{eq:vertices-KMR}) for multi-Pomeron
vertices which is mainly responsible for the slow energy rise of $\sigma_{pp}^{{\rm tot}}$
in \cite{kmr08}.%
\footnote{This can be seen also from Eq.~(\ref{eq:evol-omega-1}): with the
r.h.s.~of the equation vanishing in the dense limit, the $y$-dependence
of $\Omega_{k(j)}$ comes solely from the diffusion in impact parameter
space.%
}

On the other hand, the cross sections for diffractive processes, derived
in \cite{kmr08} from heuristic arguments, are incompatible
with the traditional RFT treatment. For example, single high mass
diffraction cross section, defined as \cite{kmr08}%
\footnote{Strictly speaking, Eq.~(\ref{eq:sigma_sd-KMR}) is designed
to give the sum of
the cross sections for single high mass diffraction and for double
(low mass + high mass) diffraction.%
}\begin{eqnarray}
\sigma_{{\rm HM}}^{{\rm SD}}(s)=\sum_{j,k}C_{j}C_{k}\int\! dy\int\! d^{2}b\, d^{2}b'\left[1-e^{-\frac{\lambda}{2}\,\Omega_{j(k)}(Y-y,\vec{b}-\vec{b}',Y,\vec{b}))}\right]^{2}\Omega_{k(j)}(y,\vec{b}',Y,\vec{b})\nonumber \\
\times\,\Delta\; e^{-\frac{\lambda}{2}\,\Omega_{j(k)}(Y-y,\vec{b}-\vec{b}',Y,\vec{b})-\frac{\lambda}{2}\,\Omega_{k(j)}(y,\vec{b}',Y,\vec{b})}\; e^{-\Omega_{jk}(s,b)}\,,\label{eq:sigma_sd-KMR}\end{eqnarray}
is in variance with the old result \cite{abr92} (Eq.~(\ref{eq:1gap-Y-2}))
already at the lowest order in the triple-Pomeron coupling. Indeed,
keeping only single Pomeron contributions to all the opacities and
omitting the RGS factor $e^{-\Omega_{jk}(s,b)}$, the integrand of
(\ref{eq:sigma_sd-KMR}) takes the form\begin{equation}
\left[1-e^{-\lambda\,\tilde{\chi}_{j}(Y-y,|\vec{b}-\vec{b}'|)}\right]^{2}\tilde{\chi}_{k}(y,b')\,\Delta\, e^{-\lambda\,\tilde{\chi}_{j}(Y-y,|\vec{b}-\vec{b}'|)-\lambda\,\tilde{\chi}_{k}(y,b')},\label{eq:sigsd-KMR-(1)}\end{equation}
which is very different from the integrand in (\ref{eq:1gap-Y-2})
for $j_{1}=j_{2}=j$. On the other hand, if we apply the AGK cutting
rules to calculate the lowest order contribution of Fig.~\ref{1-gap-fig}~(a)
and use the vertices (\ref{eq:vertices-KMR}), we obtain, keeping
the same conventions as in (\ref{eq:sigsd-KMR-(1)}),
\[
2\Delta\,\tilde{\chi}_{j}(Y-y,|\vec{b}-\vec{b}'|)\,
\tilde{\chi}_{k}(y,b')\,
\left[1-e^{-\lambda\,\tilde{\chi}_{j}(Y-y,|\vec{b}-\vec{b}'|)}\right]
e^{-\lambda\,\tilde{\chi}_{j}(Y-y,|\vec{b}-\vec{b}'|)
-2\lambda\,\tilde{\chi}_{k}(y,b')},
\]
which is still in variance to (\ref{eq:sigsd-KMR-(1)}). Given the
importance of the proper resummation of absorptive corrections for
diffractive topologies, demonstrated in the preceding Section, the
different energy behavior of single high mass diffraction cross section,
if calculated using Eq.~(\ref{eq:sigma_sd-KMR}), may not be surprising.
However, the observed difference between the high energy trend of
$\sigma_{{\rm HM}}^{{\rm SD}}$ in \cite{kmr08} compared
to our approach is at least partly caused by much stronger absorptive
effects for parton opacities in the former case, as discussed above.
The main suppression of LRG topologies is due to the eikonal RGS factor
$S_{j_{1}j_{2}kk}(s,b)$ (Eq.~(\ref{eq:rgs-fac})) in (\ref{eq:sigma-sd})
($e^{-\Omega_{jk}(s,b)}$ in (\ref{eq:sigma_sd-KMR})), which pushes
the diffractive production away to larger and larger impact parameters
with increasing energy. Using the vertices (\ref{eq:vertices-KMR}),
one obtains a much slower energy rise of parton opacities $\Omega_{j(k)}$,
hence, of the total opacity $\Omega_{jk}$, resulting in a \textsl{higher
probability for a LRG to survive}. Such an assumption is supported
by the much slower approach to the black disk limit in \cite{kmr08}.
Indeed, one can see from Tables~\ref{Flo:sigmas-tevatron}, \ref{Flo:sigmas-lhc}
that in their case $\sigma^{{\rm el}}/\sigma^{{\rm tot}}\simeq0.23$
at $\sqrt{s}=14$ TeV, compared to our values
 $\sigma^{{\rm el}}/\sigma^{{\rm tot}}\sim0.29\div0.30$.

Concerning the treatment of Refs.~\cite{glm08}, we remark
that the approach is based on a resummation of Pomeron loop diagrams
only, which, as demonstrated in Section~\ref{sec:Numerical-results},
is insufficient for cross section calculations. The contributions
of net-like graphs, more precisely of \textsl{selected unitarity cuts}
of fan-like diagrams, are included when calculating diffractive cross
sections but neglected in the calculation of elastic scattering amplitude.
In addition, the treatment takes into consideration the triple-Pomeron
vertex only, neglecting other multi-Pomeron vertices $G^{(m,n)}$
for $m+n>3$. Such an approach is known to predict total cross section
which becomes constant in the very high energy limit.

\section{Outlook\label{sec:Outlook}}

In this work, we presented the first systematic analysis of the high
energy behavior of total and diffractive proton-proton cross sections
within the RFT framework, based on all-order resummation of all important
contributions of enhanced Pomeron diagrams. In particular, we demonstrated
that absorptive corrections caused by net-like enhanced graphs and
by Pomeron loop contributions are both significant and none of those
classes can be neglected in the high energy limit. On the other hand,
we illustrated the importance of a proper resummation of absorptive
corrections for diffractively cut sub-graphs; restricting oneself
with the cut triple-Pomeron diagram only, as is often done in literature,
one comes into conflict with the $s$-channel unitarity  due to the
power-like energy rise of the corresponding contribution.

Compared to alternative approaches \cite{kmr08,kmr09,glm08}
which are based on resummations of restricted sets of enhanced diagrams,
we obtained a faster energy rise of total proton-proton cross section,
with our values of $\sigma_{pp}^{{\rm tot}}$ being some $25\div40$\%
higher at the LHC energy. On the other hand, the calculated single
high mass diffraction cross section reaches its maximum in the CERN
SPS~-~Tevatron energy range and slowly decreases at higher energies,
with $\sigma_{pp}^{{\rm SD}}\simeq 11\div12$ mb at $\sqrt{s}=14$ TeV. We
demonstrated that the different behavior of $\sigma_{pp}^{{\rm tot}}$
and $\sigma_{{\rm HM}}^{{\rm SD}}$ observed in \cite{kmr08},
with the former reaching $\sim90$ mb at the LHC energy and the latter
continuing to rise logarithmically up to very high energies is mainly
due to the specific parametrization (\ref{eq:vertices-KMR}) employed
for multi-Pomeron vertices, which leads to a critical Pomeron scheme
in the dense limit.

Still, with our analysis based on the old RFT treatment, it bears
all the drawbacks inherent to that framework: eikonal approximation
which neglects energy-momentum correlations between multiple scattering
processes \cite{bra90}, the assumption on the validity
of the AGK cutting rules, in particular, that multi-Pomeron vertices
remain unmodified by the cutting procedure, etc. Thus, it is up to
the LHC data to decide if the described approach has something in
common to the reality or  is just a solution of an abstract mathematical
problem.

\subsection*{Acknowledgments}

The author acknowledges fruitful discussions with A.~Kaidalov, V.~Khose,
U.~Maor, M.~Ryskin and the support of the European Comission under
the Marie Curie IEF program (grant 220251).

\section*{Appendix}

In the following, we are going to derive contributions of unitarity
cuts of elastic scattering diagrams, which correspond to various diffractive
topologies of hadronic final states, as depicted in Fig.~\ref{diffr-fig},
with single or multiple rapidity gaps not covered by secondary particle
production. The complete set of AGK-based cut enhanced graphs has
been derived in \cite{ost08}, the corresponding contributions being
expressed via the ones of cut net-fan sub-graphs. The principal observation
made in \cite{ost08} was that all cut enhanced diagrams can be divided
in two classes. The diagrams of the first kind are characterized by
a ``tree''-like structure of \textsl{cut Pomerons}; they consist of
cut and uncut net-fans coupled together in some ``central'' (not necessarily
unique) vertex, such that each of the cut net-fan sub-graphs is characterized
by a fan-like structure of cut Pomerons. All other graphs belong to
the second class; they contain sequences of \textsl{cut Pomerons}
arranged in a ``zigzag'' way, with subsequent Pomeron end rapidities
satisfying $y_{1}>y_{2}<y_{3}>...$; they are expressed via zigzag-like
cuts of net-fan sub-graphs.%
\footnote{More detailed discussion and examples of diagrams of both kinds can
be found in \cite{ost06a,ost08}.%
} Most importantly, such diagrams give zero contribution to the total
cross section and negligible corrections for diffractive ones. Thus,
in our analysis we can safely restrict ourselves with the tree-like
cut graphs only. In the following, we shall start with the full set
of the corresponding diagrams, derived in \cite{ost08}, selecting
from them the ones which lead to the desirable diffractive topologies
of final states. For certain graphs, this will require to replace
the general cut net-fan contributions (with the fan-like structure
of cut Pomerons) by the ones corresponding to a rapidity gap between
a given (say, projectile) proton and the nearest cut multi-Pomeron
vertex.

\begin{figure}[htb]
\begin{centering}
\includegraphics[width=15cm,height=6cm]{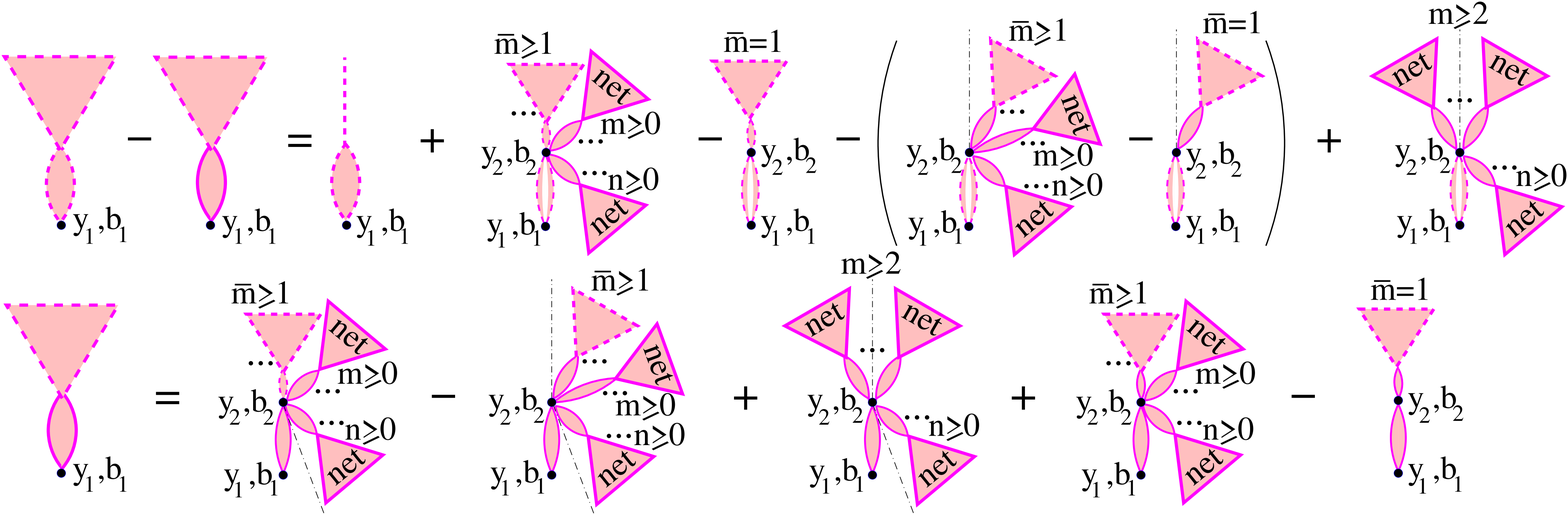}
\par\end{centering}

\caption{Recursive representations for cut net-fan diagrams characterized by
a fan-like structure of cuts. The top line of the Figure defines the
contribution $2\hat{\chi}_{jk}^{{\rm fan}}$ of the subset of graphs,
in which the handle of the fan is cut (the symbolic drawing for the
cut handle is explained in Fig.~\ref{fig:cut loop}), whereas the
bottom line gives the one of the diagrams, where the cut plane (indicated
by dot-dashed lines) goes aside the vertex $(y_{1},b_{1})$, $2\tilde{\chi}_{jk}^{{\rm fan}}$.
\label{fan-cut-fig}}

\end{figure}

The complete set of AGK-based cuts of net-fan diagrams, characterized
by the fan-like structure of cuts, is generated by the Schwinger-Dyson
equations of Fig.~\ref{fan-cut-fig} \cite{ost08}. %
\begin{figure}[htb]
\begin{centering}
\includegraphics[width=7cm,height=3.5cm]{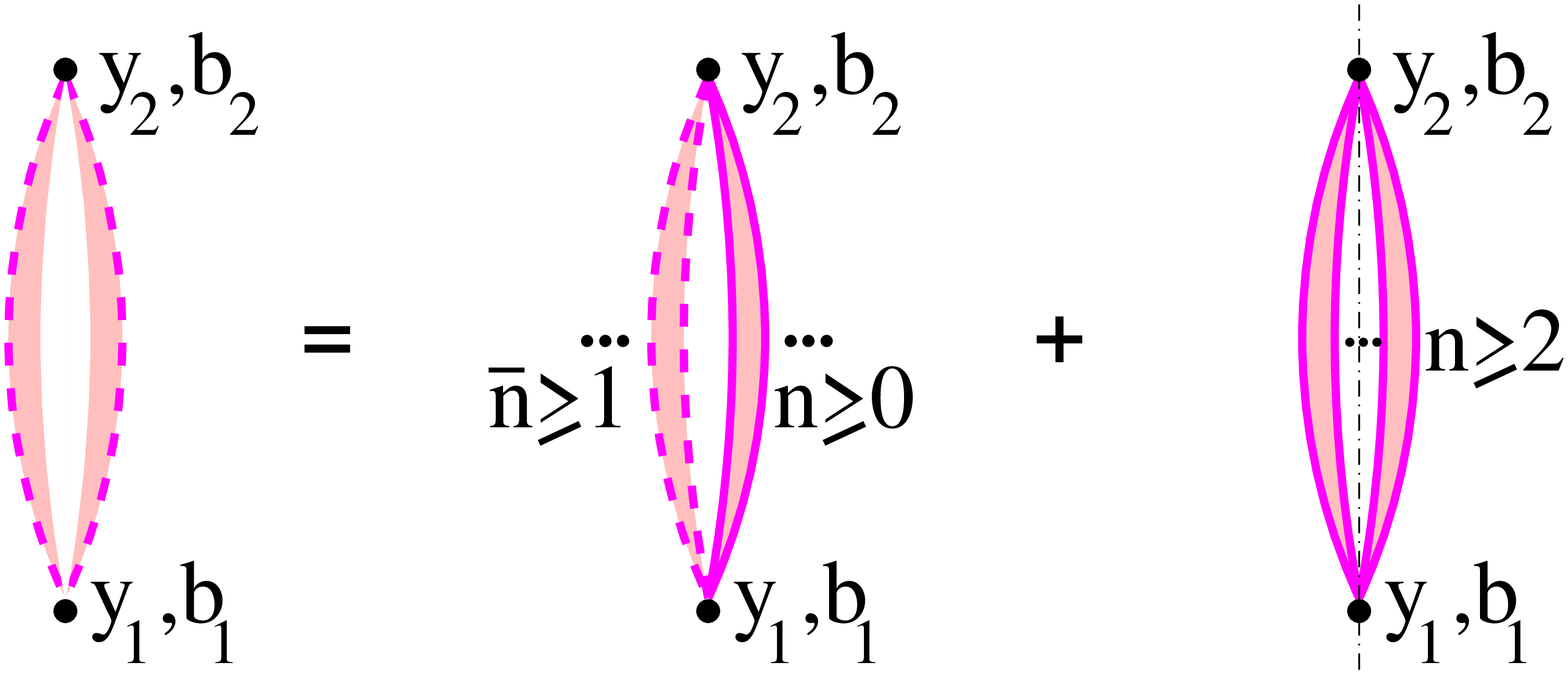}
\par\end{centering}

\caption{Cutting a general 2-point sequence of Pomerons and Pomeron loops,
exchanged between two cells of the Pomeron net (vertices $(y_{1},b_{1})$
and $(y_{2},b_{2})$ in the picture), one obtains two kinds of contributions:
i) with $\bar{n}\geq1$ cut and $n\geq0$ uncut irreducible 2-point
loop sequences (1st graph in the r.h.s.); ii) with the cut plane passing
between $n\geq2$ uncut irreducible 2-point loop sequences (2nd graph
in the r.h.s.).\label{fig:cut loop}}

\end{figure}
The top line of the Figure defines the contribution $2\hat{\chi}_{jk}^{{\rm fan}}$
of the subset, in which the handle of the fan is cut, i.e.~with the
cut plane passing through the original vertex $(y_{1},b_{1})$. In
such a case, one or a number of irreducible 2-point sequences of Pomerons
and Pomeron loops, connected to the vertex $(y_{1},b_{1})$, is cut,
as depicted in the 1st graph in the r.h.s.~of Fig.~\ref{fig:cut loop},
or, alternatively, the cut plane passes between those sequences as
in the 2nd graph in the r.h.s.~of Fig.~\ref{fig:cut loop}. In turn,
the bottom line of Fig.~\ref{fan-cut-fig} gives the contribution
$2\tilde{\chi}_{jk}^{{\rm fan}}$ of other fan-like cuts of net-fans,
where the handle of the fan is uncut, i.e.~when the cut plane passes
aside the vertex $(y_{1},b_{1})$.

Concerning the diagrams in the top line of Fig.~\ref{fan-cut-fig},
the first graph in the r.h.s.~corresponds to all possible AGK-based
cuts of the single 2-point sequence of Pomerons and Pomeron loops
exchanged between the vertex $(y_{1},b_{1})$ and the projectile proton
(the 1st graph in the r.h.s.~of Fig.~\ref{freve}). The 2nd diagram
describes the development of the cut Pomeron net, with the vertex
$(y_{2},b_{2})$ coupling together $\bar{m}\geq1$ cut projectile
net-fans, each one characterized by the fan-like structure of cuts,
and any numbers $m,n\geq0$ of uncut projectile and target net-fans.
There one has to subtract the contribution of the 3rd graph corresponding
to the Pomeron self-coupling ($\bar{m}=1$; $m,n=0$) and the ones
of the next two diagrams which correspond to a configuration of non-AGK
type, where in all the $\bar{m}$ cut net-fans connected to the vertex
$(y_{2},b_{2})$ the handles of the fans remain uncut and are situated
on the same side of the cut plane, together with all the $m$ uncut
projectile net-fans. Finally, in the last graph in the top line of
the Figure the cut plane passes between $m\geq2$ uncut projectile
net-fans with at least one of them remained on either side of the
cut, such that a rapidity gap is formed between the projectile proton
and the vertex $(y_{2},b_{2})$.

The first three diagrams in the r.h.s.~of the bottom line of Fig.~\ref{fan-cut-fig}
are similar to their corresponding counterparts in the top line, with
the difference that the handle of the fan is now uncut. Therefore,
there are $n\geq1$ uncut target net-fans connected to the vertex
$(y_{2},b_{2})$, such that at least one of them is positioned on
the opposite side of the cut plane with respect to the vertex $(y_{1},b_{1})$.
In turn, the 4th graph describes the situation when the cut plane
goes aside the vertex $(y_{2},b_{2})$. In such a case, there are
$\bar{m}\geq1$ projectile net-fans coupled to that vertex, which
are cut in a fan-like way and have their handles uncut and positioned
on the same side of the cut plane, together with any numbers $m\geq0$
of projectile and $n\geq0$ of target uncut net-fans. There one has
to subtract the Pomeron self-coupling ($\bar{m}=1$; $m,n=0$) described
by the last diagram in the line.

As demonstrated in \cite{ost08}, the total contribution $2\bar{\chi}_{jk}^{{\rm fan}}\equiv2\hat{\chi}_{jk}^{{\rm fan}}+2\tilde{\chi}_{jk}^{{\rm fan}}$
of fan-like cuts of net-fans coincides with twice the uncut one: \begin{eqnarray}
2\bar{\chi}_{jk}^{{\rm fan}}(y_{1},\vec{b}_{1}|Y,\vec{b})=2\chi_{jk}^{{\rm net}}(y_{1},\vec{b}_{1}|Y,\vec{b})\,,\label{equiv}\end{eqnarray}
whereas for the one of the subset with a cut handle one obtains the
recursive equation\begin{eqnarray}
2\hat{\chi}_{jk}^{{\rm fan}}(y_{1},\vec{b}_{1}|Y,\vec{b})=2\chi_{j}^{{\rm loop}}(y_{1},b_{1})+2G\int_{\xi}^{y_{1}-\xi}\! dy_{2}\int\! d^{2}b_{2}\;\left(1-e^{-\chi^{{\rm loop}}(y_{1}-y_{2},|\vec{b}_{1}-\vec{b}_{2}|)}\right)\nonumber \\
\times\left[\left(1-e^{-\hat{\chi}_{jk}^{{\rm fan}}(y_{2},\vec{b}_{2}|Y,\vec{b})}\right)e^{-2\chi_{kj}^{{\rm net}}(Y-y_{2},\vec{b}-\vec{b}_{2}|Y,\vec{b})}-\hat{\chi}_{jk}^{{\rm fan}}(y_{2},\vec{b}_{2}|Y,\vec{b})\right].\label{eq: fan-fan}\end{eqnarray}

However, in the following we shall need to choose elastic, respectively
inelastic, intermediate hadronic states for diffractive topologies,
as discussed in Section~\ref{sec:High-mass-diffraction}. Thus, we
have to obtain the contribution of fan-like cuts of net-fans $2\bar{\chi}_{jk_{1}k_{2}}^{{\rm fan}}$
of Fig.~\ref{fan-cut-fig}, considering different elastic scattering
eigenstates $|k_{1}\rangle$ and $|k_{2}\rangle$ to the left and
to the right of the cut plane for the partner (here, target) proton.
In addition, considering the graphs with an uncut handle (bottom line
of Fig.~\ref{fan-cut-fig}), we have to distinguish the contributions
when the vertex $(y_{1},b_{1})$ is positioned to the left ($\tilde{\chi}_{jk_{1}k_{2}}^{{\rm fan(l)}}$)
and to the right ($\tilde{\chi}_{jk_{1}k_{2}}^{{\rm fan(r)}}$) of
the cut. Proceeding as in Ref.~\cite{ost08}, we obtain instead of
(\ref{equiv}-\ref{eq: fan-fan})\begin{eqnarray}
2\bar{\chi}_{jk_{1}k_{2}}^{{\rm fan}}(y_{1},\vec{b}_{1}|Y,\vec{b})=\chi_{jk_{1}}^{{\rm net}}(y_{1},\vec{b}_{1}|Y,\vec{b})+\chi_{jk_{2}}^{{\rm net}}(y_{1},\vec{b}_{1}|Y,\vec{b})\,.\label{eq: equiv-new}\\
2\hat{\chi}_{jk_{1}k_{2}}^{{\rm fan}}(y_{1},\vec{b}_{1}|Y,\vec{b})=2\chi_{j}^{{\rm loop}}(y_{1},b_{1})+2G\int_{\xi}^{y_{1}-\xi}\! dy_{2}\int\! d^{2}b_{2}\;\left(1-e^{-\chi^{{\rm loop}}(y_{1}-y_{2},|\vec{b}_{1}-\vec{b}_{2}|)}\right)\nonumber \\
\times\left[\left(1-e^{-\hat{\chi}_{jk_{1}k_{2}}^{{\rm fan}}(y_{2},\vec{b}_{2}|Y,\vec{b})}\right)e^{-\chi_{k_{1}j}^{{\rm net}}(Y-y_{2},\vec{b}-\vec{b}_{2}|Y,\vec{b})-\chi_{k_{2}j}^{{\rm net}}(Y-y_{2},\vec{b}-\vec{b}_{2}|Y,\vec{b})}-\hat{\chi}_{jk_{1}k_{2}}^{{\rm fan}}(y_{2},\vec{b}_{2}|Y,\vec{b})\right]\label{eq:fan0(12)}\\
\tilde{\chi}_{jk_{1}k_{2}}^{{\rm fan(l)}}(y_{1},\vec{b}_{1}|Y,\vec{b})=\chi_{jk_{1}}^{{\rm net}}(y_{1},\vec{b}_{1}|Y,\vec{b})-\hat{\chi}_{jk_{1}k_{2}}^{{\rm fan}}(y_{1},\vec{b}_{1}|Y,\vec{b})\label{eq:fan-l}\\
\tilde{\chi}_{jk_{1}k_{2}}^{{\rm fan(r)}}(y_{1},\vec{b}_{1}|Y,\vec{b})=\chi_{jk_{2}}^{{\rm net}}(y_{1},\vec{b}_{1}|Y,\vec{b})-\hat{\chi}_{jk_{1}k_{2}}^{{\rm fan}}(y_{1},\vec{b}_{1}|Y,\vec{b})\,.\label{eq:fan-r}\end{eqnarray}

\begin{figure}[t]
\begin{centering}
\includegraphics[width=15cm,height=6cm]{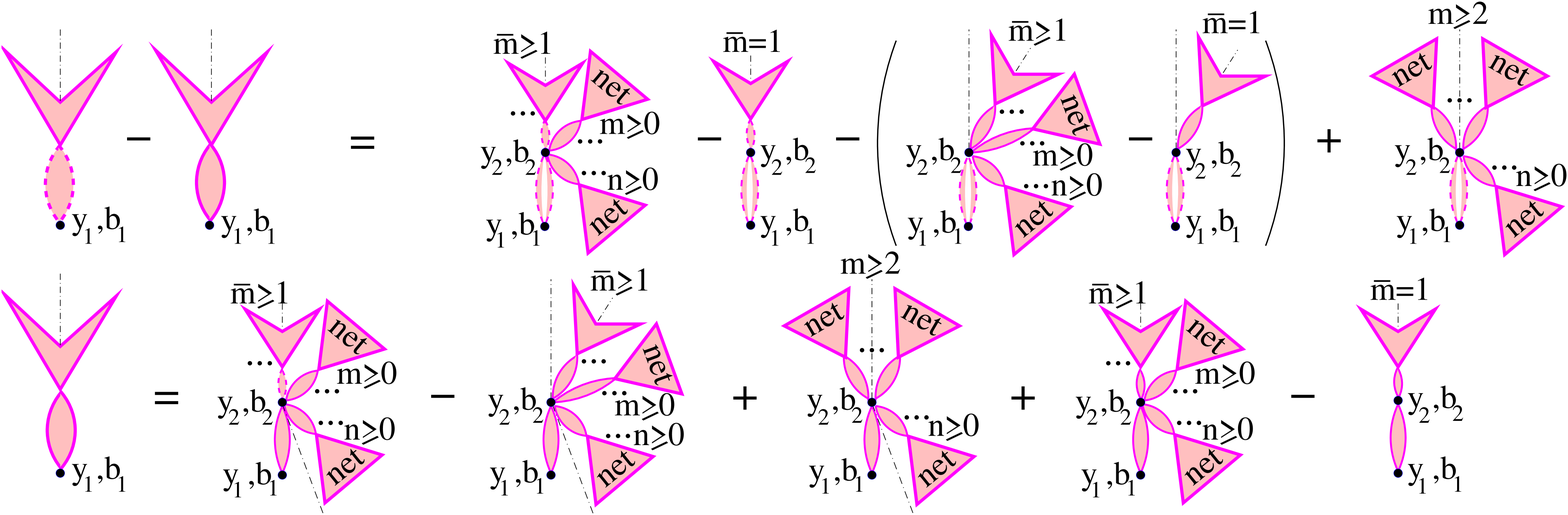}
\par\end{centering}

\caption{Recursive representations for diffractive cuts of net-fan diagrams:
the contribution $2\bar{\chi}_{j_{1}j_{2}k_{1}k_{2}}^{{\rm D}}-\tilde{\chi}_{j_{1}j_{2}k_{1}k_{2}}^{{\rm D(l)}}-\tilde{\chi}_{j_{1}j_{2}k_{1}k_{2}}^{{\rm D(r)}}$
of the graphs with the handle of the fan being cut (top line) and
the one with the uncut handle $\tilde{\chi}_{j_{1}j_{2}k_{1}k_{2}}^{{\rm D(l)}}+\tilde{\chi}_{j_{1}j_{2}k_{1}k_{2}}^{{\rm D(r)}}$
(bottom line). \label{fan-difr-fig}}

\end{figure}

Let us now derive contributions of diffractive cuts of net-fans, which
correspond to having a rapidity gap between the projectile proton%
\footnote{For definiteness, we speak here about diffractive cuts of the projectile
net-fan.%
} and all the secondary particles produced. The corresponding recursive
representations shown is Fig.~\ref{fan-difr-fig} %
 are easily obtained from the ones of Fig.~\ref{fan-cut-fig} by
merely replacing the contributions of general fan-like cuts of net-fan
graphs by the diffractive ones (shown as open ``forks''), each corresponding
to a rapidity gap of size $y_{{\rm gap}}$ or larger. Considering
now generally different elastic scattering eigenstates both for the
projectile and the target protons, we get\begin{eqnarray}
2\bar{\chi}_{j_{1}j_{2}k_{1}k_{2}}^{{\rm D}}\!\left(y_{1},\vec{b}_{1},y_{{\rm gap}}|Y,\vec{b}\right)=\frac{G}{2}\int_{\max(y_{{\rm gap}},\xi)}^{y_{1}-\xi}\! dy_{2}\int\! d^{2}b_{2}\;\left[1-e^{-\chi^{{\rm loop}}(y_{1}-y_{2},|\vec{b}_{1}-\vec{b}_{2}|)}\right]\nonumber \\
\times\left\{ \left(e^{-\chi_{k_{1}j_{1}}^{{\rm net}}(Y-y_{2},\vec{b}-\vec{b}_{2}|Y,\vec{b})}+e^{-\chi_{k_{2}j_{2}}^{{\rm net}}(Y-y_{2},\vec{b}-\vec{b}_{2}|Y,\vec{b})}\right)\left[\left(1-e^{-\chi_{j_{1}k_{1}}^{{\rm net}}(y_{2},\vec{b}_{2}|Y,\vec{b})}\right)\right.\right.\nonumber \\
\times\left.\left(1-e^{-\chi_{j_{2}k_{2}}^{{\rm net}}(y_{2},\vec{b}_{2}|Y,\vec{b})}\right)+\left(e^{2\bar{\chi}_{j_{1}j_{2}k_{1}k_{2}}^{{\rm D}}(y_{2},\vec{b}_{2},y_{{\rm gap}}|Y,\vec{b})}-1\right)e^{-\chi_{j_{1}k_{1}}^{{\rm net}}(y_{2},\vec{b}_{2}|Y,\vec{b})-\chi_{j_{2}k_{2}}^{{\rm net}}(y_{2},\vec{b}_{2}|Y,\vec{b})}\right]\nonumber \\
-4\bar{\chi}_{j_{1}j_{2}k_{1}k_{2}}^{{\rm D}}(y_{2},\vec{b}_{2},y_{{\rm gap}}|Y,\vec{b})+\left(e^{-\chi_{k_{1}j_{1}}^{{\rm net}}(Y-y_{2},\vec{b}-\vec{b}_{2}|Y,\vec{b})}-e^{-\chi_{k_{2}j_{2}}^{{\rm net}}(Y-y_{2},\vec{b}-\vec{b}_{2}|Y,\vec{b})}\right)\nonumber \\
\times\left.\left[\left(e^{\tilde{\chi}_{j_{1}j_{2}k_{1}k_{2}}^{{\rm D(l)}}(y_{2},\vec{b}_{2},y_{{\rm gap}}|Y,\vec{b})}-1\right)e^{-\chi_{j_{1}k_{1}}^{{\rm net}}(y_{2},\vec{b}_{2}|Y,\vec{b})}-\left(e^{\tilde{\chi}_{j_{1}j_{2}k_{1}k_{2}}^{{\rm D(r)}}(y_{2},\vec{b}_{2},y_{{\rm gap}}|Y,\vec{b})}-1\right)e^{-\chi_{j_{2}k_{2}}^{{\rm net}}(y_{2},\vec{b}_{2}|Y,\vec{b})}\right]\right\} \label{eq:chi-fand-tot}\\
\tilde{\chi}_{j_{1}j_{2}k_{1}k_{2}}^{{\rm D(l)}}(y_{1},\vec{b}_{1},y_{{\rm gap}}|Y,\vec{b})=\frac{G}{2}\int_{\max(y_{{\rm gap}},\xi)}^{y_{1}-\xi}\! dy_{2}\int\! d^{2}b_{2}\;\left[1-e^{-\chi^{{\rm loop}}(y_{1}-y_{2},|\vec{b}_{1}-\vec{b}_{2}|)}\right]\nonumber \\
\times\left\{ e^{-\chi_{k_{1}j_{1}}^{{\rm net}}(Y-y_{2},\vec{b}-\vec{b}_{2}|Y,\vec{b})}\left(1-e^{-\chi_{k_{2}j_{2}}^{{\rm net}}(Y-y_{2},\vec{b}-\vec{b}_{2}|Y,\vec{b})}\right)\left[\left(1-e^{-\chi_{j_{1}k_{1}}^{{\rm net}}(y_{2},\vec{b}_{2}|Y,\vec{b})}\right)\right.\right.\nonumber \\
\times\left(1-e^{-\chi_{j_{2}k_{2}}^{{\rm net}}(y_{2},\vec{b}_{2}|Y,\vec{b})}\right)+\left(e^{2\bar{\chi}_{j_{1}j_{2}k_{1}k_{2}}^{{\rm D}}(y_{2},\vec{b}_{2},y_{{\rm gap}}|Y,\vec{b})}-1\right)e^{-\chi_{j_{1}k_{1}}^{{\rm net}}(y_{2},\vec{b}_{2}|Y,\vec{b})-\chi_{j_{2}k_{2}}^{{\rm net}}(y_{2},\vec{b}_{2}|Y,\vec{b})}\nonumber \\
-\left.\left(e^{\tilde{\chi}_{j_{1}j_{2}k_{1}k_{2}}^{{\rm D(r)}}(y_{2},\vec{b}_{2},y_{{\rm gap}}|Y,\vec{b})}-1\right)e^{-\chi_{j_{2}k_{2}}^{{\rm net}}(y_{2},\vec{b}_{2}|Y,\vec{b})}\right]+\left(e^{\tilde{\chi}_{j_{1}j_{2}k_{1}k_{2}}^{{\rm D(l)}}(y_{2},\vec{b}_{2},y_{{\rm gap}}|Y,\vec{b})}-1\right)\nonumber \\
\times\left.e^{-\chi_{j_{1}k_{1}}^{{\rm net}}(y_{2},\vec{b}_{2}|Y,\vec{b})-\chi_{k_{1}j_{1}}^{{\rm net}}(Y-y_{2},\vec{b}-\vec{b}_{2}|Y,\vec{b})}\left(1+e^{-\chi_{k_{2}j_{2}}^{{\rm net}}(Y-y_{2},\vec{b}-\vec{b}_{2}|Y,\vec{b})}\right)-2\tilde{\chi}_{j_{1}j_{2}k_{1}k_{2}}^{{\rm D(l)}}(y_{2},\vec{b}_{2},y_{{\rm gap}}|Y,\vec{b})\right\} \label{eq:chi-fand-l}\\
\tilde{\chi}_{j_{1}j_{2}k_{1}k_{2}}^{{\rm D(r)}}(y_{1},\vec{b}_{1},y_{{\rm gap}}|Y,\vec{b})=\tilde{\chi}_{j_{2}j_{1}k_{2}k_{1}}^{{\rm D(l)}}(y_{1},\vec{b}_{1},y_{{\rm gap}}|Y,\vec{b})\,,\label{eq:chi-fand-r}\end{eqnarray}
where $2\bar{\chi}_{j_{1}j_{2}k_{1}k_{2}}^{{\rm D}}$ is the total
contribution of diffractively cut net-fan graphs, whereas $\tilde{\chi}_{j_{1}j_{2}k_{1}k_{2}}^{{\rm D(l)}}$
and $\tilde{\chi}_{j_{1}j_{2}k_{1}k_{2}}^{{\rm D(r)}}$ are the ones
with the handle of the fan being uncut and positioned respectively
to the left and to the right from the cut plane.

\begin{figure}[t]
\begin{centering}
\includegraphics[width=15cm,height=6cm]{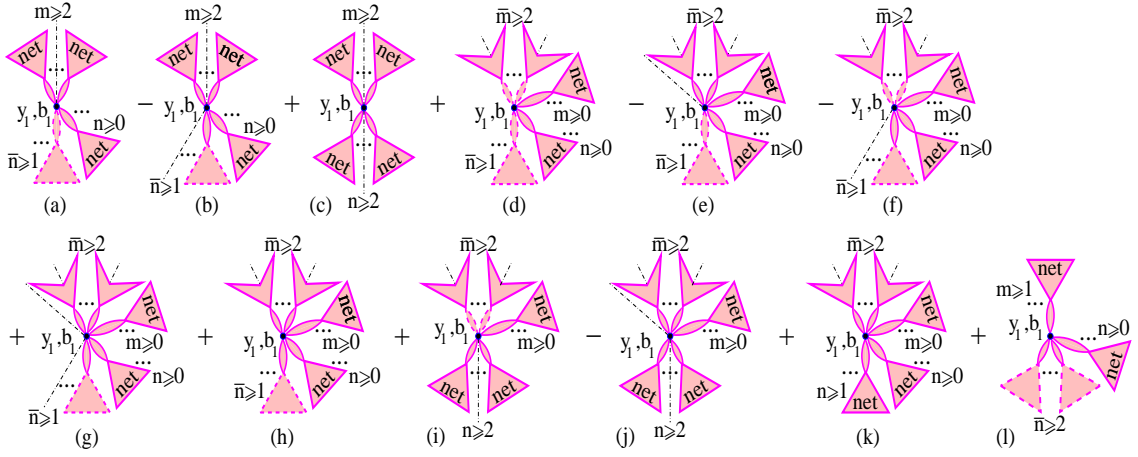}
\par\end{centering}

\caption{Complete set of irreducible cut diagrams which produce a LRG between
the quasi-elastically scattered projectile proton and all the secondary
particles produced. \label{fig: tree-dif1}}

\end{figure}

Now, starting from the full set of irreducible tree-like cut enhanced
diagrams derived in \cite{ost08} (Figs.~15-17,~19 of the paper),
we can obtain the complete subsets of cut enhanced graphs, which produce
rapidity gaps in the forward or/and in the backward direction by merely
replacing the contributions of fan-like cuts of net-fans $2\bar{\chi}^{{\rm fan}}$,
$\tilde{\chi}^{{\rm fan}}$ in the corresponding diagrams by the ones
of diffractive cuts $2\bar{\chi}^{{\rm D}}$, $\tilde{\chi}^{{\rm D(l/r)}}$,
like we did when deriving the representation of Fig.~\ref{fan-difr-fig}
for the latter. This way, for the sum of the two contributions $2\chi_{j_{1}j_{2}k}^{1-{\rm gap}}(Y,b,y_{{\rm gap}})+2\chi_{j_{1}j_{2}kk}^{2-{\rm gap}}(Y,b,y_{{\rm gap}},\xi)$
corresponding to a LRG of size $y_{{\rm gap}}$ or larger between
the quasi-elastically scattered projectile proton and all other particles
produced (and, possibly, an additional one in the backward direction)
we obtain the set of graphs of Fig.~\ref{fig: tree-dif1}.%

The diagram in Fig.~\ref{fig: tree-dif1}~(a) has explicitly the
$ABC$-like diffractive structure of Fig.~\ref{diffr-fig}~(a),
with the blocks $A$ and $B$ being represented by any numbers ($\geq1$)
of uncut projectile net-fans and the block $C$ containing $\bar{n}\geq1$
cut and $n\geq0$ uncut target net-fans. Here we have to subtract
the non-AGK type configurations of the graph in Fig.~\ref{fig: tree-dif1}~(b),
which corresponds to the situation when all the $\bar{n}$ cut target
net-fans have their handles uncut and positioned on the same side
of the cut, together with all the $n$ uncut target net-fans. The
set of diagrams generated by the graphs of Fig.~\ref{fig: tree-dif1}~(a,b)
contains all the lowest order ones of Fig.~\ref{1-gap-fig} and of
Fig.~\ref{2-gap-fig}~(f-h), dominating the diffractive contribution
$2\chi^{1-{\rm gap}}$. Applying the Reggeon diagram technique and
taking into account (\ref{eq: equiv-new}-\ref{eq:fan-r}), we obtain
for them\begin{eqnarray}
G\int_{\xi}^{\min(Y-\xi,Y-y_{{\rm gap}})}\! dy_{1}\int\! d^{2}b_{1}\;\left(1-e^{-\chi_{j_{1}k}^{{\rm net}}(Y-y_{1},\vec{b}-\vec{b}_{1}|Y,\vec{b})}\right)\left(1-e^{-\chi_{j_{2}k}^{{\rm net}}(Y-y_{1},\vec{b}-\vec{b}_{1}|Y,\vec{b})}\right)\nonumber \\
\times\left[1-e^{-\hat{\chi}_{kj_{1}j_{2}}^{{\rm fan}}\!(y_{1},\vec{b}_{1}|Y,\vec{b})}-\frac{1}{2}\left(1-e^{-\chi_{kj_{1}}^{{\rm net}}(y_{1},\vec{b}_{1}|Y,\vec{b})}\right)\left(1-e^{-\chi_{kj_{2}}^{{\rm net}}(y_{1},\vec{b}_{1}|Y,\vec{b})}\right)\right].\label{eq:1gap-main}\end{eqnarray}

In turn, the set of diagrams generated by the graph in Fig.~\ref{fig: tree-dif1}~(c)
contains the ones of Fig.~\ref{2-gap-fig}~(a-e) and describes the
production of a low mass diffractive state at central rapidities,
with the contribution\begin{eqnarray}
\frac{G}{2}\int_{\xi}^{\min(Y-\xi,Y-y_{{\rm gap}})}\! dy_{1}\int\! d^{2}b_{1}\;\left(1-e^{-\chi_{j_{1}k}^{{\rm net}}(Y-y_{1},\vec{b}-\vec{b}_{1}|Y,\vec{b})}\right)\left(1-e^{-\chi_{j_{2}k}^{{\rm net}}(Y-y_{1},\vec{b}-\vec{b}_{1}|Y,\vec{b})}\right)\nonumber \\
\times\left(1-e^{-\chi_{kj_{1}}^{{\rm net}}(y_{1},\vec{b}_{1}|Y,\vec{b})}\right)\left(1-e^{-\chi_{kj_{2}}^{{\rm net}}(y_{1},\vec{b}_{1}|Y,\vec{b})}\right).\label{eq:DPE-LM}\end{eqnarray}

All the other graphs in Fig.~\ref{fig: tree-dif1} are much less
important, the corresponding contributions being proportional to the
third or higher power of the triple-Pomeron coupling. Indeed, the
diagrams in Fig.~\ref{fig: tree-dif1}~(d-k) contain $\bar{m}\geq2$
diffractively cut projectile net-fans whereas the one in Fig.~\ref{fig: tree-dif1}~(l)
contains $\bar{n}\geq2$ cut target net-fans with uncut handles; each
of these sub-graphs contains at least one internal multi-Pomeron vertex.
By consequence, the graphs of Fig.~\ref{fig: tree-dif1}~(d-l) are
sub-dominant at large impact parameters whereas at small ones they
are suppressed by exponential factors (see the discussion in Section
\ref{sub:Simplest-enhanced-graphs}). Calculating their contributions
and adding them to (\ref{eq:1gap-main}-\ref{eq:DPE-LM}), we obtain\begin{eqnarray}
2\chi_{j_{1}j_{2}k}^{1-{\rm gap}}(Y,b,y_{{\rm gap}})
+2\chi_{j_{1}j_{2}kk}^{2-{\rm gap}}(Y,b,y_{{\rm gap}},\xi)
=G\int_{\xi}^{\min(Y-\xi,Y-y_{{\rm gap}})}\! dy_{1}\int\! d^{2}b_{1}\nonumber \\
\times\left\{ \left[\left(1-e^{-\chi_{j_{1}k}^{{\rm net}}(Y-y_{1},\vec{b}-\vec{b}_{1}|Y,\vec{b})}\right)
\left(1-e^{-\chi_{j_{2}k}^{{\rm net}}(Y-y_{1},\vec{b}-\vec{b}_{1}|Y,\vec{b})}\right)
+\left(e^{2\bar{\chi}_{j_{1}j_{2}kk}^{{\rm D}}(Y-y_{1},\vec{b}-\vec{b}_{1},y_{{\rm gap}}|Y,\vec{b})}\right.
\right.\right.\nonumber \\
-\left.\left.1-2\bar{\chi}_{j_{1}j_{2}kk}^{{\rm D}}(Y-y_{1},\vec{b}-\vec{b}_{1},y_{{\rm gap}}|Y,\vec{b})\right)
e^{-\chi_{j_{1}k}^{{\rm net}}(Y-y_{1},\vec{b}-\vec{b}_{1}|Y,\vec{b})
-\chi_{j_{2}k}^{{\rm net}}(Y-y_{1},\vec{b}-\vec{b}_{1}|Y,\vec{b})}\right]\nonumber \\
\times\left(1-e^{-\hat{\chi}_{kj_{1}j_{2}}^{{\rm fan}}(y_{1},\vec{b}_{1}|Y,\vec{b})}\right)
-\frac{1}{2}\left(e^{-\chi_{kj_{1}}^{{\rm net}}(y_{1},\vec{b}_{1}|Y,\vec{b})}
-e^{-\chi_{kj_{2}}^{{\rm net}}(y_{1},\vec{b}_{1}|Y,\vec{b})}\right)\nonumber \\
\times\left[\left(e^{\tilde{\chi}_{j_{1}j_{2}kk}^{{\rm D(l)}}(Y-y_{1},\vec{b}-\vec{b}_{1},y_{{\rm gap}}|Y,\vec{b})}
-1-\tilde{\chi}_{j_{1}j_{2}kk}^{{\rm D(l)}}(Y-y_{1},\vec{b}-\vec{b}_{1},y_{{\rm gap}}|Y,\vec{b})\right)
e^{-\chi_{j_{1}k}^{{\rm net}}(Y-y_{1},\vec{b}-\vec{b}_{1}|Y,\vec{b})}\right.\nonumber \\
-\left.\left(e^{\tilde{\chi}_{j_{1}j_{2}kk}^{{\rm D(r)}}(Y-y_{1},\vec{b}-\vec{b}_{1},y_{{\rm gap}}|Y,\vec{b})}
-1-\tilde{\chi}_{j_{1}j_{2}kk}^{{\rm D(r)}}(Y-y_{1},\vec{b}-\vec{b}_{1},y_{{\rm gap}}|Y,\vec{b})\right)
e^{-\chi_{j_{2}k}^{{\rm net}}(Y-y_{1},\vec{b}-\vec{b}_{1}|Y,\vec{b})}\right]\nonumber \\
+\left(1-e^{-\chi_{j_{1}k}^{{\rm net}}(Y-y_{1},\vec{b}-\vec{b}_{1}|Y,\vec{b})}\right)
\left(e^{\tilde{\chi}_{kj_{1}j_{2}}^{{\rm fan(l)}}(y_{1},\vec{b}_{1}|Y,\vec{b})}-1
-\tilde{\chi}_{kj_{1}j_{2}}^{{\rm fan(l)}}(y_{1},\vec{b}_{1}|Y,\vec{b})\right)
e^{-\chi_{kj_{1}}^{{\rm net}}(y_{1},\vec{b}_{1}|Y,\vec{b})}\nonumber \\
+\left.\left(1-e^{-\chi_{j_{2}k}^{{\rm net}}(Y-y_{1},\vec{b}-\vec{b}_{1}|Y,\vec{b})}\right)
\left(e^{\tilde{\chi}_{kj_{1}j_{2}}^{{\rm fan(r)}}(y_{1},\vec{b}_{1}|Y,\vec{b})}-1
-\tilde{\chi}_{kj_{1}j_{2}}^{{\rm fan(r)}}(y_{1},\vec{b}_{1}|Y,\vec{b})\right)
e^{-\chi_{kj_{2}}^{{\rm net}}(y_{1},\vec{b}_{1}|Y,\vec{b})}\right\} \label{eq:1-gap-tot}
\end{eqnarray}

\begin{figure}[t]
\begin{centering}
\includegraphics[width=15cm,height=6cm]{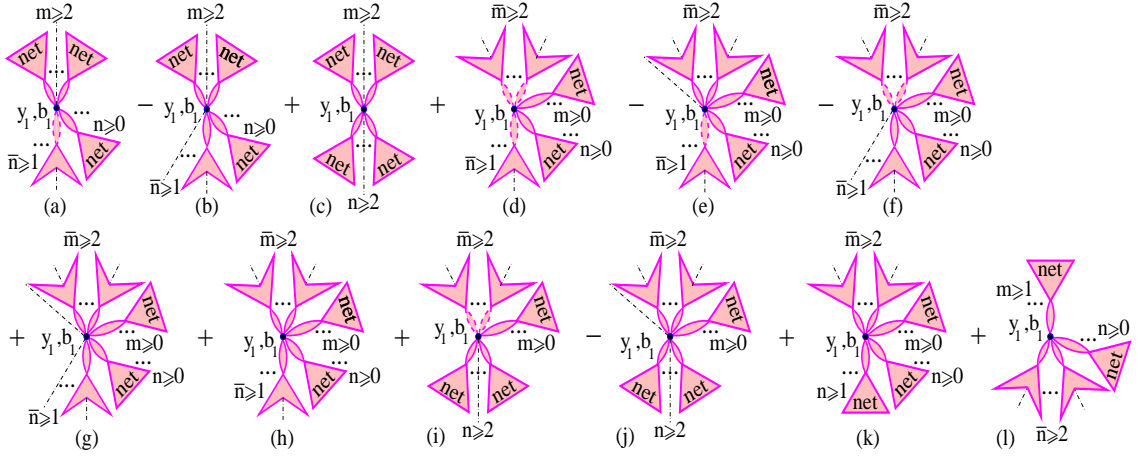}
\par\end{centering}

\caption{Complete set of irreducible cut diagrams corresponding to central
diffraction topology of the final state. \label{fig: tree-dif2}}

\end{figure}

Replacing now in Fig.~\ref{fig: tree-dif1} the contributions of
fan-like cuts of target net-fans $2\bar{\chi}^{{\rm fan}}$, $\tilde{\chi}^{{\rm fan(l/r)}}$
by the ones of the diffractive cuts $2\bar{\chi}^{{\rm D}}$, $\tilde{\chi}^{{\rm D(l/r)}}$,
we obtain the complete set of irreducible cut diagrams for central
diffraction, as depicted in Fig.~\ref{fig: tree-dif2}, %
 with the contribution\begin{eqnarray}
2\chi_{j_{1}j_{2}k_{1}k_{2}}^{2-{\rm gap}}(Y,b,y_{{\rm gap}}^{{\rm (f)}},y_{{\rm gap}}^{{\rm (b)}})=\frac{G}{2}\int_{\max(\xi,y_{{\rm gap}}^{{\rm (b)}})}^{\min(Y-\xi,Y-y_{{\rm gap}}^{{\rm (f)}})}\! dy_{1}\int\! d^{2}b_{1}\nonumber \\
\times\left\{ \left[\left(1-e^{-\chi_{j_{1}k_{1}}^{{\rm net}}(Y-y_{1},\vec{b}-\vec{b}_{1}|Y,\vec{b})}\right)\left(1-e^{-\chi_{j_{2}k_{2}}^{{\rm net}}(Y-y_{1},\vec{b}-\vec{b}_{1}|Y,\vec{b})}\right)+\left(e^{2\bar{\chi}_{j_{1}j_{2}k_{1}k_{2}}^{{\rm D}}(Y-y_{1},\vec{b}-\vec{b}_{1},y_{{\rm gap}}^{{\rm (f)}}|Y,\vec{b})}\right.\right.\right.\nonumber \\
-\left.\left.1-2\bar{\chi}_{j_{1}j_{2}k_{1}k_{2}}^{{\rm D}}(Y-y_{1},\vec{b}-\vec{b}_{1},y_{{\rm gap}}^{{\rm (f)}}|Y,\vec{b})\right)e^{-\chi_{j_{1}k_{1}}^{{\rm net}}(Y-y_{1},\vec{b}-\vec{b}_{1}|Y,\vec{b})-\chi_{j_{2}k_{2}}^{{\rm net}}(Y-y_{1},\vec{b}-\vec{b}_{1}|Y,\vec{b})}\right]\nonumber \\
\times\left[1-e^{\tilde{\chi}_{k_{1}k_{2}j_{1}j_{2}}^{{\rm D(l)}}(y_{1},\vec{b}_{1},y_{{\rm gap}}^{{\rm (b)}}|Y,\vec{b})-\chi_{k_{1}j_{1}}^{{\rm net}}(y_{1},\vec{b}_{1}|Y,\vec{b})}-e^{\tilde{\chi}_{k_{1}k_{2}j_{1}j_{2}}^{{\rm D(r)}}(y_{1},\vec{b}_{1},y_{{\rm gap}}^{{\rm (b)}}|Y,\vec{b})-\chi_{k_{2}j_{2}}^{{\rm net}}(y_{1},\vec{b}_{1}|Y,\vec{b})}\right.\nonumber \\
+\left.e^{2\bar{\chi}_{k_{1}k_{2}j_{1}j_{2}}^{{\rm D}}(y_{1},\vec{b}_{1},y_{{\rm gap}}^{{\rm (b)}}|Y,\vec{b})-\chi_{k_{1}j_{1}}^{{\rm net}}(y_{1},\vec{b}_{1}|Y,\vec{b})-\chi_{k_{2}j_{2}}^{{\rm net}}(y_{1},\vec{b}_{1}|Y,\vec{b})}\right]\nonumber \\
+\left(e^{\tilde{\chi}_{j_{1}j_{2}k_{1}k_{2}}^{{\rm D(l)}}(Y-y_{1},\vec{b}-\vec{b}_{1},y_{{\rm gap}}^{{\rm (f)}}|Y,\vec{b})}-1-\tilde{\chi}_{j_{1}j_{2}k_{1}k_{2}}^{{\rm D(l)}}(Y-y_{1},\vec{b}-\vec{b}_{1},y_{{\rm gap}}^{{\rm (f)}}|Y,\vec{b})\right)e^{-\chi_{j_{1}k_{1}}^{{\rm net}}(Y-y_{1},\vec{b}-\vec{b}_{1}|Y,\vec{b})}\nonumber \\
\times\left[1-e^{-\chi_{k_{1}j_{1}}^{{\rm net}}(y_{1},\vec{b}_{1}|Y,\vec{b})}+e^{-\chi_{k_{2}j_{2}}^{{\rm net}}(y_{1},\vec{b}_{1}|Y,\vec{b})}-e^{2\bar{\chi}_{k_{1}k_{2}j_{1}j_{2}}^{{\rm D}}(y_{1},\vec{b}_{1},y_{{\rm gap}}^{{\rm (b)}}|Y,\vec{b})-\chi_{k_{1}j_{1}}^{{\rm net}}(y_{1},\vec{b}_{1}|Y,\vec{b})-\chi_{k_{2}j_{2}}^{{\rm net}}(y_{1},\vec{b}_{1}|Y,\vec{b})}\right]\nonumber \\
+\left(e^{\tilde{\chi}_{j_{1}j_{2}k_{1}k_{2}}^{{\rm D(r)}}(Y-y_{1},\vec{b}-\vec{b}_{1},y_{{\rm gap}}|Y,\vec{b})}-1-\tilde{\chi}_{j_{1}j_{2}k_{1}k_{2}}^{{\rm D(r)}}(Y-y_{1},\vec{b}-\vec{b}_{1},y_{{\rm gap}}|Y,\vec{b})\right)e^{-\chi_{j_{2}k_{2}}^{{\rm net}}(Y-y_{1},\vec{b}-\vec{b}_{1}|Y,\vec{b})}\nonumber \\
\times\left[1+e^{-\chi_{k_{1}j_{1}}^{{\rm net}}(y_{1},\vec{b}_{1}|Y,\vec{b})}-e^{-\chi_{k_{2}j_{2}}^{{\rm net}}(y_{1},\vec{b}_{1}|Y,\vec{b})}-e^{2\bar{\chi}_{k_{1}k_{2}j_{1}j_{2}}^{{\rm D}}(y_{1},\vec{b}_{1},y_{{\rm gap}}^{{\rm (b)}}|Y,\vec{b})-\chi_{k_{1}j_{1}}^{{\rm net}}(y_{1},\vec{b}_{1}|Y,\vec{b})-\chi_{k_{2}j_{2}}^{{\rm net}}(y_{1},\vec{b}_{1}|Y,\vec{b})}\right]\nonumber \\
+2\left(1-e^{-\chi_{j_{1}k_{1}}^{{\rm net}}(Y-y_{1},\vec{b}-\vec{b}_{1}|Y,\vec{b})}\right)\left(e^{\tilde{\chi}_{k_{1}k_{2}j_{1}j_{2}}^{{\rm D(l)}}(y_{1},\vec{b}_{1},y_{{\rm gap}}^{{\rm (b)}}|Y,\vec{b})}-1-\tilde{\chi}_{k_{1}k_{2}j_{1}j_{2}}^{{\rm D(l)}}(y_{1},\vec{b}_{1},y_{{\rm gap}}^{{\rm (b)}}|Y,\vec{b})\right)\nonumber \\
\times\: e^{-\chi_{k_{1}j_{1}}^{{\rm net}}(y_{1},\vec{b}_{1}|Y,\vec{b})}+2\left(1-e^{-\chi_{j_{2}k_{2}}^{{\rm net}}(Y-y_{1},\vec{b}-\vec{b}_{1}|Y,\vec{b})}\right)\nonumber \\
\times\left.\left(e^{\tilde{\chi}_{k_{1}k_{2}j_{1}j_{2}}^{{\rm D(r)}}(y_{1},\vec{b}_{1},y_{{\rm gap}}^{{\rm (b)}}|Y,\vec{b})}-1-\tilde{\chi}_{k_{1}k_{2}j_{1}j_{2}}^{{\rm D(r)}}(y_{1},\vec{b}_{1},y_{{\rm gap}}^{{\rm (b)}}|Y,\vec{b})\right)e^{-\chi_{k_{2}j_{2}}^{{\rm net}}(y_{1},\vec{b}_{1}|Y,\vec{b})}\right\} .\label{eq:2-gap-tot}\end{eqnarray}

In a similar way one can obtain the set of irreducible cut graphs
which produce a central rapidity gap. For brevity, we give
here the corresponding contribution without derivation:
\begin{eqnarray}
2\chi_{jk}^{{\rm c-gap}}(Y,b,y_{{\rm gap}})=
\frac{G^{2}}{4}\int_{\max(\xi+y_{{\rm gap}},2\xi)}^{Y-\xi}\! dy_{1}
\int_{\xi}^{y_{1}-\max(y_{{\rm gap}},\xi)}\! dy_{2}
\int\! d^{2}b_{1}\, d^{2}b_{2}\nonumber \\
\times\left\{ \left(1-e^{-\chi^{{\rm loop}}(y_{1}-y_{2},|\vec{b}_{1}-\vec{b}_{2}|)}\right)^{2}
\left[1-e^{2\bar{\chi}_{jjkk}^{{\rm D}}(1)-2\chi_{jk}^{{\rm net}}(1)}
-2\left(e^{-\hat{\chi}_{jkk}^{{\rm fan}}(1)}
-e^{\tilde{\chi}_{jjkk}^{{\rm D(l)}}(1)-\chi_{jk}^{{\rm net}}(1)}\right)
\right]\right.\nonumber \\
\times\left[1-e^{2\bar{\chi}_{kkjj}^{{\rm D}}(2)-2\chi_{kj}^{{\rm net}}(2)}-2\left(e^{-\hat{\chi}_{kjj}^{{\rm fan}}(2)}-e^{\tilde{\chi}_{kkjj}^{{\rm D(l)}}(2)-\chi_{kj}^{{\rm net}}(2)}\right)\right]e^{-2\chi_{jk}^{{\rm net}}(2)-2\chi_{kj}^{{\rm net}}(1)}\nonumber \\
-2\left(1-e^{-\chi^{{\rm loop}}(y_{1}-y_{2},|\vec{b}_{1}-\vec{b}_{2}|)}\right)
\left[\left(1-e^{2\bar{\chi}_{jjkk}^{{\rm D}}(1)-2\chi_{jk}^{{\rm net}}(1)}\right)
e^{-\chi_{kj}^{{\rm net}}(1)}\left(1-e^{-\chi_{kj}^{{\rm net}}(1)}\right)\right.\nonumber \\
+\left.2\left(e^{-\hat{\chi}_{jkk}^{{\rm fan}}(1)}
-e^{\tilde{\chi}_{jjkk}^{{\rm D(l)}}(1)-\chi_{jk}^{{\rm net}}(1)}\right)
e^{-2\chi_{kj}^{{\rm net}}(1)}-2\left(\chi_{jk}^{{\rm net}}(1)
-\hat{\chi}_{jkk}^{{\rm fan}}(1)-\tilde{\chi}_{jjkk}^{{\rm D(l)}}(1)\right)\right]\nonumber \\
\times\left[\left(1-e^{2\bar{\chi}_{kkjj}^{{\rm D}}(2)-2\chi_{kj}^{{\rm net}}(2)}\right)
e^{-\chi_{jk}^{{\rm net}}(2)}\left(1-e^{-\chi_{jk}^{{\rm net}}(2)}\right)
+2\left(e^{-\hat{\chi}_{kjj}^{{\rm fan}}(2)}
-e^{\tilde{\chi}_{kkjj}^{{\rm D(l)}}(2)-\chi_{kj}^{{\rm net}}(2)}\right)\right.\nonumber \\
\times\left.\left.e^{-2\chi_{jk}^{{\rm net}}(2)}
-2\left(\chi_{kj}^{{\rm net}}(2)-\hat{\chi}_{kjj}^{{\rm fan}}(2)
-\tilde{\chi}_{kkjj}^{{\rm D(l)}}(2)\right)\right]\right\} .\label{eq:DPE-tot}
\end{eqnarray}
Here the arguments of the eikonals are understood as $\chi_{jk}^{{\rm net}}(i)=\chi_{jk}^{{\rm net}}(Y-y_{i},\vec{b}-\vec{b}_{i}|Y,\vec{b})$,
$\chi_{kj}^{{\rm net}}(i)=\chi_{kj}^{{\rm net}}(y_{i},\vec{b}_{i}|Y,\vec{b})$,
$i=1,2,$ and similarly for $\hat{\chi}^{{\rm fan}}$, $\bar{\chi}^{{\rm D}}$,
$\tilde{\chi}^{{\rm D(l)}}$. Keeping only the lowest order contributions
($\chi_{jk}^{{\rm net}},\hat{\chi}_{jkk}^{{\rm fan}}\rightarrow\chi_{j}^{\mathbb{P}}$;
$\chi_{kj}^{{\rm net}},\hat{\chi}_{kjj}^{{\rm fan}}\rightarrow\chi_{k}^{\mathbb{P}}$;
$\chi^{{\rm loop}}\rightarrow\chi^{\mathbb{P}}$; $\bar{\chi}^{{\rm D}},\tilde{\chi}^{{\rm D}}\rightarrow0$),
Eq.~(\ref{eq:DPE-tot}) reduces to (\ref{eq:c-gap}).

\begin{figure}[htb]
\begin{centering}
\includegraphics[width=15cm,height=4cm]{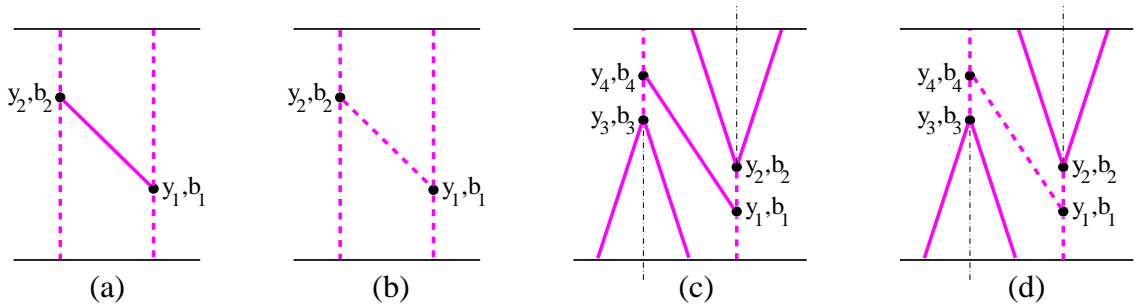}
\par\end{centering}

\caption{Examples of zigzag-like cut diagrams: lowest order ones (a,b) and
the simplest graphs which give non-zero corrections to diffractive
cross sections (c,d). \label{fig:zigzag}}

\end{figure}

Let us now briefly discuss zigzag-like cut diagrams. While it has
been shown in \cite{ost08} that their contributions to the total
cross section precisely cancel each other, it may not be obvious that
they do not contribute to various diffractive topologies of final
states. As an illustration, let us consider the two lowest order diagrams
of that kind, which are depicted in Fig.~\ref{fig:zigzag}~(a,b),%
 whose contributions are equal up to a sign. The diagram in Fig.~\ref{fig:zigzag}~(a)
provides a (negative) screening correction to the eikonal configuration
with two cut Pomerons. On the other hand, the one in Fig.~\ref{fig:zigzag}~(b)
introduces a new process, with the weight being equal to the one of
the mentioned screening contribution, and with the particle production
pattern being almost identical to the one of Fig.~\ref{fig:zigzag}~(a);
the only difference arises from the cut Pomeron exchanged between
the vertices ($y_{1},b_{1}$) and ($y_{2},b_{2}$). Thus, the combined
effect of these two graphs is to provide additional particle production
in the rapidity interval $[y_{1},y_{2}]$. However, this interval
is already covered by particles, which result from the left-most cut
Pomeron in the two graphs. Thus, the  rapidity gap structure of the
event remains unchanged. The simplest diagrams which provide non-zero
corrections to diffractive cross sections are the ones in Fig.~\ref{fig:zigzag}~(c,d).
The two contributions are again equal up to a sign, with the graph
in Fig.~\ref{fig:zigzag}~(c) providing a screening correction to
double high mass diffraction cross section, with a LRG between $y_{2}$
and $y_{3}$, and with the diagram in Fig.~\ref{fig:zigzag}~(d)
introducing additional particle production in the interval $[y_{1},y_{4}]$,
which covers the discussed rapidity gap. However, being proportional
to the 4th power of the triple-Pomeron coupling, the corresponding
contributions give a negligible correction to $\sigma_{{\rm HM}}^{{\rm DD}}$.

\end{document}